\documentclass[10pt,journal]{IEEEtran}
\pdfoutput=1

\usepackage{mathrsfs}
\usepackage{cite} 
\usepackage{url}
\usepackage{epsfig}
\usepackage{amsmath}
\usepackage{amssymb}
\usepackage{amsfonts}
\usepackage{graphicx}
\usepackage{subfigure}
\usepackage{wrapfig}
\usepackage{hyperref}
\usepackage{capt-of}
\usepackage{stfloats}
\usepackage{balance}

\newtheorem{theorem}{Theorem}
\newtheorem{lemma}{Lemma}

\newtheorem{definition}{Definition}

\DeclareMathOperator*{\argmin}{arg\,min}

\renewcommand{\thetable}{\arabic{table}}
\newcommand{\st}{\textsf{s}}
\newcommand{\ct}{\textsf{c}}

\newcommand{\ts}{\textbf{\emph{t}}}

\allowdisplaybreaks    % TO ALLOW EQNARRAY TO FLOW ACROSS PAGES

\begin{document}
\title{Competitive Selection of Ephemeral Relays in Wireless Networks}

\author{K.~P.~Naveen, 
Eitan~Altman,~\IEEEmembership{Fellow,~IEEE,} and
Anurag~Kumar,~\IEEEmembership{Fellow,~IEEE} 
\thanks{Author's addresses: K.~P.~Naveen, INRIA Saclay, Palaiseau 91120, France (email: naveenkp@inria.fr);
Anurag Kumar, Department of Electrical Communication
Engineering, Indian Institute of Science, Bangalore 560012, India (email: anurag@ece.iisc.ernet.in);
Eitan Altman, INRIA, Sophia Antipolis 06902, France (email: Eitan.Altman@inria.fr).}
\thanks{This work was done when the first author was a Ph.D student at the Department of Electrical Communication
Engineering, Indian Institute of Science, Bangalore 560012, India.}
\thanks{This work was supported by the following projects:
Indo-French Center for the Promotion of Advanced Research under Project 4000IT1,
and the IFCAM (Indo-French Centre for Applied Mathematics) program;
Department of Science and Technology (DST) through project WINSON (Wireless Networks and Techniques with
Applications to Social Needs) and a J.C.~Bose National Fellowship; and the INRIA Associates Project GANESH
(Games Optimization and Analysis of Networks: Theory and Applications).}
}
\maketitle

\begin{abstract}
We consider a setting in which two nodes (referred to as forwarders) compete
to choose a relay node from a set of relays, as they ephemerally become available 
(e.g., wake up from a sleep state).
Each relay, when it arrives, offers a 
(possibly different) ``reward'' to each forwarder.
Each forwarder's objective is to minimize a combination of the 
delay incurred  in choosing a relay and the {reward} offered by the chosen relay. 
As an example, we develop the reward structure for the specific problem of 
geographical forwarding over a network of sleep-wake cycling relays.

We study two variants of the generic relay selection problem, namely, the completely observable (CO) case where,
when a relay arrives, both forwarders get to observe both rewards, and  
the partially observable (PO) case where each forwarder can only observe its own reward.
Formulating the problem as a two person stochastic game, 
we characterize solution in terms of Nash Equilibrium Policy Pairs (NEPPs).
For the CO case we provide a general structure of the NEPPs. For the PO case 
we prove that there exists an NEPP within the class of threshold policy pairs.

We then consider the particular application of geographical forwarding of packets 
in a shared network of sleep-wake cycling wireless relays. 
For this problem, for a particular reward structure, using realistic parameter values corresponding to
TelosB wireless mote, we numerically compare the 
performance (in terms of cost to both forwarders) of the various NEPPs and draw the following key insight: 
even for moderate separation between the two forwarders, the performance of the various NEPPs is close to
the performance of a simple strategy where each forwarder behaves as if the 
other forwarder is not present. We  also conduct simulation experiments to study the end-to-end
performance of the simple forwarding policy.
\end{abstract}


\section{Introduction}
We are concerned in this paper with a class of resource allocation problems in wireless networks, 
in which competing nodes need to acquire a resource, such as a physical radio relay (see the geographical forwarding example later in 
Section~\ref{geographical_example_section}) or 
a channel (as in a cognitive radio network \cite{yao-etal12competition-channel-selection,niyato-hossain08competitive-spectrum-sharing}), 
when a sequence of such resources ``arrive'' over time, and are available only 
fleetingly for acquisition. 
In this paper, formulating such a problem for two nodes as a stochastic game, we 
consider the completely observable and partially observable cases, and provide characterizations of the 
Nash Equilibrium Policy Pairs (NEPP). We provide numerical results, and insights therefrom, for a specific
reward structure derived from the problem of geographical forwarding in 
sleep-wake cycling networks.

\textbf{\emph{The Geographical Forwarding Context:}} With the increasing importance of 
``smart'' utilization of our limited resources (e.g., energy and clean water) there is a need for instrumenting our buildings 
and campuses with wireless sensor networks. As awareness grows and sensing technologies emerge, new applications 
will be implemented. While each application will require different sensors and back-end analytics, the availability of a 
common wireless network infrastructure will promote the quick deployment of new applications. One approach 
for building such an infrastructure, say, in a large building setting, would be to deploy a large number of  
relay nodes, and employ the idea of geographical forwarding.  If the phenomena to be monitored are slowly varying over time, 
the traffic on the network can be assumed to be light. In addition, such applications are \emph{delay tolerant,} 
thus accommodating the approach of \emph{opportunistic} geographical forwarding over \emph{sleep-wake cycling} 
networks \cite{kim-etal11optimal-anycast,naveen-kumar12relay-selection-TMC}.

Sleep-wake cycling is an approach whereby, to conserve the relay battery power, 
their radios are kept turned OFF, while coming ON periodically to provide opportunities for packet 
forwarding. The problem of forwarding in such a setting was explored in \cite{kim-etal11optimal-anycast,naveen-kumar12relay-selection-TMC}, 
where the formulation was limited to a 
\emph{single} alarm packet flowing through the network. Whereas the emphasis in \cite{kim-etal11optimal-anycast} 
was to develop an 
end-to-end optimal forwarding algorithm, thus requiring a global organization step, in \cite{naveen-kumar12relay-selection-TMC},
which is our prior work on this problem, we sought a locally-optimal forwarding heuristic.
End-to-end forwarding was achieved by applying the local heuristic at each forwarding step. 
We found that, over certain range of operation, the performance obtained by the heuristic 
is comparable with the 
optimal solution provided by \cite{kim-etal11optimal-anycast}.  

In the setting discussed above, even though the 
traffic is light, there is still a chance that there is more than one 
forwarder seeking a relay from among a set of potential relays.
There then arises the problem of assigning the 
relays, as they wake-up, to one or the other of the forwarders.  This, thus, is an extension of 
the local forwarding problem discussed in \cite{naveen-kumar12relay-selection-TMC}.
Formally, the local forwarding problem we consider in this paper is
the following. There are two forwarders each of which has to choose
a relay node to forward its packet to. The relays are waking up sequentially over time. 
Whenever a relay wakes up, each forwarder first evaluates the relay based on a 
reward metric (which could be a function of the progress, towards the sink, made by the relay,
and the power required to get the packet to the relay \cite{naveen-kumar12relay-selection-TMC}),
and then decides whether to compete (with the other forwarder) for this relay
or continue to wait for further relays to wake-up. Such a geographical forwarding setting 
will serve as an example application of the stochastic game formulation developed in this paper.

\textbf{\emph{Outline and Our Contributions:}} We will describe a general
system model in Section~\ref{system_model_section}, following which, in Section~\ref{geographical_example_section},
we will discuss a geographical forwarding problem as an example.
Related work will be presented in Section~\ref{related_work_section}.
In Sections~\ref{completely_observable_case_section} and 
\ref{partially_observable_case_section}
we will study two variants of the problem (of progressive complexity), namely, one where
\emph{complete information} is available to both forwarders 
and one with only \emph{partial information}. 
We will use stochastic game theory to obtain solution in terms
of (stationary) Nash Equilibrium Policy Pairs (NEPPs). 
We will briefly study a cooperative setting in Section~\ref{cooperative_case_section}, and obtain
the Pareto optimal performance curve which provides a benchmark for the NEPPs.
The following are our main technical contributions:
\begin{itemize}
\item For the problem with complete information we obtain results illustrating
the structure of NEPPs (Theorem~\ref{nepp_structure_theorem})
\item For the partial information case we prove the existence of a NE strategy 
(for a certain Bayesian game) within the class of threshold strategies
(Theorem~\ref{NE_existance_theorem}). This result will enable
us to construct NEPPs for this case.
\item In Section~\ref{numerical_work_section} we provide a simulation study of 
the use of our formulation in the context of geographical forwarding. Using realistic parameters from the popular TelosB wireless 
mote, we make the following interesting observation:
even for moderate separation between the two forwarders, the performance
of all the NEPPs is close to the performance of a \emph{simple strategy} where each forwarder
behaves as if it is alone. 
\end{itemize}
We will finally draw our conclusions in Section~\ref{conclusion_section}. 
For the ease of presentation we have moved most of the proofs to the Appendix.

\section{System Model}
\label{system_model_section}
Let $\mathscr{F}_1$ and $\mathscr{F}_2$ denote the two competing nodes (i.e., players in 
game theoretic terms), referred to as the \emph{forwarders}. We will assume that there
are an infinite number of \emph{relay nodes} (or \emph{resources} in general) that are arriving sequentially at times $\{W_k:k\ge0\}$, 
which are the {points of a Poisson process} of rate $\frac{1}{\tau}$. Thus, the \emph{inter-``arrival''} times between 
successive relays, $U_k:=W_k-W_{k-1}$, are i.i.d.\ (independent and identically distributed) 
exponential random variables of mean $\tau$. We will refer to the relay that arrives at the instant $W_k$ as the \emph{$k$-th relay}.
Further, the $k$-th relay is only ephemerally available at the instant $W_k$.

When a relay arrives, either of the forwarders can compete for it, 
thereby obtaining a \emph{reward}.
Let $R_{\rho,k}$, $\rho=1,2$, denote the \emph{reward} offered by the $k$-th relay 
to $\mathscr{F}_\rho$ (an example reward structure will be discussed in Section~\ref{geographical_example_section}). 
The rewards $R_{\rho,k}$ ($\rho=1,2$; $k\ge1$) can take values
from a finite set $\mathcal{R}=\{r_1,r_2,\cdots,r_n\}$, where $r_1=-\infty$
and $r_i<r_j$ for $i<j$. The reward pairs $(R_{1,k},R_{2,k})$ are i.i.d.\ across
$k$, with their common joint p.m.f.\ (probability mass function) being $p_{R_1,R_2}(\cdot,\cdot)$,
For notational simplicity we will denote $p_{R_1,R_2}(r_i,r_j)$ 
as simply $p_{i,j}$. 
Further, let $p^{(1)}_i$ and $p^{(2)}_j$ denote the marginal p.m.f.s of $R_{1,k}$ and
$R_{2,k}$, respectively. Thus, $p^{(1)}_i=\sum_{j=1}^n p_{i,j}$ and $p^{(2)}_j=\sum_{i=1}^n p_{i,j}$.

\textbf{\emph{Actions and Consequences:}}
First we will study (in Section~\ref{completely_observable_case_section}) 
a \emph{completely observable case} where the reward pair
$(R_{1,k},R_{2,k})$ is revealed to both the forwarders. 
Later, in Section~\ref{partially_observable_case_section}, we will consider a more 
involved (albeit more practical) \emph{partially observable case} where only $R_{1,k}$ is revealed 
to $\mathscr{F}_1$, and $R_{2,k}$ is revealed to $\mathscr{F}_2$.
However in either case, each time a relay arrives, the two forwarders have to independently 
choose between one of the following actions: 
\begin{itemize}
\item \textsf{s}: \emph{stop} and forward the packet to the current relay, or
\item \textsf{c}: \emph{continue} to wait for further relays to arrive.
\end{itemize}

Suppose both forwarders choose to stop, then with probability (w.p.) $\nu_1$,
$\mathscr{F}_1$ gets the relay in which case $\mathscr{F}_2$ has to continue alone, 
while with the remaining probability
($\nu_2=1-\nu_1$) $\mathscr{F}_2$ gets the relay and $\mathscr{F}_1$ continues alone. 
$\nu_\rho$ ($\rho=1,2$) could be thought of as 
the probability that $\mathscr{F}_\rho$ will win the contention
when both forwarders attempt simultaneously.  For mathematical tractability we will 
assume that the forwarders make their decision instantaneously at 
the relay arrival instants. Further, if a relay is not chosen by either 
forwarder (i.e., if both forwarders choose to continue)
we will assume that the relay disappears and is not available for further use.

\textbf{\emph{System State and Forwarding Policy:}}
For the CO case,
$(R_{1,k},R_{2,k})$ can be regarded as the \emph{state} of the system at stage $k$,
provided both forwarders have not \emph{terminated} (i.e., chosen a relay) yet. 
When one of the forwarder, say $\mathscr{F}_2$, terminates,
we will represent the system state as $(R_{1,k},\textbf{\emph{t}})$.
Similarly, let $(\ts,R_{2,k})$ and $(\ts,\ts)$
represents the state of the system when only $\mathscr{F}_1$ has terminated and when
both forwarders have terminated, respectively.
Formally, we can define the \emph{state space} to be 
\begin{eqnarray}
\label{SS_CO_equn}
\mathcal{X}=\Big\{(r_i,r_j),(r_i,\ts),(\ts,r_j),(\ts,\ts):r_i,r_j\in\mathcal{R}\Big\}.
\end{eqnarray}

Given a discrete set $\mathcal{S}$, let 
$\Delta(\mathcal{S})$ denote the set of all p.m.f.s\ on $\mathcal{S}$. 
We now have the following definition.
\begin{definition}
\label{policy_defn}
A \emph{forwarding policy} $\pi$ is a mapping,
$\pi:\mathcal{X}\rightarrow \Delta(\{\mbox{\textsf{s},\textsf{c}}\})$, such that 
$\mathscr{F}_1$ (or $\mathscr{F}_2$) using $\pi$ will choose action $\textsf{s}$ or $\textsf{c}$
according to the p.m.f.\  $\pi(x_k)$ when 
the state of the system at stage $k\ge1$ is $x_k\in\mathcal{X}$.  
A \emph{policy pair}
$(\pi_1,\pi_2)$ is a tuple of policies such that $\mathscr{F}_1$ 
uses $\pi_1$ and $\mathscr{F}_2$ uses $\pi_2$.
\end{definition}

Note that we have restricted to the class of stationary policies only.
We will denote this class of policies as $\Pi_S$. 

\textbf{\emph{Problem Formulation:}}
Suppose the forwarders use a policy pair $(\pi_1,\pi_2)$, and 
let $x\in\mathcal{X}$ be the state of the system at stage $1$. 
Let $K_\rho$, $\rho=1,2$, 
denote the (random) stage at which $\mathscr{F}_\rho$ forwards its packet. 
Then, the delay incurred by $\mathscr{F}_\rho$ ($\rho=1,2$), starting from 
the instant $W_1=U_1$ (first relay's arrival instant), is 
$D_{K_\rho}=U_2+\cdots+U_{K_\rho}$, and the reward
accrued is $R_{\rho,K_\rho}$. Let $\mathbb{E}_{\pi_1,\pi_2}^x[\cdot]$ denote the expectation operator
corresponding to the probability law, $\mathbb{P}_{\pi_1,\pi_2}^x$, governing
the system dynamics when the policy pair used is $(\pi_1,\pi_2)$ and the initial state is $x$.
Then, the expected total cost incurred by $\mathscr{F}_\rho$ is
\begin{eqnarray}
\label{cost1_equn}
J^{(\rho)}_{\pi_1,\pi_2}(x)=\mathbb{E}_{\pi_1,\pi_2}^x\Big[D_{K_\rho} - \eta_\rho R_{\rho,K_\rho}\Big],
\end{eqnarray}
where $\eta_\rho>0$ is the multiplier used to trade-off between delay and reward. 

\begin{definition}
\label{NEPP_defn}
We say that a policy pair $(\pi_1^*,\pi_2^*)$ is a Nash equilibrium policy pair (NEPP)
if, for all $x\in\mathcal{X}$, $J^{(1)}_{\pi_1^*,\pi_2^*}(x)\le J^{(1)}_{\pi_1,\pi_2^*}(x)$ 
for any policy $\pi_1\in\Pi_S$, and 
$J^{(2)}_{\pi_1^*,\pi_2^*}(x)\le J^{(2)}_{\pi_1^*,\pi_2}(x)$ for any policy $\pi_2\in\Pi_S$. 
Thus, a unilateral deviation
from an NEPP is neither beneficial for  $\mathscr{F}_1$ nor for $\mathscr{F}_2$.
\end{definition}

Our objective will be to characterize the solution in terms of NEPPs.

\section{Geographical Forwarding Example}
\label{geographical_example_section}
\begin{figure}[t]
\centering
\includegraphics[scale=0.45]{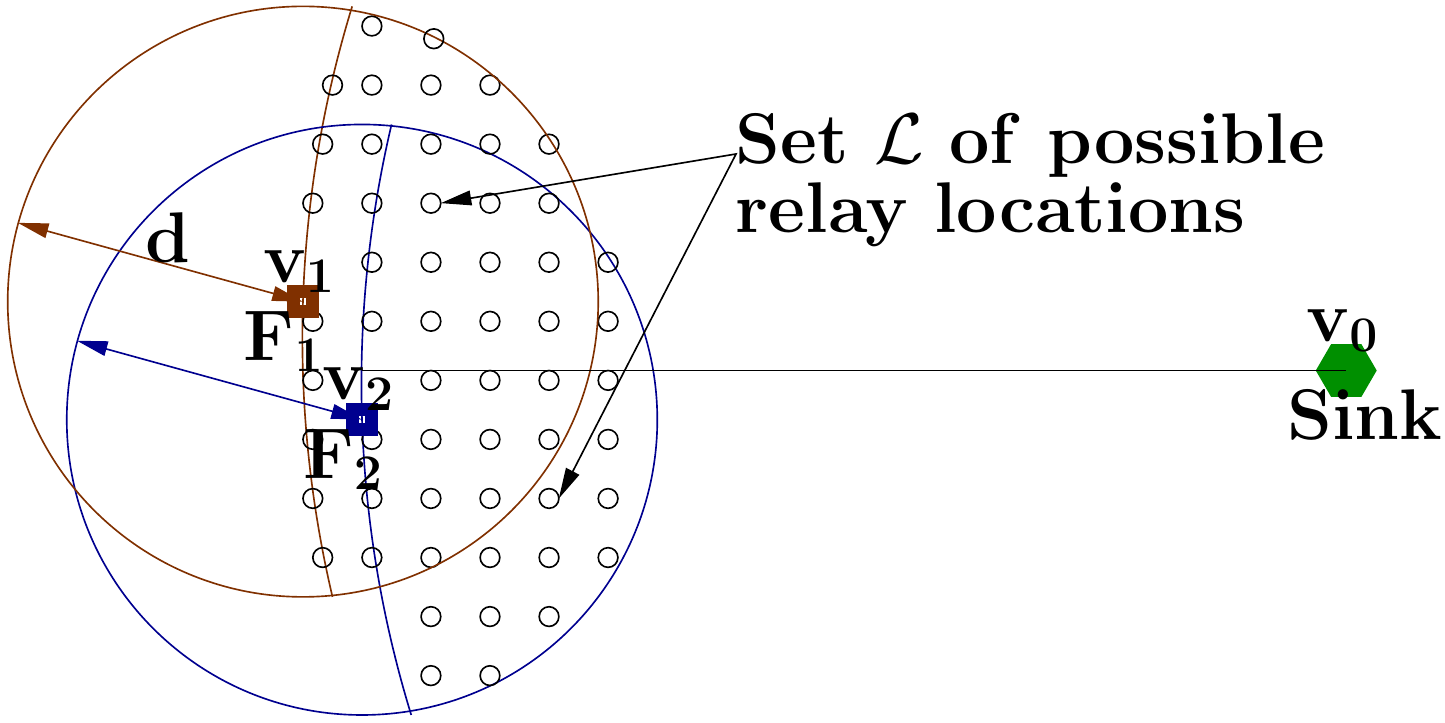}
\caption{\label{one_hop_figure} One-hop forwarding scenario: $v_0$, $v_1$, and $v_2$ are the locations of the 
sink, $\mathscr{F}_1$, and $\mathscr{F}_2$, respectively; $d$ is the range of each forwarder. 
Possible relay locations are shown as $\circ$.} 
\vspace{-4mm}
\end{figure}

Before proceeding further, in this section, as a motivating example, we will 
construct a reward structure corresponding to the problem of 
\emph{geographical forwarding}\footnote{{Geographical forwarding}
\cite{akkaya-younis05routing-survey,mauve-etal01survey-position-routing}, also
known as location based routing, is a forwarding technique where the assumption is that each node knows 
its location as well as the location of the sink node.}
in sleep-wake cycling wireless networks. 
Let  $\mathscr{F}_1$ and $\mathscr{F}_2$ actually represent two forwarding nodes in a wireless network.
As shown in Fig.~\ref{one_hop_figure}, let  
$v_1$ and $v_2$ denote their respective locations.
A sink node is located at $v_0$. Let $d$ denote the range 
of both the forwarders.
Given any location $\ell\in\Re^2$, we define 
the \emph{progress}, $Z_{\rho}(\ell)$, made by location $\ell$ with respect to (w.r.t.) 
$\mathscr{F}_\rho$ as
\begin{eqnarray}
Z_{\rho}(\ell)&=& \parallel v_{\rho}-v_0\parallel - \parallel \ell-v_0\parallel
\end{eqnarray}
where $\parallel \cdot\parallel$ denotes the Euclidean norm. Thus, $Z_{\rho}(\ell)$ is simply the difference between 
$\mathscr{F}_\rho$-to-sink and $\ell$-to-sink distances. A positive value of $Z_{\rho}(\ell)$ implies that 
location $\ell$ is closer to the sink than $\mathscr{F}_\rho$. Now, define 
the forwarding region, $\mathcal{L}_\rho$, of $\mathscr{F}_\rho$  
as the set of all locations 
that lie within the range of $\mathscr{F}_\rho$ and make non-negative progress 
w.r.t.\ $\mathscr{F}_\rho$, i.e., denoting $D_\rho(\ell) = \parallel \ell-v_\rho \parallel$ to be the distance
between $\ell$ and $\mathscr{F}_\rho$,
\begin{eqnarray}
\mathcal{L}_\rho &=& \Big\{\ell: D_\rho(\ell) \le d, Z_\rho(\ell)\ge 0 \Big\}.
\end{eqnarray}
Let $\mathcal{L}=\mathcal{L}_1 \cup \mathcal{L}_2$ denote the combined forwarding region of 
the two forwarders. As depicted in Fig.~\ref{one_hop_figure}, we will discretize $\mathcal{L}$
into a grid of finite set of $m$ locations $\{\ell_1,\ell_2,\cdots,\ell_m\}$. Thus, from here on,
whenever we refer to a location $\ell$ we mean it to be one of the above $m$ locations.

\textbf{\emph{Sleep-Wake Process:}}
Without loss of generality, we will assume that at time $0$ each forwarder is holding 
an alarm packet which has to forwarded to a downstream \emph{relay node} (i.e., a node 
in its forwarding region).
Since the relays are sleep-wake cycling, each forwarder has 
to wait until a ``good'' relay wakes up (the goodness of a relay will be based on the reward metric
to be discussed in this section). 

A practical approach for sleep-wake cycling is the 
\emph{asynchronous periodic sleep-wake process}  \cite{kim-etal11optimal-anycast,naveen-kumar12relay-selection-TMC}, 
where each relay $i$ wakes up at the periodic instants
$\{T_i+kT:k\ge0\}$ with $\{T_i\}$ being i.i.d.\ (independent and identically distributed) 
uniform on $[0,T]$ ($T$ is referred to as the sleep-wake cycling period).
Now, for dense networks where $N$ is large, if $T$ scales with $N$
such that $\frac{N}{T}=\frac{1}{\tau}$ as $N\rightarrow\infty$, then the aggregate point
process of relay wake-up instants converges to a Poisson process of 
rate $\frac{1}{\tau}$ \cite{cinlar75stochastic-processes}.
This observation motivates us to
model the \emph{aggregate} point process of wake-up instants of relays as a Poisson point process.
Furthermore, the Poisson point process assumption renders our problem
analytically tractable, leading to interesting structural results. 

Thus, formally, we model the sleep-wake cycling by assuming that 
there are an {infinite number of relays} waking up (within the combined forwarding region $\mathcal{L}$)
sequentially at the times $\{W_k:k\ge0\}$ which are the 
{points of a Poisson process} of rate $\frac{1}{\tau}$ (thus, a new relay wakes up at
each instant $W_k$). 
Let $L_k\in\mathcal{L}$ denote the location of the $k$-th relay (i.e., the relay waking up at the instant $W_k$). 
The locations $\{L_k:k\ge1\}$ are i.i.d.\ random variables with their common p.m.f.\ (probability
mass function) being $q$, i.e., $\mathbb{P}(L_k=\ell)=q_\ell$.

\textbf{\emph{Channel Model:}}
We will use the following standard model to obtain the transmission power
required by $\mathscr{F}_\rho$ to achieve an SNR (signal to noise ratio)
constraint of $\Gamma$ at the $k$-th relay:
\begin{eqnarray}
\label{power_equn}
P_{\rho,k} = \frac{\Gamma N_0}{G_{\rho,k}} {\left(\frac{D_\rho(L_k)}{d_{ref}}\right)}^{\xi}
\end{eqnarray}
where, $N_0$ is the receiver noise variance, 
$D_\rho(L_k)$ is the distance between $\mathscr{F}_\rho$ and the $k$-th relay whose location is $L_k$,
$G_{\rho,k}$ is the gain of the channel between
$\mathscr{F}_\rho$ and the $k$-th relay, $\xi$ is the path-loss attenuation factor, and 
$d_{ref}$ is the far-field reference distance beyond which the above expression is valid 
\cite{kumar-etal08wireless-networking,tse-vishwanath05wireless-communication} 
(our discretization of $\mathcal{L}$ is such that the distance between 
$\mathscr{F}_\rho$ and any $\ell\in\mathcal{L}$ is more than $d_{ref}$).

We will assume that the set of channel gains $\{G_{\rho,k}: k\ge1, \rho=1,2\}$ are i.i.d.\ taking 
values from a finite set $\mathcal{G}$. Also, let $P_{max}$ denote the maximum transmit power
with which the two forwarders can transmit, i.e., if $P_{\rho,k}>P_{max}$ then $\mathscr{F}_\rho$
cannot forward its packet to the $k$-th relay. Further, we assume that the range $d$ 
(recall Fig.~\ref{one_hop_figure}) is such that
if the $k$-th relay is outside the range of $\mathscr{F}_\rho$ (i.e., $D_\rho(L_k)>d$), then
for any $G_{\rho,k}\in\mathcal{G}$, $P_{\rho,k}>P_{max}$, so that  $\mathscr{F}_\rho$
cannot forward to a relay outside its range. Transmitting to a relay inside its range is 
possible, however, provided the channel gain is good enough so that the power required is
less than $P_{max}$.

\textbf{\emph{Relay Rewards:}}
Finally, combining progress and power, we will define the reward offered by
the $k$-th relay to $\mathscr{F}_\rho$ as,
\begin{eqnarray}
\label{reward_new_equn}
R_{\rho,k} &=& \left\{\begin{array}{ll}
                \frac{{Z_{\rho}(L_k)}^a}{P_{\rho,k}^{(1-a)}} & \mbox{ if } P_{\rho,k}\le P_{max} \\
                -\infty & \mbox{ otherwise, }
               \end{array}\right.
\end{eqnarray}
where $a\in[0,1]$ is the parameter used to trade-off between progress and power. The reward being 
inversely proportional to power is clear because it is advantageous to use low power to get
the packet across; $R_{\rho,k}$ is made proportional to $Z_{\rho}(L_k)$ to promote progress
towards the sink while choosing a relay for the next hop.

We will use the above reward structure for conducting numerical and simulation experiments
in Section~\ref{numerical_work_section}. However, it is important to note that all our 
analysis in the subsequent sections hold for the general model introduced in Section~\ref{system_model_section}.

\section{Related Work}
\label{related_work_section}
We will first make an important comparison with our prior work on the topic,
before proceeding to discuss general literature from the area of geographical 
forwarding in wireless networks. Our problem can also be considered as a variant
of the \emph{asset selling problem} studied in the operations research literature;
we will discuss related work from this field as well. Finally, we survey literature from the 
area of stochastic games. 

\textbf{\emph{Our Prior Work:}} Problem of relay selection, but by a single forwarder (i.e.,
the non-competitive version), has been extensively studied by us, starting from 
a simple model where the number of relays is exactly known to the forwarder 
to the one where only a belief is known \cite{naveen-kumar12relay-selection-TMC}. 
We have also studied a variant with channel probing
where the relay rewards are not immediately revealed to the forwarder; instead the forwarder can 
choose to learn the reward values by paying an additional cost \cite{naveen-kumar15channel-probing}. 

The basic version of our model \cite[Section~6]{naveen-kumar12relay-selection-TMC} 
comprises only one forwarder and a finite number of relay $N$; however, in the basic
model we allow for the forwarder to recall a previous relay unlike here where recalling is not allowed.
For this basic model, the solution is completely in terms of a single threshold
$\alpha$: forward to the first relay whose reward is more than $\alpha$;
at the last stage choose the best relay irrespective of its reward value.
From \cite[Section~6]{naveen-kumar12relay-selection-TMC}, we further know that the value of 
$\alpha$ does not dependent on $N$, and hence
the solution to the version of the basic model with infinite number of relays, 
should still be same. 
Furthermore, in the infinite horizon model there is no advantage in recalling the best relay since 
there is no last stage.
Thus, one can argue that the solution to the \emph{infinite horizon basic relay selection model, without recall,}
should also be characterized by the same threshold $\alpha$. 
Here, we will formally show that this is in fact the solution for one forwarder when the 
other forwarder has already terminated (Lemma~\ref{competitive_fixed_point_lemma}). However when both the forwarders are present,
the solution is more involved (studied in Section~\ref{states_ij_section}). 
Thus, the competitive
model studied here is a generalization of the aforementioned version of the
basic relay selection model. 

\textbf{\emph{Geographical Forwarding:}}
The problem of choosing a next-hop relay arises in the context of \emph{geographical forwarding};
\emph{geographical forwarding} \cite{akkaya-younis05routing-survey,mauve-etal01survey-position-routing} 
is a forwarding technique where the prerequisite  is that the nodes 
know their respective locations as well as the sink's location. 
The method of geographical forwarding was already envisioned in 
the 80's in the context of routing in packet radio networks (PRNs) 
\cite{takagi-kleinrock84optimaltransmission,hou-etal86rangecontrol}.
One of the simplest geographical forwarding technique is the greedy algorithm where each node forwards to
a neighbor in its communication region which makes maximum
progress towards the sink.
This greedy algorithm is referred to as the \emph{MFR (Max Forward within Radius)}
routing in \cite{takagi-kleinrock84optimaltransmission}.
Akin to MFR is the \emph{NFP (Nearest
with Forward Progress)} proposed in \cite{hou-etal86rangecontrol} where a node with a positive progress, and
closest to the transmitting node  is chosen. 
A generalization of MFR and NFP routing is to randomly choose any neighbor
which makes a positive progress towards the sink \cite{nelson-kleinrock84spatial-capacity}.

More recently, there are work that apply geographical forwarding for routing in 
sleep-wake cycling networks. 
For instance, 
Zorzi and Rao in \cite{zorzi-rao03geographicrandom} propose an 
algorithm called GeRaF (Geographical Random Forwarding) which, at each forwarding stage,
chooses the relay making the largest progress. For a sleep-wake cycling network,
Liu et al.\ in \cite{liu-etal07CMAC} propose a relay selection approach as a part of
CMAC, a protocol for geographical packet forwarding.
Under CMAC, node $i$ chooses an $r_0$ that minimizes the
expected normalized latency (which is the average ratio of
one-hop delay and progress). Akin to the relay selection problem is the problem of channel selection
\cite{chaporkar-proutiere08joint-probing,chang-liu07channel-probing} where a transmitter, given
several channels, has to choose one for its transmissions. 
Analogous to rewards in our case, the transmitter's decision is based on 
the throughput the transmitter can achieve on a channel.
Links to more literature on similar work from the context of wireless networks
can be found in \cite{naveen-kumar12relay-selection-TMC}.
However all these work do not consider the competitive scenario like ours.

\textbf{\emph{Asset Selling Problem:}}
Finally, our relay selection problem can be considered to be equivalent to the \emph{asset selling problem},
which is a class of the optimal stopping problems
studied in the operations research literature (other examples of 
 stopping problems include the \emph{secretary problem} \cite{freeman83secretary-problem}, \emph{bandit problem} 
\cite{bradt-etal56sequential-design}, etc). 
The basic asset selling problem \cite[Section~4.4]{bertsekas05optimal-control}
\cite{karlin62selling-asset} comprises
a single seller (analogous to a forwarder in our model) and a sequence of i.i.d.\ offers (rewards in our case). 
The seller's objective is to choose an offer so as to maximize 
a combination of the offer value and the number of previous offers rejected. 
Over the years, several variants of
the basic problem have been studied. For instance, 
In \cite{david-levi04optimal-selection}, David and Levi consider a model in
which the offers arrive at the points of a renewal process.
Kang in \cite{kang05optimal-stopping} 
has considered a model where a cost has to be paid to recall
the previous best offer; see \cite{kang05optimal-stopping}  for further references to literature on models with
uncertain recall. Variants with unknown offer (or reward) distribution, or one where a parameter of the offer distribution
is unknown have been studied in \cite{albright77generalized-house-selling,rosenfield-etal83selling-asset}.

Our competitive model here can be considered as a game variant of the basic asset selling problem,
where the two forwarders are analogous to the sellers and the reward values are analogous to the
offers. Although one game variant has been
studied by Nakagami in \cite{nakagami99game-asset-selling}, 
the specific cost structure in our problem enables us to prove 
results such as the existence of Nash equilibrium policy pair within the class of
threshold rules (Theorem~\ref{NE_existance_theorem}). Further, we also  study a completely observable case
which is not considered in \cite{nakagami99game-asset-selling}.

Similarly, literature is available on the game version of the secretary problem
\cite{nicole-etal06secretary-problem,david-krzysztof05secretary-problem}, 
but these consider the simpler case where the reward offered by an arriving secretary (or resource)
to both players is the same. Moreover, the objective in the secretary problem is to maximize the probability of 
choosing the best secretary (resource), which is in contrast to our setting (asset selling) which involves a trade-off between 
selection delay and reward. Further, a partially observable scenario is not studied in these work. 

\textbf{\emph{Stochastic Games:}}
Stochastic games can be considered as a generalization of Markov decision processes (MDPs), in 
the sense that a stochastic game comprises multiple agents (in contrast to a single agent in an MDP),
who jointly control the state of the system while individually incurring a cost
in doing so. Several references \cite{filar-vrieze96competitive-mdp,fink64stochastic-game,
thuijsman-vrieze98stochastic-games,altman96stochastic-games,
altman-etal08constrained-stochastic-games,
raghavan-filar91stochastic-games-survey}
are available on the topic starting from the seminal work by Shapley \cite{shapley53stochastic-games}.
However, most of these work 
study either discounted or average cost objectives, unlike
our problem which falls within the realm of total-cost transient stochastic games (or stopping games 
\cite[Part~III]{nowak-szajowski07advances-dynamic-games}). 
Our formulation can be alternatively thought of as a quitting game \cite{solan-vohra01quitting-games}. However,
we have introduced state transitions and state dependent quitting cost which are 
not considered in the model studied in \cite{solan-vohra01quitting-games}.

In summary, to the best of our knowledge, 
the model proposed in this paper along with the structural results we have derived,
are new contributions to the field of stopping games.

\section{Completely Observable (CO) Case}
\label{completely_observable_case_section}
For the CO model we assume that the reward pair, $(R_{1,k},R_{2,k})$,
of the $k$-th relay is entirely revealed to both the forwarders. Recalling the geographical forwarding 
example in Section~\ref{geographical_example_section}, this case would model
the scenario where the reward is simply the progress, $Z_{\rho}(L_k)$, the relay makes towards the sink,
i.e., if $a=1$ in (\ref{reward_new_equn}). Thus, observing the location $L_k$ of the $k$-th relay,
both forwarders (assuming that both a-priori know the locations $v_1, v_2$ and $v_0$; see the following remark) 
can entirely compute $(R_{1,k},R_{2,k})$.

\emph{Remark:} Justification for knowing the locations is as follows.
All the nodes are equipped with GPS (Global Positioning System) devices, using which each node 
can know its own location. Next, the sink being a fixed node, its location is already
made available to all the nodes before deployment.
Finally, each forwarder's knowledge of the other's location can be 
acquired when both forwarders broadcast control packets in 
response to the control packet transmitted by the first relay.

We will now proceed to formulate the completely observable case as a stochastic game. 
Using a key theorem from the book by Filar and Vrieze on \emph{Competitive Markov Decision
Processes} \cite{filar-vrieze96competitive-mdp},
we will characterize the structure of NEPPs. 

\subsection{Stochastic Game Formulation}
\label{stochastic_game_subsection}
Limiting ourselves to the case of finite set of states and finite action sets,
formally a stochastic game can be represented by 
a tuple $(\mathcal{N},\mathcal{X},\{\mathcal{A}_\rho\},T,\{g_\rho\})$ where,
\begin{itemize}
\item $\mathcal{N}$ is the set of agents or players,
\item $\mathcal{X}$ is the finite set of system states,
\item $\mathcal{A}=\times_{\rho\in\mathcal{N}}\mathcal{A}_\rho$ is the joint-action space with
$A_\rho$ representing the finite action set of agent $\rho$,
\item $T:\mathcal{X}\times\mathcal{A}\rightarrow\Delta(\mathcal{X})$ (the set of all p.m.f.s on $\mathcal{X}$)
is the probability transition 
kernel, i.e., $T(x'|x,a)$ is the probability that the next state is $x'$ given that the
current state is $x$ and the current joint-action is $a=(a_\rho: \rho\in\mathcal{N})$, 
\item $g_\rho:\mathcal{X}\times\mathcal{A}\rightarrow\Re$ is the (expected) 
one-step-cost function of agent $\rho$. 
\end{itemize}
We will now identify each of these components for our problem. The two forwarders,
$\mathscr{F}_1$ and $\mathscr{F}_2$, are the players (i.e., $\mathcal{N}=\{F_1,F_2\}$), and 
$\mathcal{X}$ in (\ref{SS_CO_equn}) is the state space.
The action sets are $\mathcal{A}_1=\mathcal{A}_2=\{\mbox{\textsf{s},\textsf{c}}\}$. 

\textbf{\emph{Transition Probabilities:}}
Recall that $p_{i,j}$ is the joint p.m.f of $(R_{1,k},R_{2,k})$, $p_i^{(1)}$ and $p_j^{(2)}$
are the marginal p.m.f.s of $R_{1,k}$ and $R_{2,k}$, respectively, and $\nu_\rho$ ($\rho=1,2$) is the probability that
$\mathscr{F}_\rho$ will win the contention if both forwarders cooperate.
Now, the transition probability
when the current state is of the form $x=(r_i,r_j)$ can be written as,
\begin{eqnarray}
T(x'|x,a)= 
\left\{\begin{array}{cl}
p_{i',j'}    & \mbox{ if } a=(\textsf{c},\textsf{c}), x'=(r_{i'},r_{j'})\\
p^{(1)}_{i'} & \mbox{ if } a=(\textsf{c},\textsf{s}), x'=(r_{i'},\ts)\\
p^{(2)}_{j'} & \mbox{ if } a=(\textsf{s},\textsf{c}), x'=(\ts,r_{j'})\\
\nu_2 p^{(1)}_{i'} & \mbox{ if }  a=(\textsf{s},\textsf{s}), x'=(r_{i'},\ts)\\
\nu_1 p^{(2)}_{j'} & \mbox{ if } a=(\textsf{s},\textsf{s}), x'=(\ts,r_{j'})\\
0 & \mbox{ otherwise. } 
\end{array}\right.
\end{eqnarray}
Note that when the joint-action is $(\textsf{s},\textsf{s})$, $\nu_2 p^{(1)}_{i'}$
is the probability that $\mathscr{F}_2$ gets the current relay and the reward offered by the next
relay to $\mathscr{F}_1$ is $r_{i'}$. Similarly, $\nu_1 p^{(2)}_{j'}$ is the probability (again 
when the joint-action is $(\mbox{\textsf{s},\textsf{s}})$) 
that $\mathscr{F}_1$ gets the relay and the reward value
of the next relay to $\mathscr{F}_2$ is $r_{j'}$.

Next, when the state is of the form $x=(r_i,\ts)$ (i.e., $\mathscr{F}_2$ has already
terminated) the transition probabilities depend only on the action $a_1$ of $\mathscr{F}_1$ and is
given by,
\begin{eqnarray}
T(x'|x,a)= 
\left\{\begin{array}{cl}
p^{(1)}_{i'} & \mbox{ if } a_1=\textsf{c}, x'=(r_{i'},\ts)\\ 
1 & \mbox{ if } a_1=\textsf{s}, x'=(\ts,\ts)\\
0 & \mbox{ otherwise. }      
\end{array}\right.
\end{eqnarray}
Similarly one can write the expression for $T(x'|x,a)$  when the state is $x=(\ts,r_j)$.
Finally, the state $(\ts,\ts)$ is absorbing so that 
$T((\ts,\ts)|(\ts,\ts),a)=1$.

\textbf{\emph{One-Step Costs:}} 
The one-step costs should be
such that, for any policy pair 
$(\pi_1,\pi_2)$, the sum of all one-step costs incurred by $\mathscr{F}_\rho$ ($\rho=1,2$) should equal 
the total cost in (\ref{cost1_equn}). 
With this in mind, in Table~\ref{one-step-cost-table} we write the pair of one-step costs, $(g_1(x,a),g_2(x,a))$,
incurred by $\mathscr{F}_1$ and $\mathscr{F}_2$ for different joint-actions, $a=(a_1,a_2)$,
when the current state is $x=(r_i,r_j)$.

\begin{table}[h!]
\centering
\begin{tabular}{|c||c|}
\hline
 $a=(a_1,a_2)$   & $(g_1(x,a),g_2(x,a))$\\
\hline
\hline
$(\textsf{c},\textsf{c})$ & $(\tau,\tau)$\\
\hline
$(\textsf{c},\textsf{s})$ & $(\tau,-\eta_2 r_j)$\\
\hline
$(\textsf{s},\textsf{c})$ & $(-\eta_1 r_i,\tau)$\\
\hline
$(\textsf{s},\textsf{s})$ & $(-\eta_1 r_i,\tau)$ w.p. $\nu_1$\\
			  & $(\tau,-\eta_2 r_j)$ w.p. $\nu_2$\\
\hline
\end{tabular}
\caption{\label{one-step-cost-table} One-step costs when $x=(r_i,r_j)$.}
\vspace{-6mm}
\end{table}

From Table~\ref{one-step-cost-table} we see that if the joint action is $(\ct,\ct)$
then both forwarders continue incurring a 
cost of $\tau$ which is the average time until the next relay arrives. 
When one of the forwarder, say $\mathscr{F}_2$, chooses to stop (i.e., the joint action is $(\ct,\st)$) then 
$\mathscr{F}_2$, forwarding its packet to the chosen relay, incurs a terminating cost of $-\eta_2 r_j$,
while $\mathscr{F}_1$ simply continues incurring an average waiting delay of $\tau$. Analogous is the case 
whenever the joint action is $(\st,\ct)$. Finally, if both forwarders compete (i.e., the case $(\st,\st)$), 
then with probability $\nu_\rho$,	
$\mathscr{F}_\rho$ gets the relay incurring the terminating cost while the other forwarder has to continue.

\begin{table}[h!]
\begin{minipage}[b]{0.5\linewidth}
\centering
\begin{tabular}{|c||c|}
\hline
$a_1$ & $(g_1(x,a),g_2(x,a))$\\
\hline
\hline
$\textsf{c}$ & $(\tau,0)$\\
\hline
$\textsf{s}$ & $(-\eta_1 r_i,0)$\\
\hline
\end{tabular} 
\caption{\label{S1_along_table} $x=(r_i,\ts)$}
\end{minipage}
\begin{minipage}[b]{0.4\linewidth}
\centering
\begin{tabular}{|c||c|}
\hline
$a_2$ & $(g_1(x,a),g_2(x,a))$\\
\hline
\hline
$\textsf{c}$ & $(0,\tau)$\\
\hline
$\textsf{s}$ & $(0,-\eta_2 r_j)$\\
\hline
\end{tabular} 
\caption{\label{S2_alone_table} $x=(\ts,r_j)$}
\end{minipage}
\vspace{-6mm}
\end{table}

When the state is of the form $(r_i,\ts)$ the cost incurred by $\mathscr{F}_2$ is $0$ for
any joint-action $a$, and further the one-step cost incurred by 
$\mathscr{F}_1$ depends only on the action $a_1$ of $\mathscr{F}_1$. Analogous situation holds
for $\mathscr{F}_2$
when the state is $(\ts,r_j)$. 
These costs are given in Table~\ref{S1_along_table} and \ref{S2_alone_table}, respectively.
Finally, the cost incurred by both the forwarders once the termination
state $(\ts,\ts)$ is reached is $0$.

Now, given a policy pair $(\pi_1,\pi_2)$ (recall Definition~\ref{policy_defn})
and an initial state $x\in\mathcal{X}$, let $\{X_k: k\ge1\}$ denote the sequence
of (random) states traversed by the system, and let $\{(A_{1,k},A_{2,k}):k\ge1\}$
denote the sequence of joint-actions. 
The total cost in (\ref{cost1_equn}) 
can now be expressed as 
the sum of all the one-step costs as follows:
\begin{eqnarray}
\label{total_cost1_equn}
J^{(\rho)}_{\pi_1,\pi_2}(x) 
&=& \sum_{k=1}^\infty \mathbb{E}^x_{\pi_1,\pi_2}\Big[g_\rho(X_k,(A_{1,k},A_{2,k}))\Big].   
\end{eqnarray}

\subsection{Characterization of NEPPs}
\label{NEPP_characterization_ri_section}
\noindent
\textbf{\emph{States of the form}} $(r_i,\ts)$, $(\ts,r_j)$

\label{states_phi_section}
Once the system enters a state of the form $(r_i,\ts)$,
since only $\mathscr{F}_1$ is present in the system, we essentially have an 
MDP problem where $\mathscr{F}_1$ is attempting to optimize its cost. Formally, if 
$(\pi_1^*,\pi_2^*)$ is an NEPP then it can be argued\footnote{Using the definition of an NEPP
and the fact that the costs and the state transitions do not depend on the policy of the other
forwarder anymore.} that
 $J^{(1)}_{\pi_1^*,\pi_2^*}(r_i,\ts)$
is the optimal cost to $\mathscr{F}_1$ with $\pi_1^*$ being an optimal policy; the cost incurred by $\mathscr{F}_2$ is $0$
and $\pi_2^*$ can be arbitrary, but for simplicity we fix 
$\pi_2^*(r_i,\ts)=\textsf{s}$ for all $i\in[n]$.
Hence $J^{(1)}_{\pi_1^*,\pi_2^*}(\cdot,\ts)$
satisfies the following Bellman optimality equation: 
\begin{eqnarray}
\label{bellman_J1_equn}
J^{(1)}_{\pi_1^*,\pi_2^*}(r_i,\ts)&=&\min\Big\{-\eta_1 r_i,D^{(1)}\Big\}, 
\end{eqnarray}
where 
\begin{eqnarray}
\label{d1_cost_equn}
D^{(1)} &=& \tau+\sum_{i'} p^{(1)}_{i'}J^{(1)}_{\pi_1^*,\pi_2^*}(r_{i'},\ts)
\end{eqnarray}
is the expected \emph{cost of continuing alone in the system}
($\tau$ is the one-step cost and the 
remaining term is the future cost-to-go). $-\eta_1 r_i$ in the $\min$-expression
above is the cost of stopping.
Thus, denoting $\frac{D^{(1)}}{-\eta_1}$ by $\alpha^{(1)}$,
whenever the state is of the form $(r_i,\ts)$ an optimal policy is as follows:
\begin{eqnarray}
\label{S1_policy_equn}
\pi_1^*(r_i,\ts) &=& 
\left\{ \begin{array}{l l}
\textsf{s} & \mbox{ if }   r_i\ge \alpha^{(1)} \\
\textsf{c} & \mbox{ otherwise}.
\end{array}\right.
\end{eqnarray}

\emph{Remark:} As mentioned earlier (recall the discussion in related work), the solution to
the basic relay selection problem, comprising a single forwarder (say only $\mathscr{F}_1$) and a
finite number of relays $N$, is characterized in terms of a 
single threshold $\alpha$. Furthermore, from our earlier work \cite[Section~6]{naveen-kumar12relay-selection-TMC} 
we know that $\alpha$ is the unique fixed point of 
\begin{eqnarray}
\label{beta1_equn}
\beta^{(1)}(x)=\mathbb{E}\Big[\max\{x,R_1\}\Big]-\frac{\tau}{\eta_1},
\end{eqnarray}
where the expectation in the above expression is w.r.t.\ the p.m.f.\ $p^{(1)}$ of $R_1$. 
Here we will show that $\alpha^{(1)}$
is the fixed point of $\beta^{(1)}$, formalizing our earlier claim that  
the competitive model with only one forwarder and the infinite horizon basic model are equivalent.
Although this result can be deduced by showing the equivalence between our {competitive model with a single forwarder}
and the {infinite horizon version of the asset selling problem}, we prove it here for completeness.

\begin{lemma}
\label{competitive_fixed_point_lemma}
$\alpha^{(1)}$ is the unique fixed point of $\beta^{(1)}(x)$ ($x\in(-\infty,r_n]$) in (\ref{beta1_equn}).
\end{lemma}
\begin{IEEEproof}
We will first show that $\beta^{(1)}$ is a contraction mapping. Then, from the
Banach fixed point theorem \cite{patayyfixed-point} it follows 
that there exists a unique fixed point $\alpha^*$
of $\beta^{(1)}$. Next, through an induction argument we will prove that
$J^{(1)}_{\pi_1^*,\pi_2^*}(r_i,\ts)=\min\Big\{-\eta_1 r_i, -\eta_1 \alpha^*\Big\}$.
Finally, substituting for $J^{(1)}_{\pi_1^*,\pi_2^*}(r_i,\ts)$ 
in $\alpha^{(1)}=\frac{D^{(1)}}{-\eta_1}$ (recall $D^{(1)}$ from (\ref{d1_cost_equn}))
and simplifying, we obtain the desired
result. Details of the proof are available in Appendix~\ref{competitive_fixed_point_lemma_appendix}.
\end{IEEEproof}

Similarly, when the state is of the form $(\ts,r_j)$
(i.e., $\mathscr{F}_1$ has already terminated),
if $(\pi_1^*,\pi_2^*)$ is an NEPP then, $J^{(1)}_{\pi_1^*,\pi_2^*}(\ts,r_j)=0$ and
$\pi_1^*(\ts,r_j)=\textsf{s}$, while 
$J^{(2)}_{\pi_1^*,\pi_2^*}(\ts,r_j)$ satisfies 
% $J^{(2)}_{\pi_1^*,\pi_2^*}(\ts,r_j)=\min\Big\{-\eta_2 r_j, D^{(2)} \Big\}$
\begin{eqnarray}
\label{bellman_J2_equn}
J^{(2)}_{\pi_1^*,\pi_2^*}(\ts,r_j)&=&\min\Big\{-\eta_2 r_j, D^{(2)} \Big\},
\end{eqnarray}
where $D^{(2)}=\tau+\sum_{j'}p^{(2)}_{j'} J^{(2)}_{\pi_1^*,\pi_2^*}(\ts,r_{j'})$.
Further, $\alpha^{(2)}=\frac{D^{(2)}}{-\eta_2}$, is the unique fixed point of 
$\beta^{(2)}(x)=\mathbb{E}\Big[\max\{x,R_2\}\Big]-\frac{\tau}{\eta_2}$, where now
the expectation is w.r.t.\ the p.m.f.\ $p^{(2)}$ of $R_2$.
Finally, an optimal policy $\pi_2^*$ is such that 
\begin{eqnarray}
\label{S2_policy_equn}
\pi_2^*(\ts,r_j) &=& 
\left\{ \begin{array}{l l}
\textsf{s} & \mbox{ if }  r_j \ge \alpha^{(2)} \\
\textsf{c} & \mbox{ otherwise}.
\end{array}\right.
\end{eqnarray}

\noindent
\textbf{\emph{States of the form}} $(r_i,r_j)$
\label{states_ij_section}

This is the more interesting case where both forwarders are present in the system and
are competing to choose a relay.
When the state is of the form $(r_i,r_j)$, 
if $\mathscr{F}_1$ decides to continue while $\mathscr{F}_2$ chooses to stop (i.e., the joint-action is 
$(\textsf{c},\textsf{s})$), then $\mathscr{F}_2$ terminates by incurring a cost of $-\eta_2 r_j$
so that the next state is of the form $(r_{i'},\ts)$. 
Hence the expected total  cost incurred by
$\mathscr{F}_1$, if it uses the policy in (\ref{S1_policy_equn}) from the next stage
onwards, is $D^{(1)}$ (recall (\ref{d1_cost_equn})). 
Similarly, if the joint-action is $(\textsf{s},\textsf{c})$ then $\mathscr{F}_1$
terminates incurring a cost of $-\eta_1 r_i$, and $\mathscr{F}_2$ incurs a cost
of $D^{(2)}$ if it uses the policy in  (\ref{S2_policy_equn}) from the next stage onwards.

If both forwarders decide to stop (joint-action is $(\textsf{s},\textsf{s})$) 
then with probability $\nu_1$, $\mathscr{F}_1$ gets the
relay in which case $\mathscr{F}_2$ continues alone, and with probability $\nu_2$ it is vice versa.
Thus, the expected cost incurred by $\mathscr{F}_1$ is, 
\begin{eqnarray}
\label{e1_cost_equn}
E^{(1)}(r_i) 
&=& \nu_1 (-\eta_1 r_i)+\nu_2 D^{(1)},
\end{eqnarray}
and that by $\mathscr{F}_2$ is, 
\begin{eqnarray}
\label{e2_cost_equn}
E^{(2)}(r_j) 
&=& \nu_1 D^{(2)} + \nu_2 (-\eta_2 r_j).
\end{eqnarray}

Finally, if both forwarders choose to continue (i.e., if the joint-action is
$(\textsf{c},\textsf{c})$) then the next state is again of the form $(r_{i'},r_{j'})$.
Thus if $(\pi_1,\pi_2)$ is the policy pair  used from the next
stage onwards then the expected costs incurred by $\mathscr{F}_1$ and $\mathscr{F}_2$ are, respectively, 
\begin{eqnarray}
\label{c1_cost_equn}
C^{(1)}_{\pi_1,\pi_2} &=& \tau + \sum_{i',j'}p_{i',j'}J^{(1)}_{\pi_1,\pi_2}(r_{i'},r_{j'})\\
\label{c2_cost_equn}
C^{(2)}_{\pi_1,\pi_2} &=& \tau + \sum_{i',j'}p_{i',j'}J^{(2)}_{\pi_1,\pi_2}(r_{i'},r_{j'}).
\end{eqnarray}

We are now ready to state the following main theorem (adapted from \cite{filar-vrieze96competitive-mdp}),
which relates the ``NEPPs of the stochastic game'' with the 
``Nash equilibrium strategies of a certain static game'' played at a stage.
The various cost terms described above are used to construct
this static game. We state the theorem below with the understanding
that for states of the form $x=(r_i,\ts)$ and 
$x=(\ts,r_j)$, $\pi_1^*(x)$ and $\pi_2^*(x)$ are as in 
(\ref{S1_policy_equn}) and (\ref{S2_policy_equn}), respectively.
\begin{theorem}
\label{filar_theorem}
Given a policy pair, $(\pi_1^*,\pi_2^*)$, for each state $x=(r_i,r_j)$ construct the static game given in 
Table~\ref{bimatrix_game_table}.
\begin{table}[h]
\centering
\begin{tabular}{|c||c|c|}
\hline
	   & \textsf{c} & \textsf{s} \\
\hline
\hline
\textsf{c} & $C^{(1)}_{\pi_1^*,\pi_2^*},C^{(2)}_{\pi_1^*,\pi_2^*}$ & $D^{(1)},-\eta_2 r_j$ \\
\hline
\textsf{s} & $-\eta_1 r_i,D^{(2)}$ & 
$E^{(1)}(r_i),E^{(2)}(r_j)$\\
\hline 
\end{tabular} 
\vspace{1mm}
\caption{\label{bimatrix_game_table} Static stage game.}
\vspace{-5mm}
\end{table}

\noindent
Then the following statements are equivalent:
\begin{enumerate}
\item[a)] 
$(\pi_1^*,\pi_2^*)$ is an NEPP.

\item[b)]
For each $x=(r_i,r_j)$, 
$(\pi_1^*(x),\pi_2^*(x))$ is a 
\emph{Nash equilibrium (NE) strategy}
for the game in Table~\ref{bimatrix_game_table}.
Further, the expected payoff pair at this NE strategy is,
$(J^{(1)}_{\pi_1^*,\pi_2^*}(x),J^{(2)}_{\pi_1^*,\pi_2^*}(x))$.
\end{enumerate}
\end{theorem}
\begin{IEEEproof}
Although the proof of this theorem is along the lines of the proof of Theorem~4.6.5 in 
\cite{filar-vrieze96competitive-mdp},
however some additional efforts are required since
the proof in \cite{filar-vrieze96competitive-mdp} is for the case where the costs are discounted,
while ours is a total cost undiscounted stochastic game. Further, the presence 
of a cost-free absorption state for each player renders our problem \emph{transient}
by which we mean, when the policy of one player is fixed the problem of obtaining the optimal policy
for the other player is a stopping problem \cite{bertsekas-tsitsiklis91stochastic-shortest-path}. Using this property we have
modified the proof of \cite[Theorem~4.6.5]{filar-vrieze96competitive-mdp} appropriately so that the result holds for our case.
For details, see Appendix~\ref{filar_theorem_appendix}. 
\end{IEEEproof}

\emph{Discussion:} In this discussion for simplicity we will omit $(\pi_1^*,\pi_2^*)$ 
from all the associated notations. Now, Theorem~\ref{filar_theorem} can be seen as an extension 
of the Bellman optimality equation in (\ref{bellman_J1_equn}), where to obtain 
$J^{(1)}(r_i,\ts)$ we require the cost term 
$D^{(1)}$ in (\ref{d1_cost_equn}), which in turn depends on the function
$J^{(1)}(\cdot,\ts)$. This essentially suggests that
$J^{(1)}(\cdot,\ts)$ is the fixed point of the Bellman equation 
in (\ref{bellman_J1_equn}).  
Similarly, here we see that, given the cost
pair $(C^{(1)},C^{(2)})$, one can obtain 
$(J^{(1)}(x),J^{(2)}(x))$ by solving the game in 
Table~\ref{bimatrix_game_table}. However, computing $(C^{(1)},C^{(2)})$ itself will require
the function pair $(J^{(1)}(\cdot),J^{(2)}(\cdot))$,
thus suggesting that $(J^{(1)}(\cdot),J^{(2)}(\cdot))$
has to be fixed point of a mapping which involves computing the payoff pair of 
the static game in Table~\ref{bimatrix_game_table}. Furthermore, analogous to computing the minimum in (\ref{bellman_J1_equn})
to obtain the optimal action, here, by 
computing the NE strategies of the game in Table~\ref{bimatrix_game_table} we obtain the solution to our stochastic game.

Assuming that the cost pair $(C^{(1)}_{\pi_1^*,\pi_2^*},C^{(2)}_{\pi_1^*,\pi_2^*})$ is
given to us, we now proceed to obtain all the NE strategies of the game in Table~\ref{bimatrix_game_table}.
We will first require the following key lemma.
\begin{lemma}
\label{costs_ordered_corollary}
For an NEPP, $(\pi_{1}^*,\pi_2^*)$, the various costs are ordered as follows:
\begin{eqnarray}
\label{costs_ordered_equn}
D^{(1)}\le C^{(1)}_{\pi_1^*,\pi_2^*} \mbox{ and } D^{(2)}\le C^{(2)}_{\pi_1^*,\pi_2^*}.
\end{eqnarray}
\end{lemma}
\begin{IEEEproof}
See Appendix~\ref{costs_ordered_lemma_appendix}.
\end{IEEEproof}

\begin{figure}[t]
\centering
\includegraphics[scale=0.5]{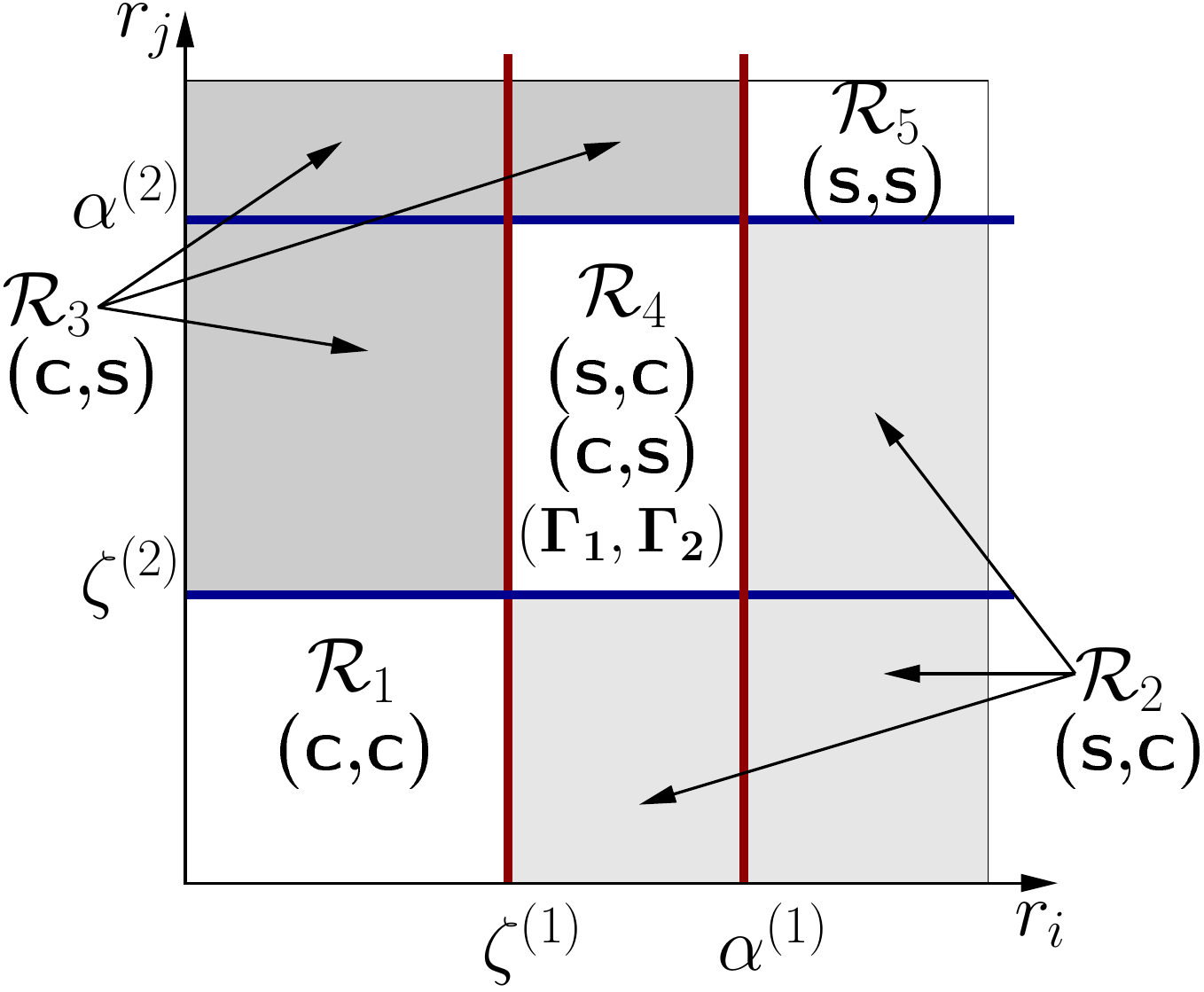}
\caption{\label{partition_figure_nosub} Illustration of the various regions
along with the NE strategies corresponding to these regions.}
\vspace{-6mm}
\end{figure}

%%%%%%%%%%%%%%%%%%%%%%%%%%%%%%%%%%%%%%%%%%%%%%%%%%%%%%%%%%%%%%%%%%%%%%%%%%%%%%%%%%
\begin{figure*}[b]
\vspace{-4mm}
\verb1--------------------------------------------------------------------------------------1
\begin{eqnarray}
\label{c1_fixed_equn}
\mathcal{T}_1(\mathbf{C}) 
&=&  \tau + \sum_{(i,j)\in\mathcal{R}_1(\mathbf{C})}p_{i,j} C^{(1)} + 
\sum_{(i,j)\in\mathcal{R}_2(\mathbf{C})\cup\mathcal{R}_4(\mathbf{C})}p_{i,j} (-\eta_1 r_i) % \nonumber \\
 +  \sum_{(i,j)\in\mathcal{R}_3(\mathbf{C})}p_{i,j} D^{(1)} + 
 \sum_{(i,j)\in\mathcal{R}_5(\mathbf{C})}p_{i,j} E^{(1)}(r_i)
\end{eqnarray}
\begin{eqnarray}
\label{c2_fixed_equn}
\mathcal{T}_2(\mathbf{C})  
&=&\tau + \sum_{(i,j)\in\mathcal{R}_1(\mathbf{C})}p_{i,j} C^{(2)} + 
\sum_{(i,j)\in\mathcal{R}_2(\mathbf{C})\cup\mathcal{R}_4(\mathbf{C})}p_{i,j} D^{(2)} 
+\sum_{(i,j)\in\mathcal{R}_3(\mathbf{C})}p_{i,j} (-\eta_2 r_j) + 
\sum_{(i,j)\in\mathcal{R}_5(\mathbf{C})}p_{i,j} E^{(2)}(r_j).
\end{eqnarray}
\end{figure*}
%%%%%%%%%%%%%%%%%%%%%%%%%%%%%%%%%%%%%%%%%%%%%%%%%%%%%%%%%%%%%%%%%%%%%%%%%%%%%%%%%%%%

{\emph{Discussion:}}
The above lemma becomes intuitive once we recall that 
$D^{(1)}$ is the optimal cost incurred by $\mathscr{F}_1$ if it is alone in the system, while
$C^{(1)}_{\pi_1^*,\pi_2^*}$ is the cost incurred if $\mathscr{F}_2$ is also
present, and competing with $\mathscr{F}_1$ in choosing a relay. One would 
expect $\mathscr{F}_1$ to incur a lower cost without the competing forwarder.

For notational simplicity, from here on,
we will denote the costs 
$C^{(1)}_{\pi_1^*,\pi_2^*}$ and  $C^{(2)}_{\pi_1^*,\pi_2^*}$ 
as simply $C^{(1)}$ and $C^{(2)}$.
We will write $\mathbf{C}$ for the pair $(C^{(1)},C^{(2)})$.
An important consequence of Lemma~\ref{costs_ordered_corollary} is that,
while solving the game in Table~\ref{bimatrix_game_table}, it
is sufficient to only consider cost pairs, $(C^{(1)},C^{(2)})$,
which are ordered as in the lemma; the other cases (e.g., $D^{(1)}>C^{(1)}$ or
$D^{(2)}>C^{(2)}$) cannot occur, and hence need not be considered. 
Further, for convenience
let us denote the thresholds $\frac{C^{(1)}}{-\eta_1}$ and $\frac{C^{(2)}}{-\eta_2}$
by $\zeta^{(1)}$ and $\zeta^{(2)}$, respectively (recall that we already have,
$\alpha^{(1)}=\frac{D^{(1)}}{-\eta_1}$ and $\alpha^{(2)}=\frac{D^{(2)}}{-\eta_2}$).
Then, the solution (i.e., the NE strategies) to the game in Table~\ref{bimatrix_game_table},
for each $(r_i,r_j)$ pair, is as depicted in Fig.~\ref{partition_figure_nosub}.

We see that the thresholds $(\alpha^{(1)},\zeta^{(1)})$ and $(\alpha^{(2)},\zeta^{(2)})$
partition the reward pair set, $\{(r_i,r_j):i,j\in[n]\}$, into $5$ regions 
($\mathcal{R}_1,\cdots,\mathcal{R}_5$)\footnote{These regions depend on the cost pair $\mathbf{C}$;
for simplicity we neglect $\mathbf{C}$ in their notation. However, we will invoke this dependency when required.}
such that 
the NE strategy (strategies) corresponding to each region are different. For instance, 
for any $(r_i,r_j)\in\mathcal{R}_1$, $(\ct,\ct)$ (i.e., both forwarders continue)
is the only NE strategy, while within $\mathcal{R}_2$, $(\st,\ct)$ is the 
NE strategy, and so on. All regions contain a unique pure NE strategy except for 
$\mathcal{R}_4$ where $(\st,\ct)$, $(\ct,\st)$, and the mixed strategy 
$(\Gamma_1,\Gamma_2)$ ($\Gamma_\rho$ is the probability with which 
$\mathscr{F}_\rho$ chooses $\st$) are all NE strategies.
The expression for $\Gamma_1$ is
\begin{eqnarray}
\Gamma_{1} = \frac{-\eta_2 r_j -C^{(2)}}{\Big(-\eta_2 r_j -C^{(2)}\Big)- \Big(E^{(2)}(r_j)-D^{(2)}\Big)}.
\end{eqnarray}
Analogously one can write the expression for $\Gamma_2$.
For details on how 
to solve the game in Table~\ref{bimatrix_game_table} to obtain the various regions, 
see Appendix~\ref{obtaining_NE_appendix}. Finally, we summarize the observations made thus far in the form 
of the following theorem.

\begin{theorem}
\label{nepp_structure_theorem}
The NE strategies of the game in Table~\ref{bimatrix_game_table} are completely characterized by the threshold pairs
$(\alpha^{(\rho)},\zeta^{(\rho)})$, $\rho=1,2$ as follows (recall Fig.~\ref{partition_figure_nosub} for illustration):
\begin{itemize}
\item If $r_i$ is less than $\zeta^{(1)}$, then the NE strategy recommends $\textsf{c}$ for $\mathscr{F}_1$ 
irrespective of the reward value $r_j$ of $\mathscr{F}_2$. 
\item On the other hand, if $r_i$ is more than $\alpha^{(1)}$, then the NE strategy recommends
action  $\textsf{s}$ for $\mathscr{F}_1$ irrespective of the value of $r_j$
(note that this is exactly the action $\mathscr{F}_1$ would
choose if it was alone in the system; see the discussion following (\ref{S1_policy_equn})). 
\item Finally, the presence of the competing forwarder $\mathscr{F}_2$ 
is felt by $\mathscr{F}_1$ only when its reward value $r_i$ is between $\zeta^{(1)}$ and $\alpha^{(1)}$, in which case
the NE strategies are: 
$(\textsf{s},\textsf{c})$ if $r_j<\zeta^{(2)}$; 
$(\textsf{s},\textsf{c})$, $(\textsf{c},\textsf{s})$ and $(\Gamma_1,\Gamma_2)$ if $\zeta^{(2)}\le r_j\le \alpha^{(2)}$;
and $(\textsf{c},\textsf{s})$ if $r_j>\alpha^{(2)}$.
\end{itemize} 
Analogous results hold for $\mathscr{F}_2$. 
\end{theorem}

\subsection{Constructing NEPPs from NE strategies}
\label{NEPP_construction_section}
The cost terms $D^{(1)}$ and $D^{(2)}$ 
can be easily computed by solving the optimality equations (\ref{bellman_J1_equn}) and 
(\ref{bellman_J2_equn}), respectively. Alternatively, we can first compute the fixed points of 
$\beta^{(1)}(\cdot)$ and $\beta^{(2)}(\cdot)$ to obtain $\alpha^{(1)}$ and $\alpha^{(2)}$,
respectively (recall Lemma~\ref{competitive_fixed_point_lemma}). 
Then, $D^{(1)}=-\eta_1\alpha^{(1)}$ and $D^{(2)}=-\eta_2\alpha^{(2)}$. 

The costs $C^{(1)}$ and $C^{(2)}$ (in (\ref{c1_cost_equn}) and (\ref{c2_cost_equn}))
depend on the particular NEPP used, i.e., require the cost terms 
$J^{(1)}_{\pi_1^*,\pi_2^*}(r_i,r_j)$ and 
$J^{(2)}_{\pi_1^*,\pi_2^*}(r_i,r_j)$ for all $(r_i,r_j)$ to compute them.
Conversely, Part-(b)  of Theorem~\ref{filar_theorem} suggests that 
$J^{(1)}_{\pi_1^*,\pi_2^*}(r_i,r_j)$ (respectively, $J^{(2)}_{\pi_1^*,\pi_2^*}(r_i,r_j)$) 
can be obtained by computing the expected 
cost incurred by $\mathscr{F}_1$ (respectively, $\mathscr{F}_2$) at a NE strategy of the game in Table~\ref{bimatrix_game_table},
which in turn requires the terms $C^{(1)}$ and $C^{(2)}$. Hence, to obtain $(C^{(1)},C^{(2)})$
we proceed by expressing $(C^{(1)},C^{(2)})$ as the fixed point of a mapping $\mathcal{T}$ which can then be used
to compute these costs.

Suppose $(\pi_1^*,\pi_2^*)$ is a NEPP 
such that for all $x=(r_i,r_j)\in\mathcal{R}_4(\mathbf{C})$ the 
NE strategy  $(\pi_1^*(x),\pi_2^*(x))$ is $(\textsf{s},\textsf{c})$.
Then using part~2(b) of Theorem~\ref{filar_theorem} we can write,
\begin{eqnarray}
\label{J1_regions_equn}
J^{(1)}_{\pi_1^*,\pi_2^*}(r_i,r_j) = 
\left\{\begin{array}{cl}
C^{(1)} & \mbox{ if } (r_i,r_j)\in\mathcal{R}_1(\mathbf{C})\\
-\eta_1 r_i & \mbox{ if } (r_i,r_j)\in\mathcal{R}_2(\mathbf{C})\\
D^{(1)} & \mbox{ if } (r_i,r_j)\in\mathcal{R}_3(\mathbf{C})\\
-\eta_1 r_i & \mbox{ if } (r_i,r_j)\in\mathcal{R}_5(\mathbf{C})\\
E^{(1)}(r_i) & \mbox{ if } (r_i,r_j)\in\mathcal{R}_4(\mathbf{C}).
\end{array}\right.
\end{eqnarray}
Using the above in (\ref{c1_cost_equn}), $C^{(1)}$ can be
written as $C^{(1)}=\mathcal{T}_1(\mathbf{C})$ where the function $\mathcal{T}_1(\mathbf{C})$
is as in (\ref{c1_fixed_equn}) 
(where for simplicity, we have used $(i,j)$ instead of $(r_i,r_j)$).
Similarly, $C^{(2)}$ can be expressed as $C^{(2)}=\mathcal{T}_2(\mathbf{C})$; see (\ref{c2_fixed_equn}).
Thus, $\mathbf{C}$ is a fixed point of the mapping 
$\mathcal{T}(\mathbf{C}):= (\mathcal{T}_1(\mathbf{C}),\mathcal{T}_2(\mathbf{C}))$.

We do not have results showing that $\mathcal{T}$ indeed has a fixed point
or equivalently that an NEPP $(\pi_1^*,\pi_2^*)$ always exists,\footnote{This equivalence 
can be easily shown by first using $(\pi_1^*,\pi_2^*)$ in part-(a) of Theorem~\ref{filar_theorem}
to conclude that part-(b) holds, and then simply from the definition of $\mathcal{T}$ it 
will follow that it has a fixed point. For the other direction, given a fixed point 
$\mathbf{C}$ of  $\mathcal{T}$, one can easily
obtain the corresponding NEPP $(\pi_1^*,\pi_2^*)$ by constructing the various regions as
shown in Fig.~\ref{partition_figure_nosub}.} although such a result holds for 
the discounted stochastic game \cite[Theorem~4.6.4]{filar-vrieze96competitive-mdp}
(recall that ours is a transient stochastic game).
However, in our numerical results section (Section~\ref{numerical_work_section})
we were able to numerically  obtain $\mathbf{C}$ by iteration.
Thus, we begin with an initial $\mathbf{C}(0)$ such that $C^{(1)}(0)<D^{(1)}$ and $C^{(2)}(0)<D^{(2)}$,
and inductively iterate to obtain $\mathbf{C}(k)=\mathcal{T}(\mathbf{C}(k-1))$ until
convergence is achieved. Finally, given a fixed point 
$\mathbf{C}$, we obtain the corresponding NEPP
$(\pi_1^*,\pi_2^*)$ by constructing the various regions as in Fig.~\ref{partition_figure_nosub}. 

\textbf{\emph{Other NEPPs:}} Recall that to obtain $(C^{(1)},C^{(2)})$
we had restricted $(\pi_1^*,\pi_2^*)$ to use NE strategy $(\textsf{s},\textsf{c})$
whenever $(r_i,r_j)\in\mathcal{R}_4(\mathbf{C})$. 
We can similarly obtain NEPPs $({\pi_1^\circ},{\pi_2^\circ})$
and $({\pi_1^{\Box}},{\pi_2^{\Box}})$ (whose corresponding cost pairs are
$\mathbf{C}_\circ$ and $\mathbf{C}_\Box$) 
by restricting to the NE strategies $(\textsf{c},\textsf{s})$ and 
$(\Gamma_1,\Gamma_{2})$ whenever 
$(r_i,r_j)\in\mathcal{R}_4(\mathbf{C}_\circ)$ and
$(r_i,r_j)\in\mathcal{R}_4(\mathbf{C}_\Box)$, respectively.
In Section~\ref{numerical_work_section}
we will numerically compare the performances of all these various 
NEPPs. 

\section{Partially Observable Case}
\label{partially_observable_case_section}
Let us first formally introduce a finite \emph{location set} $\mathcal{L}$.
Let $L_k$ denote the \emph{location} of the $k$-th relay. The locations $\{L_k:k\ge1\}$ are i.i.d with
their common p.m.f.\ being $(q_{\ell}:\ell\in\mathcal{L})$. Recall that for the PO case we assume that only $R_{\rho,k}$
is revealed to $\mathscr{F}_\rho$ ($\rho=1,2$). In addition, we will assume that $L_k$ is revealed to both the forwarders.

Recalling the geographical forwarding example from Section~\ref{geographical_example_section}, the PO case corresponds to the scenario
where, in addition to $L_k$, the gains $G_{\rho,k}$ are required to compute $R_{\rho,k}$,
i.e., if $a<1$ in (\ref{reward_new_equn}). Hence, $\mathscr{F}_1$ not knowing $G_{2,k}$
cannot compute $R_{2,k}$. However, knowing the channel gain distribution  (recall that the gains are identically 
distributed) it is possible for $\mathscr{F}_1$ to compute the probability distribution of $R_{2,k}$ given $L_k$. 
Similarly, $\mathscr{F}_2$ can compute the distribution of $R_{1,k}$ given $L_k$. Further, since the gains, $(G_{1,k},G_{2,k})$, are independent,
it follows that $R_{1,k}$ and $R_{2,k}$ are independent given $L_k$ (but unconditionally they may be dependent). 

Formally, given that 
$L_k=\ell$, we will assume the following \emph{independence condition}:
\begin{eqnarray}
\label{independence_assumption_equn}
p_{R_1,R_2|L_k}(r_i,r_j|\ell)=p_{R_1|L_k}(r_i|\ell) p_{R_2|L_k}(r_j|\ell). 
\end{eqnarray}
For simplicity, we will
 denote the conditional p.m.f.s
$p_{R_1|L_k}(r_i|\ell)$ and $p_{R_2|L_k}(r_j|\ell)$, $i,j\in[n]$, by $p^{(1)}_{i|\ell}$ and $p^{(2)}_{j|\ell}$,
respectively.

\emph{Remark:} Usually for a model with partial observations the belief that $\mathscr{F}_1$
will maintain about $R_{2,k}$ will simply be the conditional distribution $p_{R_2|R_1}(r_j|r_i)$ $=\frac{p_{i,j}}{p^{(1)}_i}$.
However, we have exploited the particular structure in our reward expression to 
come up with the independence condition in (\ref{independence_assumption_equn}).
This condition will enable us to prove a key result later  which 
is otherwise not possible (see the remark following Lemma~\ref{best_response_PO_lemma}).
Finally, all our subsequent results will hold for a more general model 
wherever the independence condition in (\ref{independence_assumption_equn}) will hold. 

We will now proceed to formulate our partially observable model as a 
partially observable stochastic game (POSG). We will first formally describe the problem setting and then briefly 
discuss POSGs, before proceeding to our main results.  

\vspace{-2mm}
\subsection{Problem Formulation}
The actual state space of the system continues to be $\mathcal{X}$ (see~(\ref{SS_CO_equn})).
However, each forwarder now gets to observe only its part of the actual state (i.e., only its reward value) 
along with the relay's location. Thus, when the $k$-th relay arrives, 
and if both forwarders are still competing then the observations of $\mathscr{F}_1$ and $\mathscr{F}_2$
are of the form $(r_i,\ell)$ and $(\ell,r_j)$, respectively, where $(r_i,r_j)$ is the actual state,
 $L_k=\ell$ is the location of the $k$-th relay. 
Suppose $\mathscr{F}_2$ has already terminated before stage $k$ then\footnote{As mentioned earlier, $\mathscr{F}_1$
will come to know about $\mathscr{F}_2$'s termination by listening to the exchange of control packets
between $\mathscr{F}_2$ and the chosen relay just before termination.} the location information is 
no more required by $\mathscr{F}_1$, and hence we will denote its observation as $(r_i,\ts)$ which is
simply the system state. Finally, when $\mathscr{F}_1$ terminates we use $\ts$ to denote its subsequent observations.
Thus, we can write the \emph{observation space} of $\mathscr{F}_1$ as,
\vspace{-1mm}
\begin{eqnarray}
\mathcal{O}_1 
&=& \Big\{(r_i,\ell),(r_i,\ts),\ts:i\in[n],\ell\in[m]\Big\}.
\end{eqnarray}
Similarly, the observation space of $\mathscr{F}_2$ is given by
\begin{eqnarray}
\mathcal{O}_2 
&=& \Big\{(\ell,r_j),(\ts,r_j),\ts:j\in[n],\ell\in[m]\Big\}.
\end{eqnarray} 

\begin{definition}
We will modify\footnote{In this section we 
will apply overline to most of the symbols in order to distinguish them from the 
corresponding symbols that have already appeared in Section~\ref{completely_observable_case_section}.} 
 the definition of a policy pair, $(\overline{\pi}_1,\overline{\pi}_2)$
(see Definition~\ref{policy_defn}),  
such that
$\overline{\pi}_1:\mathcal{O}_1\rightarrow\{\textsf{s},\textsf{c}\}$
and $\overline{\pi}_2:\mathcal{O}_2\rightarrow\{\textsf{s},\textsf{c}\}$.
Thus, the decision to stop or continue by $\mathscr{F}_1$ and $\mathscr{F}_2$, when the $k$-th relay arrives
is based on their respective observations $o_{1,k}\in\mathcal{O}_1$ and 
$o_{2,k}\in\mathcal{O}_2$. 
\end{definition}

\emph{{Remark:}} Note that we have restricted the PO policies to be deterministic (and as before stationary),
i.e., $\overline{\pi}_1(o_1)$ is either \textsf{s} or \textsf{c} without mixing between
the two. Let $\Pi_D$ denote the set of all such deterministic policies. 
Restricting to $\Pi_D$ is primarily to simplify the analysis. However, it is not
immediately clear if a partially observable NEPP (to be formally defined very soon) 
should even exist within the class $\Pi_D$.
Our main result is to construct a Bayesian stage game and prove that 
this game contains pure strategy (or deterministic) NE vectors using which PO-NEPPs in $\Pi_D$
can be constructed.

Let $\{(O_{1,k},O_{2,k})$: $k\ge1\}$, denote
the sequence of joint-observation at stage $k$, and let $\{X_k: k\ge1\}$ as before denote the
sequence of states. Then
the expected cost incurred by $\mathscr{F}_\rho$, $\rho=1,2$, when the PO policy pair used is 
$(\overline{\pi}_1,\overline{\pi}_2)$, and when its initial 
observation is $o_\rho$, can be written as
\begin{eqnarray}
\label{PO_total_cost1_equn}
G^{(\rho)}_{\overline{\pi}_1,\overline{\pi}_2}(o_\rho)
= \sum_{k=1}^
\infty \mathbb{E}^{o_\rho}_{\overline{\pi}_1,\overline{\pi}_2}\bigg[g_\rho(X_k,(A_{1,k},A_{2,k}))\bigg],
\end{eqnarray}
where $A_{1,k}=\overline{\pi}_1(O_{1,k})$ and $A_{2,k}=\overline{\pi}_2(O_{2,k})$. 

Similar to the completely observable case, the objective for the partially 
observable (PO) case is to characterize PO-NEPPs which are defined as follows:
\begin{definition}
\label{PO_NEPP_defn}
We say that a PO policy pair $(\overline{\pi}_1^*,\overline{\pi}_2^*)$ is a PO-NEPP
if $G^{(1)}_{\overline{\pi}_1^*,\overline{\pi}_2^*}(o_1)\le G^{(1)}_{\overline{\pi}_1,\overline{\pi}_2^*}(o_1)$
for all $o_1\in\mathcal{O}_1$ and PO policy $\overline{\pi}_1\in\Pi_D$, and 
$G^{(2)}_{\overline{\pi}_1^*,\overline{\pi}_2^*}(o_2)\le G^{(2)}_{\overline{\pi}_1^*,\overline{\pi}_2}(o_2)$
where $o_2\in\mathcal{O}_2$ and $\overline{\pi}_2\in\Pi_D$.
\end{definition}

We will end this section with the expressions for the various cost terms corresponding to a
PO-NEPP, which are analogues of the cost terms in 
Section~\ref{completely_observable_case_section}.

\textbf{\emph{Various Cost Terms:}}
Recall the expression for $D^{(1)}$ from (\ref{d1_cost_equn}). Given a 
NEPP $(\pi_1^*,\pi_2^*)$, $D^{(1)}$ is the 
cost incurred by $\mathscr{F}_1$ if it continues alone. Similar expressions
can be written for a PO-NEPP $(\overline{\pi}_1^*,\overline{\pi}_2^*)$:
\begin{eqnarray}
\overline{D}^{(1)} &=& \tau + \sum_{i'}p^{(1)}_{i'} 
{G}^{(1)}_{\overline{\pi}_{1}^*,\overline{\pi}_2^*}(r_{i'},\ts).
\end{eqnarray}
Similarly, for $\mathscr{F}_2$, the cost of continuing alone is 
\begin{eqnarray}
\overline{D}^{(2)} &=& \tau + \sum_{j'}p^{(2)}_{j'} 
{G}^{(2)}_{\overline{\pi}_1^*,\overline{\pi}_2^*}(\ts,r_{j'}).
\end{eqnarray}	
The following lemma will be useful.
\begin{lemma}
\label{PO_costs_same_lemma}
Let $(\pi_1^*,\pi_2^*)$ be an NEPP and $(\overline{\pi}_1^*,\overline{\pi}_2^*)$ be a PO-NEPP
then
$J^{(1)}_{\pi_1^*,\pi_2^*}(r_i,\ts)=G^{(1)}_{\overline{\pi}_1^*,\overline{\pi}_2^*}(r_i,\ts)$ and
$J^{(2)}_{\pi_1^*,\pi_2^*}(\ts,r_j)=G^{(2)}_{\overline{\pi}_1^*,\overline{\pi}_2^*}(\ts,r_j)$. 
\end{lemma}
\begin{IEEEproof}
Whenever $\mathscr{F}_1$ is alone in the system, all its observations (which are of the form
$(r_i,\ts)$ until $\mathscr{F}_1$ terminates) are exactly the actual states 
traversed by the system. Hence the problem of obtaining 
$G^{(1)}_{\overline{\pi}_1^*,\overline{\pi}_2^*}(r_i,\ts)$
is identical to the MDP problem of obtaining $J^{(1)}_{\pi_1^*,\pi_2^*}(r_i,\ts)$ in Section~\ref{states_phi_section},
 so that
$G^{(1)}_{\overline{\pi}_1^*,\overline{\pi}_2^*}(r_i,\ts)$ satisfies the Bellman 
equation in (\ref{bellman_J1_equn}). Since the solution to (\ref{bellman_J1_equn})
is unique \cite{bertsekas-tsitsiklis91stochastic-shortest-path}
we obtain $J^{(1)}_{\pi_1^*,\pi_2^*}(r_i,\ts)=G^{(1)}_{\overline{\pi}_1^*,\overline{\pi}_2^*}(r_i,\ts)$. Similarly it follows that 
$J^{(2)}_{\pi_1^*,\pi_2^*}(\ts,r_j)=G^{(2)}_{\overline{\pi}_1^*,\overline{\pi}_2^*}(\ts,r_j)$.
\end{IEEEproof}

{\emph{Discussion:}} An immediate consequence of the above lemma is that $\overline{D}^{(1)}=D^{(1)}$
and $\overline{D}^{(2)}=D^{(2)}$. Further, if $(\overline{\pi}_1^*,\overline{\pi}_2^*)$ is a
PO-NEPP then for states of the form $(r_i,\ts)$, $\overline{\pi}_1^*(r_i,\ts)$
is same as $\pi_1^*(r_i,\ts)$ in (\ref{S1_policy_equn}). Similarly, for states
of the form $(\ts,r_j)$, $\overline{\pi}_2^*(\ts,r_j)$ is 
same as that in (\ref{S2_policy_equn}). 

However, the analogues of the cost terms $C^{(1)}_{\pi_1,\pi_2}$
and $C^{(2)}_{\pi_1,\pi_2}$ (recall (\ref{c1_cost_equn}) and (\ref{c2_cost_equn})) are different for
the partially observable case. The expressions for these are,
\begin{eqnarray}
\label{PO_c1_cost_equn}
\overline{C}^{(1)}_{\overline{\pi}_1,\overline{\pi}_2} &=& 
\tau + \sum_{\ell',i'}q_{\ell'}\cdot p^{(1)}_{i'|\ell'} \cdot 
G^{(1)}_{\overline{\pi}_1,\overline{\pi}_2}(r_{i'},\ell'),
\end{eqnarray}
\begin{eqnarray}
\label{PO_c2_cost_equn}
\overline{C}^{(2)}_{\overline{\pi}_1,\overline{\pi}_2} &=& \tau + \sum_{\ell',j'}q_{\ell'}\cdot p^{(2)}_{j'|\ell'} \cdot
G^{(2)}_{\overline{\pi}_1,\overline{\pi}_2}(\ell',r_{j'}).
\end{eqnarray}

Finally, similar to the result in Lemma~\ref{costs_ordered_corollary}, we can show that for 
a PO-NEPP $(\overline{\pi}_1^*,\overline{\pi}_2^*)$,
\begin{eqnarray}
\label{PO_costs_ordered_equn} 
\overline{D}^{(1)}\le\overline{C}^{(1)}_{\overline{\pi}_1^*,\overline{\pi}_2^*} \mbox{ and } 
\overline{D}^{(2)}\le\overline{C}^{(2)}_{\overline{\pi}_1^*,\overline{\pi}_2^*}.
\end{eqnarray}
The proof of these is along exactly the same lines as the proof of Lemma~\ref{costs_ordered_corollary}. 
We do not repeat it for brevity. 

\subsection{Partially Observable Stochastic Game (POSG)}
A POSG is a tuple $(\mathcal{N},\mathcal{X},\mathcal{O},\{\mathcal{A}_\rho\},\hat{T},\{{g}_\rho\})$,
where $\mathcal{N}$, $\mathcal{X}$, $\mathcal{A}_\rho$, and ${g}_\rho$ are as before 
(see Section~\ref{stochastic_game_subsection}), while
\begin{itemize}
\item $\mathcal{O}=\times_{\rho\in\mathcal{N}}\mathcal{O}_\rho$ is the joint-observation space, with  
$\mathcal{O}_\rho$ being the observation space of player $\rho$, and
\item $\hat{T}:\mathcal{X}\times\mathcal{O}\times\mathcal{A}\rightarrow\Delta(\mathcal{X}\times\mathcal{O})$
is the transition function where $\hat{T}(x',o'|x,o,a)$ is the probability that the next state and the 
joint-observation is $(x',o')$ conditioned on the event that
the current state, joint-observation and joint-action is $(x,o,a)$.
\end{itemize}

In the previous section we have seen that the NEPPs for a stochastic game
can be obtained by constructing a normal-form static stage game. 
Similarly for POSGs, there is work (for instance see, \cite{hansen-etal04dynamic-programming-POSG}) 
that constructs a game which is effectively played at each stage,
however, with the players not knowing the
exact state of the system the stage game now happens to be 
a \emph{Bayesian game} \cite[Chapter~9]{osborne03game-theory}.
Hence, the drawback with POSGs in general is that, at each stage $k$, each player 
needs to maintain a belief (distribution) about the
entire history of joint-observations and joint-actions,
\centerline{$((o_{1,1},o_{2,1}),(a_{1,1},a_{1,2}),\cdots, (a_{1,k-1},a_{2,k-1}),(o_{1,k},o_{2,k})$,}
(referred to as the \emph{joint-type of the Bayesian game}),
obtaining which for a general POSG is computationally intensive. 

For this reason the authors in \cite{dermed-etal11MGIIs}
have studied a restriction of POSGs referred to as,
Markov games of Incomplete information (MGII). In MGIIs the transition function $\hat{T}$
satisfies the following Markov property:
player-1's belief about the player-2's current observation, $o'_2$, is independent 
of player-2's previous observation, $o_2$, given the current state, $x'$, previous 
state, $x$, and player-1's current and previous observations, ${o'}_1$ and $o_1$, respectively, i.e.,
for two different observations $u,v\in\mathcal{O}_2$ of player-2,
$\hat{T}({o'}_2|x',x,{o'}_2,o_1,o_2=u)=
 \hat{T}({o'}_{2}|x',x,{o'}_{1},o_{1},o_{2}=v)$.
Similar Markov structure should hold for other players also.
For our case it is easy to check that the above condition is trivially satisfied, primarily
because all the associated random variables, $\{L_k\}$ and $\{(R_{1,k},R_{2,k})\}$, are i.i.d.\ across
the stage index $k$.

A major advantage with MGIIs is that the  \emph{current joint-observation constitutes the type 
of the Bayesian game to be played at that stage}. With this in mind, we will set up
a Bayesian stage game in the next section, with $(r_i,\ell)$
and $(\ell,r_j)$ constituting the type of the game at stage $k$, provided both forwarders
are still competing\footnote{When only one forwarder is present we already know that the 
solution can be obtained by solving an MDP problem as in Section~\ref{NEPP_characterization_ri_section}
(see Lemma~\ref{PO_costs_same_lemma}).} at stage $k$.

\subsection{Bayesian Stage Game}
\label{bayesian_game_section}
We are now ready to provide a solution to the partially observable case in terms
of a certain Bayesian game \cite[Chapter~9]{osborne03game-theory}
which is effectively played at any stage whenever both forwarders are contending.
For the completely observable case, given a policy pair $(\pi_1,\pi_2)$, corresponding
to each $(r_i,r_j)$ pair
we constructed the normal-form
game in Table~\ref{bimatrix_game_table}. 
However here, given a PO policy pair $(\pi_1,\pi_2)$ and given the observation 
$(r_i,\ell)$, $\mathscr{F}_1$'s belief that 
the game in Table~\ref{bimatrix_game_table} (with 
$(C^{(1)}_{\pi_1,\pi_2},C^{(2)}_{\pi_1,\pi_2})$ replaced by 
$(\overline{C}^{(1)}_{\overline{\pi}_1,\overline{\pi}_2}
\overline{C}^{(2)}_{\overline{\pi}_1,\overline{\pi}_2})$)
will be played is $p^{(2)}_{j|\ell}$, $j\in[n]$. Hence, $\mathscr{F}_1$ needs to first
compute the costs incurred for playing $\textsf{s}$ and $\textsf{c}$, averaged
over all observations $(\ell,r_j)$, $j\in[n]$, of $\mathscr{F}_2$.
We will formally develop these in the following. 

\textbf{\emph{Strategy vectors and corresponding costs:}}
Fixing the PO-policy pair to be $(\overline{\pi}_1,\overline{\pi}_2)$
(unless otherwise stated), we will refer to the subsequent development (which includes, the strategy vectors, 
various costs, best responses and NE vectors, to be discussed next) as the \emph{Bayesian game corresponding
to $(\overline{\pi}_1,\overline{\pi}_2)$}, denoted $\mathcal{G}(\overline{\pi}_1,\overline{\pi}_2)$.

\begin{definition}
For $\ell\in\mathcal{L}$ (recall that $\mathcal{L}$ is the set of possible relay locations),
we define a \emph{strategy vector}, $f_\ell$, of $\mathscr{F}_1$ as 
$f_{\ell}:\{r_i:i\in[n]\}\rightarrow\{\textsf{s},\textsf{c}\}$.
Similarly, a strategy vector $g_\ell$ of $\mathscr{F}_2$ is 
$g_\ell:\{r_j:j\in[n]\}\rightarrow\{\textsf{s},\textsf{c}\}$.
Thus, given the observation $(r_i,\ell)$ of $\mathscr{F}_1$, $f_\ell$
decides for $\mathscr{F}_1$ whether to stop or continue. 
\end{definition}

Now, given the strategy vector $g_\ell$ of $\mathscr{F}_2$, and the location information $\ell$,
$\mathscr{F}_1$'s belief that $\mathscr{F}_2$ will choose action \textsf{c} is
\begin{eqnarray}
\label{continuing_probability_equn}
\widetilde{g}_\ell&=&\sum_{j:g_\ell(r_j)=\textsf{c}} p^{(2)}_{j|\ell};
\end{eqnarray}
$(1-\widetilde{g}_\ell)$ is the probability that $\mathscr{F}_2$ will stop.
Thus, the expected cost incurred by $\mathscr{F}_1$ for playing 
\textsf{s} when its observation is $(r_i,\ell)$
and when $\mathscr{F}_2$ uses ${g}_\ell$ is
\begin{eqnarray}
\label{PO_stopping_cost1_equn}
C_{\st,g_\ell}^{(1)}(r_i) &=& \widetilde{g}_\ell (-\eta_1 r_i) + 
(1-\widetilde{g}_\ell) E^{(1)}(r_i),
\end{eqnarray}
where, recall from (\ref{e1_cost_equn}) that
 $E^{(1)}(r_i)=\nu_1(-\eta_1 r_i) +\nu_2 {D}^{(1)}$. 
The various terms in (\ref{PO_stopping_cost1_equn}) can be understood as follows: 
$\widetilde{g}_\ell$ is the probability that $\mathscr{F}_2$ will 
continue in which case $\mathscr{F}_1$ (having chosen the action $\st$) stops, incurring 
a terminating cost of $-\eta_1 r_i$, while $(1-\widetilde{g}_\ell)$ is the probability that 
$\mathscr{F}_2$ will stop in which case the expected cost is, $\nu_1(-\eta_1 r_i) +\nu_2 {D}^{(1)}$;
$\nu_1$ is the probability that $\mathscr{F}_1$ gets the relay and terminates incurring a cost of
$(-\eta_1 r_i)$,
otherwise w.p.\ $\nu_2$, $\mathscr{F}_2$ gets the relay in which case $\mathscr{F}_1$ continues alone, the expected 
cost of which is $\overline{D}^{(1)}=D^{(1)}$ (from Lemma~\ref{PO_costs_same_lemma}). 

The expected cost of continuing when $\mathscr{F}_1$'s observation is $(r_i,\ell)$ is
\begin{eqnarray}
\label{PO_continuing1_equn}
C_{\ct,{g}_\ell}^{(1)}(r_i) &=& \widetilde{g}_\ell \overline{C}^{(1)}_{\overline{\pi}_1,\overline{\pi}_2} + 
(1-\widetilde{g}_\ell) D^{(1)}.
\end{eqnarray}
From the above expression we see that the cost of continuing is 
a constant in the sense that it does not depend on the value of $r_i$.
Hence we will denote it as simply $C_{\ct,g_\ell}^{(1)}$. 
Further, note that $C_{\ct,g_\ell}^{(1)}$
depends on the PO policy pair $(\overline{\pi}_1,\overline{\pi}_2)$, but for
simplicity we have not shown this dependence in the notation for $C_{\ct,g_\ell}^{(1)}$. 

Similarly for $\mathscr{F}_2$, when its observation is $(\ell,r_j)$ 
and when $\mathscr{F}_1$ uses ${f}_\ell$, 
we have
\begin{eqnarray*}
C^{(2)}_{\st,{f}_\ell}(r_j) &=& \widetilde{f}_\ell (-\eta_2 r_j) + 
(1-\widetilde{f}_\ell) E^{(2)}(r_j) \\
C_{\ct,{f}_\ell}^{(2)} &=& \widetilde{f}_\ell \overline{c}^{(2)}_{\overline{\pi}_1^*,\overline{\pi}_2^*} + 
(1-\widetilde{f}_\ell) D^{(2)},
\end{eqnarray*}
where $\widetilde{f}_\ell=\underset{i:{f}_\ell(r_i)=\textsf{c}}{\sum} p^{(1)}_{i|\ell}$.

\begin{definition}
\label{best_response_definition}
We say that ${f}_\ell$ is the \emph{best response vector of $\mathscr{F}_1$} against the 
strategy vector ${g}_\ell$ played by $\mathscr{F}_2$, denoted 
${f}_\ell=BR_1({g}_\ell)$, if ${f}_\ell(r_i)=\textsf{s}$
iff $C^{(1)}_{\st,{g}_\ell}(r_i)\le C^{(1)}_{\ct,{g}_\ell}$. 
Note that such an ${f}_\ell$ is unique. 
Similarly, ${g}_\ell$ is the (unique) best response against ${f}_\ell$ if, 
${g}_\ell(r_j)=\textsf{s}$
iff $C^{(2)}_{\st,{f}_\ell}(r_j)\le C^{(2)}_{\ct,{f}_\ell}$. We denote this as 
${g}_\ell=BR_2({f}_\ell)$.
\end{definition}

\begin{definition}
\label{PO_NE_vector_defn}
For $\ell\in\mathcal{L}$, a pair of strategy vectors $({f}_\ell^*,{g}_\ell^*)$ is said to be a 
\emph{Nash equilibrium (NE) vector} 
for the game $\mathcal{G}(\overline{\pi}_1,\overline{\pi}_2)$ 
iff ${f}_\ell^*=BR_1({g}_\ell^*)$, and ${g}_\ell^*=BR_2({f}_\ell^*)$.
\end{definition}

As remarked earlier, it is not immediately clear whether a 
NE vector should even exist among the 
pure strategies for the game $\mathcal{G}(\overline{\pi}_1,\overline{\pi}_2)$. Our main result in the next section 
(Theorem~\ref{NE_existance_theorem}) is to provide a positive answer to this. 
In fact, we will not only prove the existence of NE vectors but also
provide a method to construct them.

We will end this section
with the following theorem which is similar to
Theorem~\ref{filar_theorem}-(b), that was 
used to obtain NEPPs. This theorem will enable us to  
construct PO-NEPPs.

\begin{theorem}
\label{filar_theorem_PO}
Given a PO policy pair $(\overline{\pi}_1^*,\overline{\pi}_2^*)$,
construct the strategy vector pair $\{({f}_\ell^*,{g}_\ell^*):\ell\in\mathcal{L}\}$ as follows: 
$f_\ell^*(r_i)=\overline{\pi}_1^*(r_i,\ell)$ and 
${g}_\ell^*(r_j)=\overline{\pi}_2^*(\ell,r_j)$ for all $i,j\in[n]$.
Now, suppose for each $\ell$, $({f}_\ell^*,{g}_\ell^*)$ is a NE vector 
for the game $\mathcal{G}(\overline{\pi}_1^*,\overline{\pi}_2^*)$ such that, 
\begin{eqnarray}
\label{min_costs1_equn}
\min\Big\{C^{(1)}_{\st,{g}_\ell^*}(r_i),C^{(1)}_{\ct,{g}_\ell^*}\Big\}&=&
G^{(1)}_{\overline{\pi}_1^*,\overline{\pi}_2^*}(r_i,\ell), \mbox{ and }\\
\label{min_costs2_equn}
\min\Big\{C^{(2)}_{\st,f_\ell^*}(r_j),C^{(2)}_{\ct,f_\ell^*}\Big\}&=&
G^{(2)}_{\overline{\pi}_1^*,\overline{\pi}_2^*}(\ell,r_j).
\end{eqnarray}
Then $(\overline{\pi}_1^*,\overline{\pi}_2^*)$ is a PO-NEPP.
\end{theorem}
\begin{IEEEproof}
See Appendix~\ref{filar_theorem_PO_appendix}.
\end{IEEEproof}

\emph{Discussion:} If $\{({f}_\ell^*,{g}_\ell^*)\}$ happens to be a NE vector, then 
from Definition~\ref{PO_NE_vector_defn} it simply follows that the LHS of (\ref{min_costs1_equn}) (resp.~(\ref{min_costs2_equn}))
is simply the cost incurred by $\mathscr{F}_1$ (resp.~$\mathscr{F}_2$) for playing the action, $f_\ell^*(r_i)$ 
(resp.~${g}_\ell^*(r_j)$), suggested by its NE vector. Thus, (\ref{min_costs1_equn})
and (\ref{min_costs2_equn}) collective say that the cost-pair obtained by playing the 
NE vector $({f}_\ell^*,{g}_\ell^*)$ in the Bayesian game $\mathcal{G}(\overline{\pi}_1^*,\overline{\pi}_2^*)$, 
is equal to the cost-pair incurred by the PO policy pair $(\overline{\pi}_1^*,\overline{\pi}_2^*)$ in the original POSG. 
Hence, this result could be thought of as the analogue of Theorem~\ref{filar_theorem}-(b) proved for the 
completely observable case. 

\textbf{\emph{Existence of a NE Vector:}}
We will fix a PO policy pair $(\overline{\pi}_1^*,\overline{\pi}_2^*)$
that satisfies the inequalities in (\ref{PO_costs_ordered_equn}). In this section
we will prove that 
there exists a NE vector for
$\mathcal{G}(\overline{\pi}_1^*,\overline{\pi}_2^*)$. 
Before proceeding to the main theorem we need the following results 
(Lemma~\ref{best_response_PO_lemma} and \ref{best_response_ordering_lemma}).
\begin{lemma}
\label{best_response_PO_lemma}
For any $\ell\in\mathcal{L}$, the best response vector, $f_\ell$, against any 
vector $g_\ell$ of $\mathscr{F}_2$ is 
a \emph{threshold vector}, i.e., there exists an $\Phi_\ell\in\{0,1,\cdots,n\}$ such that 
$f_\ell(r_i)=\textsf{s}$ iff $i>\Phi_\ell$. We refer to $\Phi_\ell$ as the 
\emph{threshold of $f_\ell$}.
Similarly, if $g_\ell$ is the best response against any vector $f_\ell$ of $\mathscr{F}_1$,
then $g_\ell$ is a threshold vector with threshold $\Psi_\ell$.
\end{lemma}
\begin{IEEEproof}
Since $r_{i'}\le r_i$ whenever $i'\le i$, we can write
$C^{(1)}_{\st,g_\ell}(r_{i'})\ge C^{(1)}_{\ct,g_\ell}(r_i)$ (see (\ref{PO_stopping_cost1_equn})). Then the proof follows by recalling 
Definition~\ref{best_response_definition}.
\end{IEEEproof}

{\emph{Remark:}} The above lemma is possible primarily because
of the independence assumption we had imposed at the beginning of
Section~\ref{partially_observable_case_section}. 
Suppose we had worked with the model where, given only $r_i$, $\mathscr{F}_1$'s
belief about $\mathscr{F}_2$'s observation is simply the conditional p.m.f.\ $p_{R_1,R_2}(r_j|r_i)$, $j\in[n]$,
then, as in (\ref{continuing_probability_equn}), we can write the expression for the
continuing probability as
\begin{eqnarray}
\widetilde{g}_{\ell,r_i}= \sum_{j:g_\ell(r_j)=\textsf{c}} p_{R_1,R_2}(r_j|r_i),
\end{eqnarray}
which is now a function of $r_i$. If we replace $\widetilde{g}_{\ell}$ in
(\ref{PO_stopping_cost1_equn}) by $\widetilde{g}_{\ell,r_i}$ it is not
possible to conclude, $C^{(1)}_{s,g_\ell}(r_{i'})\ge C^{(1)}_{s,g_\ell}(r_i)$
whenever $i'\le i$, as required for the proof of the above lemma. 

The following is an immediate consequence of Lemma~\ref{best_response_PO_lemma}: 
if $(f_\ell^*,g_\ell^*)$ is a 
NE vector then $f_\ell^*$ and $g_\ell^*$ are both threshold vectors.
Thus, we can restrict our search for NE vectors over the class of
all pairs of threshold vectors. Since a threshold vector $f_\ell$
can be equivalently represented by its threshold $\Phi_\ell$ 
we can alternatively work with the thresholds. Thus $\Phi_\ell\in\mathcal{A}_0:=\{0,1,\cdots,n\}$
represents the $n+1$ thresholds that $\mathscr{F}_1$ can use. $\Phi_\ell=0$ (respectively, $\Phi_\ell=n$) 
represents the threshold vector which, when used by $\mathscr{F}_1$, stops (respectively, continues) 
for any value of $r_i$ when the location is $\ell$.
Similarly, we will represent the $n+1$ thresholds that $\mathscr{F}_2$ can use by $\Psi_\ell\in\mathcal{A}_0$.
We will write $\Phi_\ell=BR_1(\Psi_\ell)$ whenever their corresponding threshold vectors, $f_\ell$ 
and $g_\ell$, respectively, are such that $f_\ell=BR_1(g_\ell)$. 
Similarly, we will write $\Psi_\ell=BR_2(\Phi_\ell)$ whenever 
$g_\ell=BR_2(f_\ell)$.

\begin{lemma}
\label{best_response_ordering_lemma}
(1) Let $\Psi_\ell,\Psi_\ell^{o}\in\mathcal{A}_0$ be two thresholds of $\mathscr{F}_2$ such that $\Psi_\ell<\Psi_\ell^{o}$,
then the best response of $\mathscr{F}_1$ to these are ordered as, $BR_1(\Psi_\ell)\ge BR_1(\Psi_\ell^{o})$.
(2) Similarly, if $\Phi_\ell,\Phi_\ell^{o}\in\mathcal{A}_0$ are two thresholds of $\mathscr{F}_1$ such that $\Phi_\ell<\Phi_\ell^{o}$
then $BR_2(\Phi_\ell)\ge BR_2(\Phi_\ell^{o})$.
\end{lemma}
\begin{IEEEproof}
See Appendix~\ref{best_response_ordering_lemma_appendix}.
\end{IEEEproof}

We are now ready to prove the following main theorem.
We will present the complete proof here because the proof technique 
will be required in the next section 
to construct PO-NEPPs.
\begin{theorem}
\label{NE_existance_theorem}
For every $\ell\in\mathcal{L}$, there exists a NE vector $(f_\ell^*,g_\ell^*)$ for the game
$\mathcal{G}(\overline{\pi}_1^*,\overline{\pi}_2^*)$. 
\end{theorem}
\begin{IEEEproof}
As mentioned earlier, a consequence of Lemma~\ref{best_response_PO_lemma} is that
it is sufficient to restrict our search for NE vectors within the class of all pairs
of threshold vectors. Let $\mathcal{A}_0:=\{\Phi_\ell:0\le \Phi_\ell\le n\}$ denote
the set of all $n+1$ thresholds of $\mathscr{F}_1$. Now,
for $1\le k\le n$, inductively define the sets $\mathcal{B}_k$ and $\mathcal{A}_k$
as follows:
$\mathcal{B}_k = \Big\{BR_2(\Phi_\ell): \Phi_\ell\in\mathcal{A}_{k-1} \Big\}$ and 
$\mathcal{A}_k = \Big\{BR_1(\Psi_\ell): \Psi_\ell\in\mathcal{B}_k \Big\}$.

It is easy to check that through this \emph{inductive process} we will finally end up 
with non-empty sets
$\mathcal{A}_n$ and $\mathcal{B}_n$ such that
\begin{itemize}
\item for each $\Phi_\ell\in\mathcal{A}_n$  there exists a unique $\Psi_\ell\in\mathcal{B}_n$ 
such that $\Phi_\ell=BR_1(\Psi_\ell)$, and
\item for each $\Psi_\ell\in\mathcal{B}_n$ there exists a unique 
$\Phi_\ell\in\mathcal{A}_n$ such that $\Psi_\ell=BR_2(\Phi_\ell)$.
\end{itemize}
Since best responses are unique, these would also mean that $|\mathcal{A}_n|=|\mathcal{B}_n|$.

Note that there is nothing special about this inductive process, in the sense that 
for any normal form game with two player, each of whose action set is $\mathcal{A}_0$,
this inductive process will still yield sets $\mathcal{A}_n$ and $\mathcal{B}_n$ 
satisfying the above properties whenever the best responses are unique.
However, it is possible that there exists no pair $(\Phi_\ell,\Psi_\ell)\in\mathcal{A}_n\times\mathcal{B}_n$
such that $\Phi_\ell=BR_1(\Psi_\ell)$ and $\Psi_\ell=BR_2(\Phi_\ell)$. 
For instance, $\mathcal{A}_n=\{\Phi_\ell,\Phi_\ell'\}$, $\mathcal{B}_n=\{\Psi_\ell,\Psi_\ell'\}$
and $BR_2(\Phi_\ell)=\Psi_\ell$ and $BR_2(\Phi_\ell')=\Psi_\ell'$ while $BR_1(\Psi_\ell)=\Phi_\ell'$ and 
$BR_1(\Psi_\ell')=\Phi_\ell$.
This is precisely where Lemma~\ref{best_response_ordering_lemma} will be useful, due
to which such a situation cannot arise in our case.

Now, arrange the $N=|\mathcal{A}_n|$ $(=|\mathcal{B}_n|)$ remaining thresholds in $\mathcal{A}_n$
and $\mathcal{B}_n$ as, $\Phi_{\ell,1}<\Phi_{\ell,2}<\cdots<\Phi_{\ell,N}$ and 
$\Psi_{\ell,1}<\Psi_{\ell,2}<\cdots<\Psi_{\ell,N}$, respectively. Then
$\Phi_{\ell,1}=BR_1(\Psi_{\ell,N})$, since if not then 
using Lemma~\ref{best_response_ordering_lemma} we can write
$\Phi_{\ell,1}<BR_1(\Psi_{\ell,N})\le BR_1(\Psi_{\ell,t})$ for every $t=1,2,\cdots,N$ 
contradicting the fact that $\Phi_{\ell,1}$ being in $\mathcal{A}_n$ has to be the
best response for some $\Psi_{\ell,t}\in\mathcal{B}_n$. Similarly 
$\Psi_{\ell,N}=BR_2(\Phi_{\ell,1})$, otherwise again from Lemma~\ref{best_response_ordering_lemma}
we obtain $\Psi_{\ell,N}>BR_2(\Phi_{\ell,1})\ge BR_2(\Phi_{\ell,t})$ for every $t=1,2,\cdots,N$
leading to a contradiction that $\Psi_{\ell,N}$ is not the best response 
of any $\Phi_{\ell,t}\in\mathcal{A}_n$. Thus the threshold strategy pair 
$(f_\ell^*,g_\ell^*)$ corresponding 
to the threshold pair $(\Phi_{\ell,1},\Psi_{\ell,N})$ is a NE vector. 
By an inductive argument, it can be shown that all the 
threshold vector pairs corresponding to the 
threshold pairs $(\Psi_{\ell,t},\Psi_{\ell,N-(t-1)})$, $t=1,2,\cdots,N$, are NE vectors.
\end{IEEEproof}

\vspace{-4mm}
\subsection{PO-NEPP Construction from NE Vectors}
Once we have obtained NE vectors $(f_\ell^*,g_\ell^*)$, for each $\ell\in[m]$, 
The procedure for constructing PO-NEPP 
from NE vectors
is along the same lines as the construction of NEPP from NE strategies 
(see Section~\ref{NEPP_construction_section}). 

We begin with a pair of cost terms,
$\overline{\mathbf{C}}=(\overline{C}^{(1)},\overline{C}^{(2)})$,
satisfying (\ref{PO_costs_ordered_equn}). 
Using the procedure in the proof of Theorem~\ref{NE_existance_theorem},
we obtain, for each $\ell\in\mathcal{L}$, the NE vector 
$(f^\nabla_\ell,g^\nabla_\ell)$ corresponding to the threshold pair 
$(\Phi_{\ell,1},\Psi_{\ell,N})$ ($\mathscr{F}_1$ using lowest threshold while $\mathscr{F}_2$ uses the highest). Then we
define 
\begin{eqnarray*}
G^{(1)}(r_i,\ell)&=&\min\Big\{C^{(1)}_{\st,g_\ell^\nabla}(r_i),C^{(1)}_{\ct,g_\ell^\nabla}\Big\} \\
G^{(2)}(\ell,r_j)&=&\min\Big\{C^{(2)}_{\st,f_\ell^\nabla}(r_j),C^{(2)}_{\ct,f_\ell^\nabla}\Big\}.
\end{eqnarray*}
Now recall the expressions for the costs 
$\overline{C}^{(1)}$ and 
$\overline{C}^{(2)}$ from 
(\ref{PO_c1_cost_equn}) and (\ref{PO_c2_cost_equn}). 
Compute the RHS of these expressions by replacing 
$G^{(1)}_{\overline{\pi}_1^*,\overline{\pi}_2^*}(\cdot)$ and 
$G^{(2)}_{\overline{\pi}_1^*,\overline{\pi}_2^*}(\cdot)$ by the functions
$G^{(1)}(\cdot)$ and $G^{(2)}(\cdot)$, respectively. Denote the 
computed sums as $\overline{\mathcal{T}}_1(\overline{\mathbf{C}})$
and $\overline{\mathcal{T}}_2(\overline{\mathbf{C}})$, respectively.
Suppose $\overline{\mathbf{C}}$ is such that 
$\overline{\mathbf{C}}=(\overline{\mathcal{T}}_1(\overline{\mathbf{C}}),
\overline{\mathcal{T}}_2(\overline{\mathbf{C}}))$ (we inductively iterate to obtain 
such a $\overline{\mathbf{C}}$)
then using Theorem~\ref{filar_theorem_PO} we can construct the PO-NEPP, 
$(\overline{\pi}_1^\nabla,\overline{\pi}_2^\nabla)$ 
using $(f^\nabla_\ell,g^\nabla_\ell)$ as follows: for each 
$i,j\in[n]$ and $\ell\in\mathcal{L}$,
$\overline{\pi}_1^\nabla(r_i,\ell)=f_\ell^\nabla(r_i)$ { and }
$\overline{\pi}_2^\nabla(\ell,r_j)=g_\ell^\nabla(r_j)$. 

Finally, since the threshold vector $(f_\ell^\Delta,g_\ell^\Delta)$ corresponding to 
the threshold pair $(\Phi_{\ell,N},\Psi_{\ell,1})$ 
($\mathscr{F}_1$ using highest threshold while $\mathscr{F}_2$ uses the lowest) is also a NE vector, 
one can similarly construct the PO-NEPP,
 $(\overline{\pi}_1^\Delta,\overline{\pi}_2^\Delta)$, using 
$(f_\ell^\Delta,g_\ell^\Delta)$. 

\section{Cooperative Case}
\label{cooperative_case_section}
It will be interesting to benchmark the best performance that can be 
achieved if both forwarders would \emph{cooperate} with each other. In this section, we will 
describe this case and construct a \emph{Pareto optimal} performance curve.

We will assume the completely observable case. The definition of a policy pair 
$(\pi_1,\pi_2)$ and the costs $J^{(1)}_{\pi_1,\pi_2}(x)$ and $J^{(2)}_{\pi_1,\pi_2}(x)$ will remain
as in Section~\ref{completely_observable_case_section}. However, here our objective is instead
to optimize a linear combination of the two costs. Formally, let $\gamma\in(0,1)$, then 
the problem we are interested in is,
\begin{eqnarray}
\label{cooperative_problem_equn}
\mbox{Minimize}_{(\pi_1,\pi_2)} \Big(\gamma J^{(1)}_{\pi_1,\pi_2}(x) + (1-\gamma) J^{(2)}_{\pi_1,\pi_2}(x)\Big).
\end{eqnarray}
Let  $(\pi_1^\gamma,\pi_2^\gamma)$ denote the policy pair which is optimal for the 
above problem. Then, using (\ref{c1_cost_equn}) and (\ref{c2_cost_equn}), it is
easy to show that $(\pi_1^\gamma,\pi_2^\gamma)$ is also optimal for 
\begin{eqnarray}
\label{cooperative_problem_c_equn}
\mbox{Minimize}_{(\pi_1,\pi_2)}\ \Big(\gamma C^{(1)}_{\pi_1,\pi_2} + (1-\gamma) C^{(2)}_{\pi_1,\pi_2}(x)\Big).
\end{eqnarray}
\noindent
We have the following lemma.
\begin{lemma}
\label{pareto_lemma}
The policy pair $(\pi_1^\gamma,\pi_2^\gamma)$ is \emph{Pareto optimal}, i.e.,
for any other policy $(\pi_1,\pi_2)$,
\begin{itemize}
\item[(1)] if $C^{(1)}_{\pi_1,\pi_2}<C^{(1)}_{\pi_1^\gamma,\pi_2^\gamma}$ then 
$C^{(2)}_{\pi_1^\gamma,\pi_2^\gamma}<C^{(2)}_{\pi_1,\pi_2}$, and
\item[(2)] if $C^{(2)}_{\pi_1,\pi_2}<C^{(2)}_{\pi_1^\gamma,\pi_2^\gamma}$ then 
$C^{(1)}_{\pi_1^\gamma,\pi_2^\gamma}<C^{(1)}_{\pi_1,\pi_2}$.
\end{itemize}
\end{lemma}

\begin{IEEEproof}
Available in Appendix~\ref{cooperative_case_section_appendix}.
\end{IEEEproof}

Thus, by varying $\gamma\in(0,1)$, we obtain a Pareto optimal boundary whose points are
$(C^{(1)}_{\pi_1^\gamma,\pi_2^\gamma},C^{(2)}_{\pi_1^\gamma,\pi_2^\gamma})$. Details on 
how to obtain $(\pi_1^\gamma,\pi_2^\gamma)$ is available in Appendix~\ref{cooperative_case_section_appendix}.

\section{Numerical and Simulation Results for the Geographical Forwarding Example}
\label{numerical_work_section}

%%%%%%%%%%%%%%%%%%%%%%%%%%%%%%%%%%%%%%%%%%%%%%%%%%%%%%%%%%%%%%%%%%%%%%%%%%%%%
\begin{figure*}[t!]
\centering
\subfigure[]{
\includegraphics[scale=0.3]{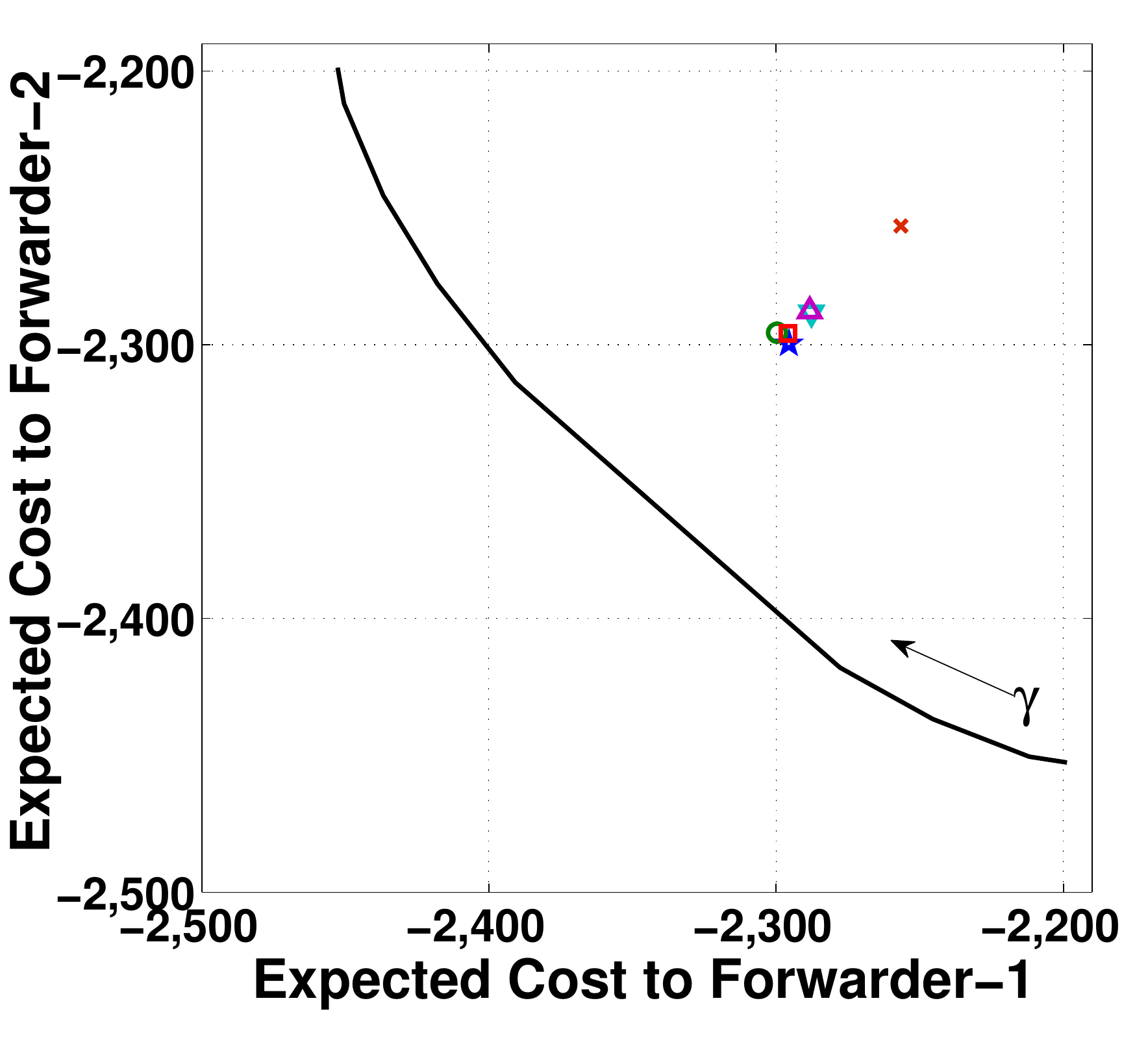}
        \label{theta0_figure}
}
\subfigure[]{
\includegraphics[scale=0.3]{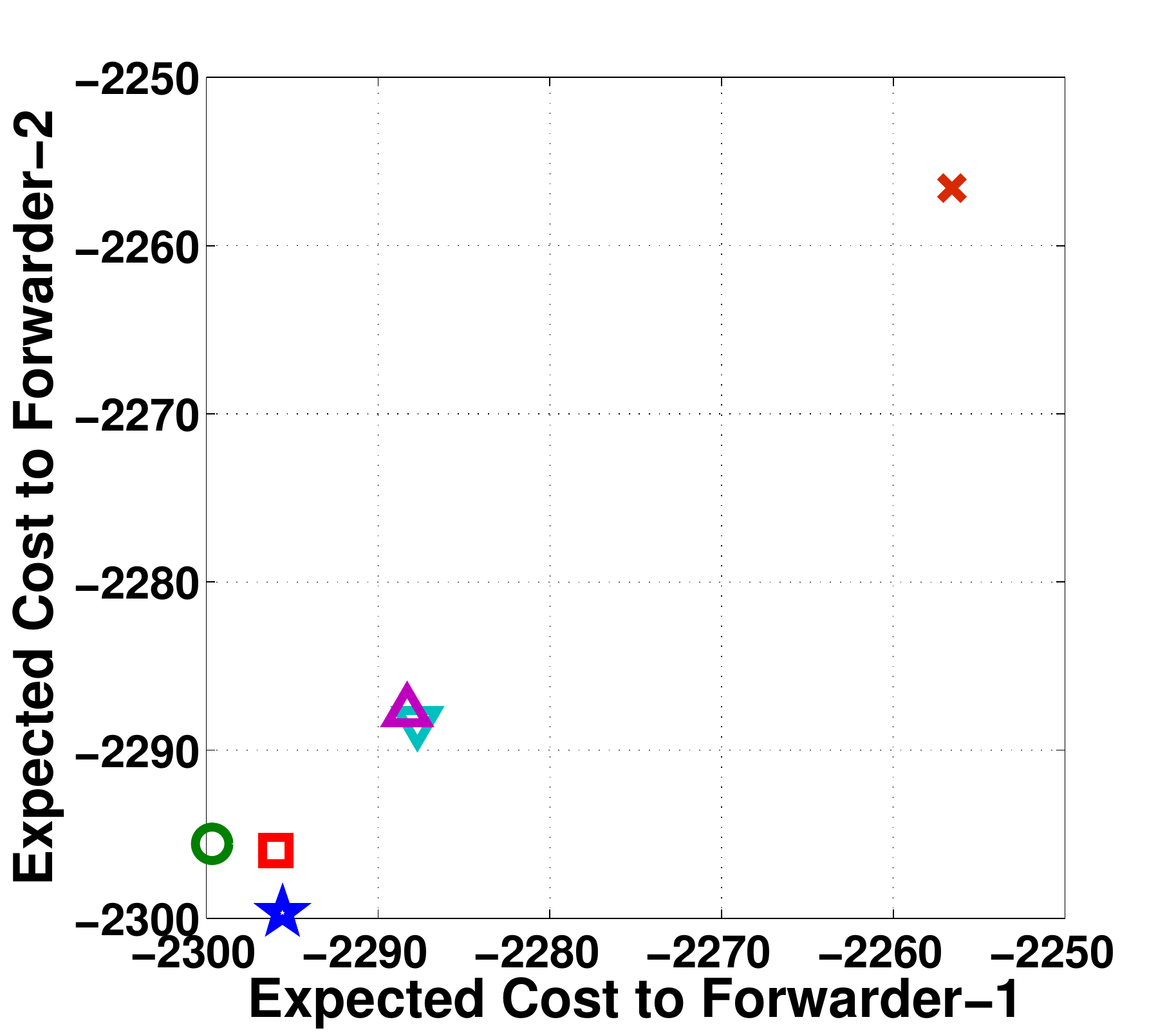}
        \label{theta0_zoomed_figure}
}
\subfigure[]{
\includegraphics[scale=0.3]{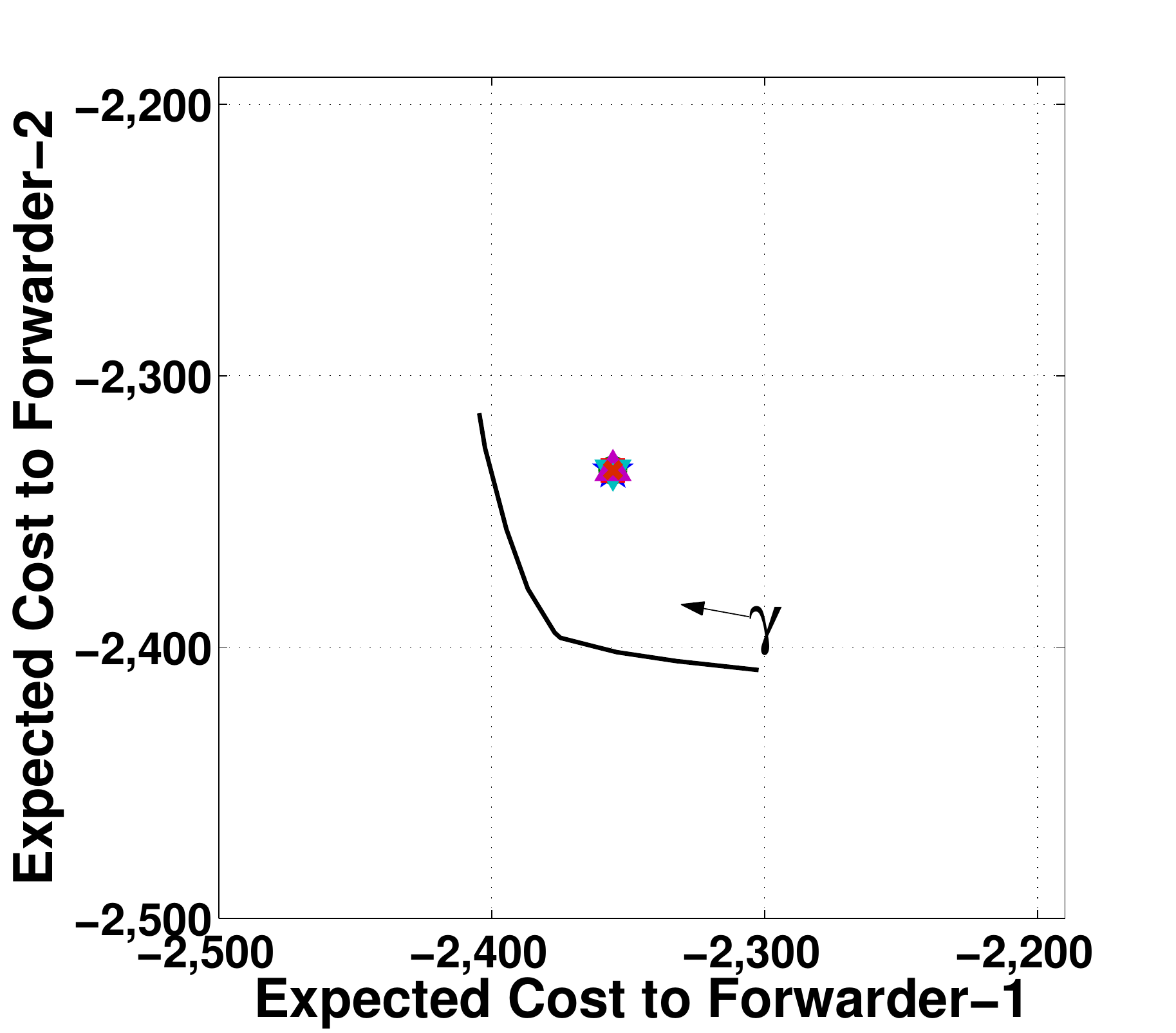}
        \label{theta3_zoomed_figure}
}
\caption{\label{PO_figures} Performance of the various NEPPs and PO-NEPPs are depicted as
points in $\Re^2$ where the first (second) coordinate is the expected cost incurred by $\mathscr{F}_1$ ($\mathscr{F}_2$). 
Fig.~\subref{theta0_figure} corresponds to the case when the distance of separation $\theta=0$ m.
A portion of Fig.~\subref{theta0_figure} is enlarged and shown in Fig.~\subref{theta0_zoomed_figure}.
Fig.~\subref{theta3_zoomed_figure} corresponds to $\theta=10$ m.}
\vspace{-6mm}
\end{figure*}
%%%%%%%%%%%%%%%%%%%%%%%%%%%%%%%%%%%%%%%%%%%%%%%%%%%%%%%%%%%%%%%%%%%%%%%%%%%%%%%%%%

\subsection{One-Hop Study}
The one-hop study can be more general, requiring only a joint p.m.f.\ $p_{i,j}$, a location p.m.f.\ $q_{\ell}$, and conditional 
p.m.f.s $p^{(1)}_{i|\ell}$ and $p_{j|\ell}^{(2)}$ (for all $i,j$ and $\ell$). However, to illustrate
the practicality of our study, we will study the geographical forwarding example described in Section~\ref{geographical_example_section}.

Recall the packet forwarding scenario illustrated in Fig.~\ref{one_hop_figure}.
We will fix the locations of $\mathscr{F}_1$ and $\mathscr{F}_2$ to be 
$v_1=[0,\frac{\theta}{2}]$ and $v_2=[0,-\frac{\theta}{2}]$,
respectively. Thus, the distance of separation between the two forwarders is $\theta$ meters (m);
we will vary $\theta$ and study the performance of the various policies. 
The range of each forwarder is $d=80$ m. The combined forwarding region is discretized 
into a uniform grid where the distance between the neighboring points is $5$ m.
Finally, the sink node is placed at $v_0=[1000,0]$. 

Next, recall the power and reward expressions from (\ref{power_equn}) and (\ref{reward_new_equn}), 
respectively. 
We have fixed $d_{ref}=5$ m, $\xi=2.5$, and $a=0.5$. For $\Gamma N_0$, which is referred to as 
the \emph{receiver sensitivity},
we use a value of $10^{-9}$ milliWatts (mW) (equivalently $-90$ dBm) specified for the 
Crossbow TelosB wireless mote \cite{telosb-datasheet}. The maximum transmit power available at 
a node is $P_{max}=1$ mW (equivalently $0$ dBm; again from the Crossbow TelosB data sheet). 
We allow for four different channel gain values: $0.4\times 10^{-3}$, $0.6\times 10^{-3}$,
$0.8\times 10^{-3}$, and $1\times 10^{-3}$, each occurring with equal probability. 
Finally, we fix $\eta_1=\eta_2=100$ (recall that $\eta_\rho$ 
is the parameter used to trade-off between delay and reward (see (\ref{cost1_equn})),
$\nu_1=1-\nu_2=0.5$ ($\nu_\rho$ is the probability that $F_\rho$ will win the contention),
and the mean inter-wake-up time $\tau=10$ milliseconds (ms).

We first set $\theta=0$ m (recall that $\theta$ is the distance between the two forwarders) and,
in Fig.~\ref{theta0_figure},  depict the performance of various NEPPs and
PO-NEPPs as pair of costs $\mathbf{C}=(C^{(1)},C^{(2)})$ where $C^{(\rho)}$ is the cost incurred by
$F_\rho$ starting from time $0$
if the particular NEPP or PO-NEPP is used. Also shown in Fig.~\ref{theta0_figure}  is the performance
of a \emph{simple policy} (the point
marked $\times$; to be describe next) along with the Pareto optimal boundary (the solid curve).
Since, from Fig.~\ref{theta0_figure} it is not easy to distinguish between the various points,
we show a section of Fig.~\ref{theta0_figure} as 
Fig.~\ref{theta0_zoomed_figure}. 
Fig.~\ref{theta3_zoomed_figure} corresponds to $\theta=10$ m.

\emph{\textbf{Various Policy Pairs:}} The description of various points seen in Fig.~\ref{PO_figures} 
is as follows (we will use $\mathbf{C}_{symbol}$ to denote the cost pair corresponding to the policy $symbol$):

\begin{itemize}
\item ${\bigstar}$,$\bigcirc$,and {$\Box$}: performances of the NEPPs that uses the NE strategies
$(\textsf{s},\textsf{c})$, $(\textsf{c},\textsf{s})$, and the mixed strategy
$(\Gamma_1,\Gamma_2)$, respectively, whenever
$(r_i,r_j)\in\mathcal{R}_4(\mathbf{C}_\bigstar)$,  
$\mathcal{R}_4(\mathbf{C}_\bigcirc)$,and $\mathcal{R}_4(\mathbf{C}_\Box)$, respectively
(recall Fig.~\ref{partition_figure_nosub}).

\item  $\nabla$ and $\Delta$: performances of the PO-NEPPs that are constructed by choosing, for each 
$\ell\in\mathcal{L}$, the thresholds $(\Phi_{\ell,1},\Psi_{\ell,N})$ and 
$(\Phi_{\ell,N},\Psi_{\ell,1})$, respectively
(recall the proof of Theorem~\ref{NE_existance_theorem}).

\item  $\times$: performance of a simple policy where each forwarder $\mathscr{F}_\rho$ ($\rho=1,2$) 
chooses $\textsf{s}$ if and only if its reward value $r_\rho\ge\alpha^{(\rho)}$.
Such a policy is optimal whenever $\mathscr{F}_\rho$ is alone in the system
(recall (\ref{S1_policy_equn}) and (\ref{S2_policy_equn})).
Thus, using the simple policy each forwarder behaves as if the competing forwarder is not present.

\item solid curve: Pareto optimal boundary obtained by $(\pi_1^\gamma,\pi_2^\gamma)$, $\gamma\in(0,1)$;
recall Section~\ref{cooperative_case_section}.
\end{itemize}

\textbf{\emph{Observations:}} 
From Fig.~\ref{theta0_zoomed_figure} 
we see that operating at NEPP $\bigstar$ is most favorable for $\mathscr{F}_2$ since 
$C_\bigstar^{(2)}$ is less than the cost to $\mathscr{F}_2$ at the other two NEPPs, $C^{(2)}_\bigcirc$ and
$C^{(2)}_\Box$. This is because whenever $(r_i,r_j)\in\mathcal{R}_4(\mathbf{C}_\bigstar)$
the joint-action $(\textsf{s},\textsf{c})$ played by $\bigstar$ fetches the least cost (of $D^{(2)}$) 
possibly by any strategy. 
In contrast, $\mathscr{F}_1$ incurs highest cost (of $-\eta_1 r_i$)
possible because of which NEPP $\bigstar$ is least favorable for $\mathscr{F}_1$.
For a similar reason, operating at NEPP $\bigcirc$ is most favorable for $\mathscr{F}_1$
while being least favorable for $\mathscr{F}_2$. 
The NEPP {${\Box}$} which chooses the mixed strategy $(\Gamma_{1},\Gamma_{2})$
whenever $(r_i,r_j)\in\mathcal{R}_4(\mathbf{C}_{\Box})$ helps to achieve a 
fairer cost to both players, however the performance at $\Box$ is slightly
farther from the Pareto boundary when compared with the other two NEPPs.

The performance at the PO-NEPPs, $\nabla$ and $\Delta$,
is worse than at the NEPPs thus exhibiting the loss in performance
due to partial information. The PO-NEPP $\nabla$ which uses the NE vector 
corresponding to the lowest-highest best response pair, $(\Phi_{\ell,1},\Psi_{\ell,N})$
(for each $\ell\in\mathcal{L}$),
provides lower cost to $\mathscr{F}_2$ than the PO-NEPP $\Delta$.
This is because, $\mathscr{F}_1$ using a lower threshold will essentially choose
an initial relay, thus leaving $\mathscr{F}_2$ alone in the system
which can now accrue a better cost. 
For a similar reason, operating at $\Delta$ leads to $\mathscr{F}_1$ 
achieving a lower cost. Finally, the simple policy $\times$
has the worst performance in comparison with all other points,
suggesting that it may not be wise to be operating using this policy pair.
However, as we increase the value of $\theta$ the performance 
of the simple policy improves, and interestingly for 
$\theta=10$ m (which is only $12.5\%$ of the forwarders' range of $80$ m)
we observe that the various points are practically indistinguishable
(note that the magnitude of the scales in plots Fig.~\ref{theta0_figure} and 
\ref{theta3_zoomed_figure} is the same).  We have  observed similar trend when 
$\eta_1=\eta_2$ and $a$ are set to different values.

{\textbf{\emph{Key Insight:}}} Thus, based on our numerical work we draw the following key insight:
even for a small distance of separation between the forwarders,
using the simple policy 
pair (where each forwarder behaves as if it is alone in the 
system) yields little (or, practically, no) loss in performance when compared with the performance of
an NEPP or a PO-NEPP; however the performance degradation of the simple policy is significant
whenever the forwarders are very close to each other. 
These observations are for the case where there are two forwarders. However, we expect a similar 
behavior for the simple policy even if there are more than two forwarders, i.e., we believe that
the simple policy performs well if the competing forwarders are moderately separated.

\vspace{-4mm}
\subsection{End-to-End Study}
Finally, in this section we use simulation to provide an evaluation 
of the end-to-end performance of local forwarding. The competitive forwarding 
policies (i.e., NEPP and PO-NEPP) are difficult to implement since their 
structure has to be evaluated for each forwarding instance along the path of a packet. However, 
based on our observations in the previous section, we study the performance of the simple policy pair.
In our prior work we have already studied the simple policy's performance 
(see \cite[Fig.~8]{naveen-kumar12relay-selection-TMC} where the simple policy is 
referred to as SF), 
but there the setting was that of the  \emph{lone packet model}
where a single alarm packet is generated which is then routed to the sink.
Here, we will generalize the lone packet setting by generating multiple packets simultaneously
across the network so that a packet, along its route, might have to compete with other packets in its vicinity
before reaching the sink. 

We first form a network by randomly placing $1000$ nodes in a square region of area $1$ Km$^2$.
A source node is placed at $[0,1000]$ followed by a sink node at the diagonally opposite corner $[1000,0]$.
Each node is allowed to {asynchronously and periodically} sleep-wake cycle
with period $T=100$ ms, i.e., each node $i$ wakes up 
and stays ON for a small duration (which we neglect, given the other time scales) at the 
periodic instants $T_i+kT$, $k\ge1$
where $\{T_i\}$ are i.i.d.\ uniform on $[0,T]$ (recall the discussion on the sleep-wake process from 
Section~\ref{system_model_section}). 

Each node $i$, assuming an 
inter-wake-up time of ${1}/{N_i}$ (where $N_i$ is the average number of 
nodes in the forwarding region of node $i$), 
obtains $\alpha_i$ which is the threshold (on reward) required to implement the simple
policy by node $i$. The values of all the other parameters required to compute the 
threshold, e.g., $P_{max}$, $\xi$, etc., remain the same as in our one-hop study.
If there is no relay whose reward value is more than
$\alpha_i$ (node $i$ will know of this
after waiting for one entire duty-cycling period $T$), node $i$, at time $T$, will simply 
forward the packet to the relay with the maximum reward (thus, as relays wake-up the best 
relay so far, is asked to wait). 

The source node generates an alarm packet at time $0$.
We introduce competition by generating additional packets at randomly chosen nodes, randomly in time
at the points of a Poisson process of rate $\lambda$. All the packets are destined for the same sink.
While forwarding, if a relay is  chosen simultaneously by more than one forwarder, 
then randomly one of them will win the contention and gets the relay to forward its packet to.
We are interested in studying, as a function of $\lambda$, the performance obtained (in terms of
end-to-end delay and the total power expended) in routing the source's packet.

\begin{figure}[t]
\centering
\includegraphics[scale=0.35]{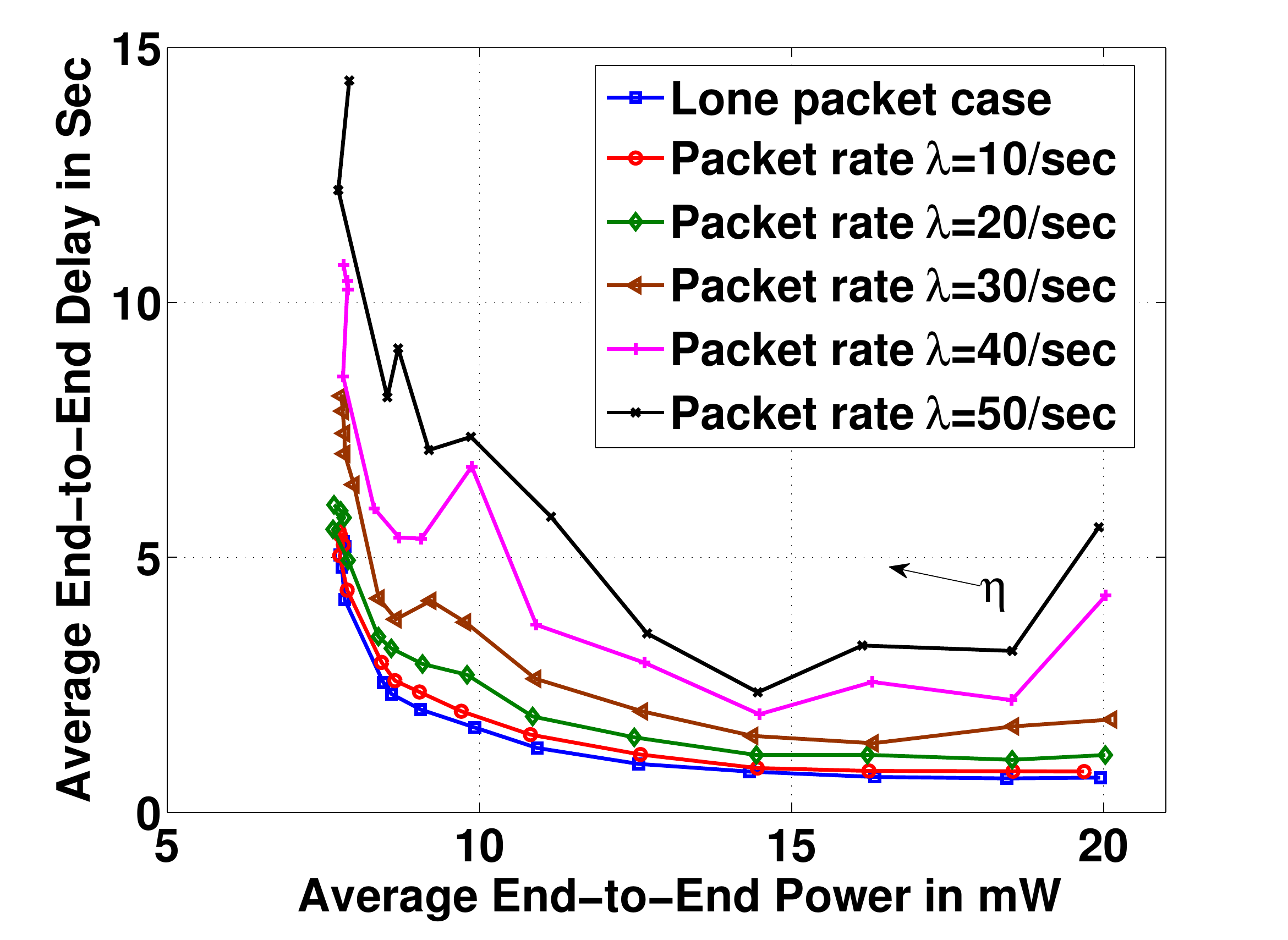}
\caption{\label{end_plot_figure} End-to-end performance 
(average power vs.\ average delay)
of the simple policy 
as the competing packet rate $\lambda$ in the network is increased.}
\vspace{-6mm}
\end{figure}

In Fig.~\ref{end_plot_figure} we have plotted, for different values of $\lambda$,
the mean end-to-end delay vs.\ the mean end-to-end power 
(averaged over packets from the source located at $(0,1000)$). 
These curves are obtained by varying $\eta$, the parameter used
to trade-off between delay and reward in the local problem.
Each data point in Fig.~\ref{end_plot_figure}
is the average of the respective quantities over $100$ alarm packets 
generated by the source node. 
Also shown in the figure is the performance
curve corresponding to the ``lone packet case'' where no additional packets are generated. 
Hence the lone packet curve is analogous to the  SF policy's performance curves in 
\cite[Fig.~8]{naveen-kumar12relay-selection-TMC}. 

Observe that,
as we increase $\lambda$ we obtain a \emph{degradation 
in performance}, i.e., increased delay and power compared with the lone packet case. 
This is because, as $\lambda$ increases, since there are more packets in the network, 
there is a larger probability that a forwarding node carrying the source's packet 
has to compete with other packets in the process of acquiring a relay. 
Also, as $\lambda$ increases, at these instances of competition, the competing 
nodes tend to be closer together. From the observations in the previous section, 
we can conclude that as $\lambda$ increases the performance of the 
simple policy will progressively degrade. 
However, the performance degradation is only marginal when the packet rate 
$\lambda\le 20$ packets/sec while being moderate for $\lambda=30$ packets/sec, thus supporting the usage of the 
simple policy for these packet rates.
For higher values of $\lambda$ (e.g., $\lambda=40$ packets/sec and beyond) the performance 
degradation is significant and hence 
there could be a benefit in using NEPPs to forward packets for these rates.

Finally, we have only presented simulation results for the simple policy, since
implementing NEPPs or PO-NEPPs for end-to-end routing has the following difficulties:
(1) for a given pair of neighboring nodes, obtaining NEPPs will require fixed point iterations,
(2) NEPPs are node pair dependent, so that all possible neighboring node pairs are required to compute the
corresponding NEPPs, since during actual forwarding a node may be competing with any of its neighbors. 
Thus, there is a large complexity involved in implementing NEPPs. In contrast, the simple policy
(being a single threshold based) is easy to implement.
Moreover, for realistic parameter values corresponding to 
TelosB wireless mote, we have seen that the performance of simple policy is good
(in comparison with the lone packet case) 
for packet rates $\lambda\le 30$ packets/sec. 

\vspace{-2mm}
\section{Conclusion}
\label{conclusion_section}
We studied the problem of competitive relay selection 
when two forwarders compete for a next-hop relay (or some resource in general).
We first considered the model where complete information is available to 
both the forwarders. We formulated the problem as a stochastic game and proceeded to obtain
solution in terms of Nash equilibrium policy pairs (NEPPs).
We were able to provide insight into the structure of NEPPs, 
which was primarily possible because
of our following key result (Lemma~\ref{costs_ordered_corollary}): ``cost of 
continuing alone'' is less than the ``cost of
continuing along with a competing forwarder''.
We next studied a partially observable case for which we constructed a Bayesian game
which is effective played at each stage. For this Bayesian game, we proved the 
existence of a Nash equilibrium strategy within the class of (pure) threshold 
vectors (Theorem~\ref{NE_existance_theorem}). The proof method of this result 
enabled us to construct NEPPs for the partial case. 
For the geographical forwarding example,
through numerical experiments we observed
that, even for moderate separation between the two forwarders, the 
performance of our simple policy 
is as good as the performance of any other NEPP/PO-NEPP.
In the context of end-to-end forwarding, 
through simulations we established (for the considered setting)
that for packet rates less than $30$ packets/second, the performance of the simple policy 
is good compared with the lone packet case.

\vspace{-2mm}
\bibliographystyle{IEEEtran}
\bibliography{IEEEabrv,stochastic_game}

\onecolumn
\appendices

\section{Proof of Lemma~\ref{competitive_fixed_point_lemma}}
\label{competitive_fixed_point_lemma_appendix}
For convenience, here in the appendix we will recall the respective Lemma/Theorem statement before providing
its proof.\\

\emph{Lemma}~\ref{competitive_fixed_point_lemma}:
$\alpha^{(1)}$ is the unique fixed point of $\beta^{(1)}(x)$ ($x\in(-\infty,r_n]$) in (\ref{beta1_equn}).

\begin{IEEEproof}
Let us recall the expression of $\beta^{(1)}(x)$:
\begin{eqnarray*}
\beta^{(1)}(x)
&=&\mathbb{E}\Big[\max\{x,R_1\}\Big]-\frac{\tau}{\eta_1},
\end{eqnarray*}
where the expection is w.r.t.\ the p.m.f.\ $p^{(1)}$ of $R_1$
(recall that $R_1$ takes values from the  set $\{r_1,r_2,\cdots,r_n\}$).

Let $m=\max\{i\in[n]: p^{(1)}_i>0\}$. 
For $x>r_m$, note that 
$\beta^{(1)}(x) = x-\frac{\tau}{\eta_1}<x$. Hence a fixed point, if any, should lie
within $(-\infty,r_{m}]$. 
Let us restrict $\beta^{(1)}(\cdot)$ to the domain $(-\infty,r_{m}]$. Then, 
since $\beta^{(1)}(x)\le r_m$ for any $x\in(-\infty,r_{m}]$, 
we have $\beta^{(1)}:(-\infty,r_{m}]\rightarrow(-\infty,r_{m}]$.
We can now proceed to show that $\beta^{(1)}(x)$
restricted to $x\in(-\infty,r_{m}]$ is a \emph{contraction mapping}, i.e., 
for any $x,x'\in(-\infty,r_{m}]$, we need to show that
\begin{eqnarray}
\label{fp_illust_equn}
\parallel\beta^{(1)}(x)-\beta^{(1)}(x')\parallel\le \kappa \parallel x-x'\parallel
\end{eqnarray}
for some $\kappa<1$. Without loss of generality let $x>x'$.
Then, 
\begin{eqnarray*}
\parallel\beta^{(1)}(x)-\beta^{(1)}(x')\parallel 
&=& \beta^{(1)}(x)-\beta^{(1)}(x') \\
&=& \mathbb{E}\Big[\max\{x,R_1\}\Big]-\mathbb{E}\Big[\max\{x',R_1\}\Big] \\
&=& \sum_{i=1}^n p^{(1)}_i \Big(\max\{x,r_i\}-\max\{x',r_i\}\Big) \\
&\overset{*}{=}& \sum_{i=1}^m p^{(1)}_i \Big(\max\{x,r_i\}-\max\{x',r_i\}\Big) \\
&\overset{o}{=}& \sum_{i=1}^{m-1} p^{(1)}_i \Big(\max\{x,r_i\}-\max\{x',r_i\}\Big)\\
&\overset{\dagger}{\le}& \sum_{i=1}^{m-1} p^{(1)}_i \Big(x-x'\Big) \\
&=& (1-p^{(1)}_m) \parallel x-x'\parallel.
\end{eqnarray*} 
In the above derivation, $*$ is because $p_i^{(1)}=0$ for $i>m$ (recall the definition of  $m$);
$o$ is because, since $x,x'\le r_m$, we have  $\Big(\max\{x,r_m\}-\max\{x',r_m\}\Big)=0$;
to obtain $\dagger$ note that, $\Big(\max\{x,r_i\}-\max\{x',r_i\}\Big)\le (x-x')$ for any $r_i$.
Thus, $\beta^{(1)}(x)$, $x\in(-\infty,r_m]$ is a contraction mapping (recall \ref{fp_illust_equn}) 
with $\kappa=(1-p^{(1)}_m)<1$ (since $p^{(1)}_m>0$ from definition).
Hence from the Banach fixed point theorem \cite{patayyfixed-point} it follows that there exists a unique fixed point
$\alpha^*\in(-\infty,r_m]$, i.e., $\alpha^*$ satisfies $\alpha^*=\beta^{(1)}(\alpha^*)$.

Now, suppose we can show that 
\begin{eqnarray}
\label{J_alpha_equn}
J^{(1)}_{\pi_1^*,\pi_2^*}(r_i,\ts)=\min\Big\{-\eta_1 r_i, -\eta_1 \alpha^*\Big\}
\end{eqnarray}
then, recalling the expression for $D^{(1)}$ from (\ref{d1_cost_equn}), we obtain
\begin{eqnarray*}
\alpha^{(1)}
&=& \frac{D^{(1)}}{-\eta_1}\\
&=& -\frac{\tau}{\eta_1}-\frac{1}{\eta_1}\sum_{i}p^{(1)}_i J^{(1)}_{\pi_1^*,\pi_2^*}(r_i,\ts) \\
&=& \mathbb{E}\Big[\max\{\alpha^*,R_1\}\Big]-\frac{\tau}{\eta_1}. \\
&=& \beta^{(1)}(\alpha^*)\\
&=& \alpha^*.
\end{eqnarray*}
Thus, $\alpha^{(1)}$ is the unique fixed point of $\beta^{(1)}(\cdot)$.

To show (\ref{J_alpha_equn}), we proceed as follows. Let $J_0(r_i)=0$ for all $r_i$,
and for $k\ge1$ define $J_k(r_i)$ inductively as 
\begin{eqnarray}
J_k(r_i)
&=& \min\Big\{-\eta_1 r_i, \tau + \mathbb{E}\Big[J_{k-1}(R_1)\Big]\Big\}. 
\end{eqnarray}
Since our problem with one player is equivalent to the optimal stopping problem studied in
\cite{bertsekas-tsitsiklis91stochastic-shortest-path},
the above iterations converge to the optimal cost, i.e.,
$\lim_{k\rightarrow\infty} J_k(r_i)=J^{(1)}_{\pi_1^*,\pi_2^*}(r_i,\ts)$.
Now, defining $\alpha_1=-\frac{\tau}{\eta_1}$, $J_1(r_i)$ can be written 
as $J_1(r_i)=\min\{-\eta_1 r_i,-\eta_1 \alpha_1\}$. Proceeding further we can write,
\begin{eqnarray*}
J_2(r_i) 
&=& \min\Big\{-\eta_1 r_i, \tau + \mathbb{E}\Big[J_{1}(R_1)\Big]\Big\} \\
&=& \min\Big\{-\eta_1 r_i, \tau + \mathbb{E}\Big[\min\{-\eta_1\alpha_1,-\eta_1R\}\Big]\Big\} \\
&=& \min\Big\{-\eta_1 r_i, -\eta_1 \beta^{(1)}(\alpha_1)\Big\} \\
&=& \min\Big\{-\eta_1 r_i, -\eta_1 \alpha_2\Big\}
\end{eqnarray*}
where $\alpha_2=\beta^{(1)}(\alpha_1)$. Similarly it can be shown that,
if $J_{k-1}(r_i)=\min\Big\{-\eta_1 r_i,-\eta_1 \alpha_{k-1}\Big\}$, then 
\begin{eqnarray}
J_{k}(r_i)=\min\Big\{-\eta_1 r_i,-\eta_1 \alpha_{k}\Big\}
\end{eqnarray}
where $\alpha_k=\beta^{(1)}(\alpha_{k-1})$. Thus $\alpha_k\rightarrow\alpha^*$.
Finally, in the above expression taking
the limit as $k\rightarrow\infty$ on both sides, and using $J_k(r_i)\rightarrow J^{(1)}_{\pi_1^*,\pi_2^*}(r_i,\ts)$
and $\alpha_k\rightarrow\alpha^*$, we obtain the desired result. 
\end{IEEEproof}

\section{Proof of Theorem~\ref{filar_theorem}}
\label{filar_theorem_appendix}
\vspace{2mm}
\emph{Theorem}~\ref{filar_theorem}:
Given a policy pair, $(\pi_1^*,\pi_2^*)$, construct the static game given in 
Table~\ref{bimatrix_game_table_appendix}.
\begin{table}[h]
\centering
\begin{tabular}{|c||c|c|}
\hline
	   & \textsf{c} & \textsf{s} \\
\hline
\hline
\textsf{c} & $C^{(1)}_{\pi_1^*,\pi_2^*},C^{(2)}_{\pi_1^*,\pi_2^*}$ & $D^{(1)},-\eta_2 r_j$ \\
\hline
\textsf{s} & $-\eta_1 r_i,D^{(2)}$ & 
$E^{(1)}(r_i),E^{(2)}(r_j)$\\
\hline 
\end{tabular} 
\vspace{1mm}
\caption{\label{bimatrix_game_table_appendix} Static stage game.}
\vspace{-4mm}
\end{table}
Then the following statements are equivalent:
\begin{enumerate}
\item[(a)] 
$(\pi_1^*,\pi_2^*)$ is an NEPP.
\item[(b)]
For any $x=(r_i,r_j)$, 
$(\pi_1^*(x),\pi_2^*(x))$ is a 
\emph{Nash equilibrium (NE) strategy}
for the game in Table~\ref{bimatrix_game_table_appendix}.
Further, the expected cost pair at this NE strategy is,
$(J^{(1)}_{\pi_1^*,\pi_2^*}(x),J^{(2)}_{\pi_1^*,\pi_2^*}(x))$.
\end{enumerate}

\begin{IEEEproof}[Proof of (a)$\implies$(b)] 
Suppose (a) is true, i.e., $(\pi_1^*,\pi_2^*)$ is an NEPP. Then, 
$\pi_1^*$ is the best response policy of $\mathscr{F}_1$ against the policy $\pi_2^*$ of $\mathscr{F}_2$.
Hence $\pi_1^*$ is optimal for the MDP problem, denoted $MDP_1(\pi_2^*)$, which is obtained by fixing
the policy $\pi_2^*$ of $\mathscr{F}_2$ (note that $MDP_1(\pi_2^*)$ is a time homogeneous MDP since
$\pi_2^*$ is stationary; recall Definition~\ref{policy_defn}).
Since (1) the states of the form $(\ts,r_j)$ are absorbing and cost free for $\mathscr{F}_1$, 
and (2) the policy of $\mathscr{F}_1$ which never stops incurs infinite cost to $\mathscr{F}_1$, it follows that
$MDP_1(\pi_2^*)$ is an optimal stopping problem 
\cite{bertsekas-tsitsiklis91stochastic-shortest-path}.
Hence, $J^{(1)}_{\pi_1^*,\pi_2^*}(x)$,  $x=(r_i,r_j)$
satisfies the following Bellman equation,
\begin{eqnarray}
\label{optimal_stopping_bellman_equn}
J^{(1)}_{\pi_1^*,\pi_2^*}(x)
&=&\min\Big\{C_\textsf{s}(x),C_\textsf{c}(x)\Big\}\nonumber\\
&=&\min\Big\{\pi_2^*(x,\textsf{c})(-\eta_1 r_i) +
\pi_2^*(x,\textsf{s}) E	^{(1)}(r_i), \nonumber \\
&&\hspace{4cm}\pi_2^*(x,\textsf{c}) C^{(1)}_{\pi_1^*,\pi_2^*} + \pi_2^*(x,\textsf{s}) D^{(1)} \Big\},
\end{eqnarray}
where $\pi_2^*(x,\textsf{c})$ (resp.\ $\pi_2^*(x,\textsf{s})$) is the probability that 
$\mathscr{F}_2$ will choose action $\textsf{c}$ (resp.\ $\textsf{s}$) when the state is $x$.
The two terms in the $\min$-expression above (denoted $C_\textsf{s}(x)$
and $C_\textsf{c}(x)$) are the expected cost to $\mathscr{F}_1$ for taking
actions $\textsf{s}$ and $\textsf{c}$, respectively. Note that
these costs are exactly the expected cost incurred by $\mathscr{F}_1$,
for playing actions $\textsf{s}$ and $\textsf{c}$, respectively,
in the static game in Table~\ref{bimatrix_game_table},
when the strategy of $\mathscr{F}_2$ is $\pi_2^*(x)$. Now $\pi_1^*$, being
optimal for $MDP_1(\pi_2^*)$, chooses action $\pi_1^*(x)\in\Delta(\{\textsf{s},\textsf{c}\})$ 
whichever gives a minimum cost or can randomize between the two if both the costs are equal.
Hence, it follows from the structure of (\ref{optimal_stopping_bellman_equn})
that $\pi_1^*(x)$ is the best
response against $\pi_2^*(x)$ for the game in Table~\ref{bimatrix_game_table}.
Further the cost to $\mathscr{F}_1$ for playing $\pi_1^*(x)$, from Table~\ref{bimatrix_game_table},
is $\min\{C_\textsf{s}(x),C_\textsf{c}(x)\}=J^{(1)}_{\pi_1^*,\pi_2^*}(x)$.

Similarly, by writing the Bellman equation corresponding to the $MDP_2(\pi_1^*)$
problem (which is obtained by fixing the policy $\pi_1^*$ of $\mathscr{F}_1$), we can conclude
that $\pi_2^*(x)$ is the best
response against $\pi_1^*(x)$ for the game in Table~\ref{bimatrix_game_table},
with the cost to player $\mathscr{F}_2$ being $J^{(2)}_{\pi_1^*,\pi_2^*}(x)$.\\

\emph{Proof of (b)$\implies$(a):} Given that the policy $(\pi_1^*,\pi_2^*)$ satisfies the 
condition in (b), let $\pi_1$ be any policy of $\mathscr{F}_1$.
Then, for any $x=(r_i,r_j)$, since $(\pi_1^*(x),\pi_2^*(x))$ is a NE strategy for the 
game in Table~\ref{bimatrix_game_table} with cost to $\mathscr{F}_1$ at equilibrium being
$J^{(1)}_{\pi_1^*,\pi_2^*}(x)$, we can write
\begin{eqnarray*}
J^{(1)}_{\pi_1^*,\pi_2^*}(x) &\le& \pi_1(x,\textsf{c})\Big(\pi_2^*(x,\textsf{c}) C^{(1)}_{\pi_1^*,\pi_2^*} +
\pi_2^*(x,\textsf{s})D^{(1)}\Big) + \\
&&\hspace{2cm}\pi_1(x,\textsf{s}) \Big(\pi_2^*(x,\textsf{c})(-\eta_1 r_i) + 
\pi_2^*(x,\textsf{s})E^{(1)}(r_i)\Big).
\end{eqnarray*}
LHS of the above expression is the cost incurred to $\mathscr{F}_1$ 
when the strategy played is $(\pi_1(x),\pi_2^*(x))$ (refer to (\ref{optimal_stopping_bellman_equn})).

Substituting for $D^{(1)}$, $E^{(1)}(r_i)$ and $C^{(1)}_{\pi_1^*,\pi_2^*}$,  
(from (\ref{d1_cost_equn}), (\ref{e1_cost_equn}) and (\ref{c1_cost_equn}), respectively)
in the above expression and then rearranging, we can write
\begin{eqnarray*}
J^{(1)}_{\pi_1^*,\pi_2^*}(x) &\le& \mathbb{E}_{\pi_1,\pi_2^*}^x\Big[g_1(X_1,(A_{1,1},A_{2,1}))\Big]
+ \mathbb{E}_{\pi_1,\pi_2^*}^x \Big[J^{(1)}_{\pi_1^*,\pi_2^*}(X_2)\Big]
\end{eqnarray*}
Observe that, $J^{(1)}_{\pi_1^*,\pi_2^*}(\cdot)$ appears on the RHS of the above expression. 
Hence, inductively applying the above inequality $K$ times, we obtain
\begin{eqnarray*}
J^{(1)}_{\pi_1^*,\pi_2^*}(x) &\le& \sum_{k=1}^K\mathbb{E}_{\pi_1,\pi_2^*}^x\Big[g_1(X_k,(A_{1,k},A_{2,k}))\Big]
+ \mathbb{E}_{\pi_1,\pi_2^*}^x \Big[J^{(1)}_{\pi_1^*,\pi_2^*}(X_{K+1})\Big]
\end{eqnarray*}
Taking limit as $K\rightarrow\infty$ in the above expression we obtain,
\begin{eqnarray}
\label{limit_ineq_equn}
J^{(1)}_{\pi_1^*,\pi_2^*}(x)\le J^{(1)}_{\pi_1,\pi_2^*}(x) + \lim_{K\rightarrow\infty}
\mathbb{E}_{\pi_1,\pi_2^*}^x \Big[J^{(1)}_{\pi_1^*,\pi_2^*}(X_{K+1})\Big].
\end{eqnarray}

Now, let $\mathcal{A}_\ts=\{(\ts,\ts)\}\cup\{(\ts,r_j):j\in[n]\}$.
$\mathcal{A}_\ts$ is the set of all states, which are entered once $\mathscr{F}_1$ terminates.
We will assume that the policy pair $(\pi_1,\pi_2^*)$ is such that 
$\mathscr{F}_1$ will eventually terminate starting from any state $x$, i.e., 
$\lim_{K\rightarrow\infty}\mathbb{P}_{\pi_1,\pi_2^*}^x(X_{K}\in\mathcal{A}_\ts)=1$,
or equivalently, for any $x'\notin\mathcal{A}_\ts$,  $\lim_{K\rightarrow\infty}\mathbb{P}_{\pi_1,\pi_2^*}^x(X_{K}=x')=0$
(otherwise, with positive probability $\mathscr{F}_1$ will continue forever incurring a delay cost
of $\tau$ at every stage yielding $J^{(1)}_{\pi_1^*,\pi_2^*}(x)=\infty$, so that 
the inequality $J^{(1)}_{\pi_1^*,\pi_2^*}(x)\le J^{(1)}_{\pi_1,\pi_2^*}(x)$ trivially holds).
Using this along with
$J^{(1)}_{\pi_1^*,\pi_2^*}(x^o)=0$ for any $x^o\in\mathcal{A}_\ts$, we 
can write
\begin{eqnarray*}
{\lim_{K\rightarrow\infty}
\mathbb{E}_{\pi_1,\pi_2^*}^x \Big[J^{(1)}_{\pi_1^*,\pi_2^*}(X_{K+1})\Big]}
&=& \lim_{K\rightarrow\infty}\Bigg(\sum_{x^o\in\mathcal{A}_\ts} 
\mathbb{P}_{\pi_1,\pi_2^*}^x(X_{K+1}=x^o)J^{(1)}_{\pi_1^*,\pi_2^*}(x^o)+\\
&&\hspace{3cm}\sum_{x'\notin\mathcal{A}_\ts}\mathbb{P}_{\pi_1,\pi_2^*}^x(X_{K+1}=x') J^{(1)}_{\pi_1^*,\pi_2^*}(x')\Bigg)\\
&\overset{*}{=}& \sum_{x'\notin\mathcal{A}_\ts}  \lim_{K\rightarrow\infty}\Big( \mathbb{P}_{\pi_1,\pi_2^*}^x(X_{K+1}=x') J^{(1)}_{\pi_1^*,\pi_2^*}(x')\Big)\\
&\overset{o}{=}& \sum_{x'\notin\mathcal{A}_\ts} \Big(\lim_{K\rightarrow\infty} \mathbb{P}_{\pi_1,\pi_2^*}^x(X_{K+1}=x')\Big)\ J^{(1)}_{\pi_1^*,\pi_2^*}(x')\\
&=& 0.
\end{eqnarray*}
Note that, in $*$
interchanging the limit and summation was possible because we have a finite sum (since our
state space is finite).
Also, since we have restricted ourselves to the class of stationary policies (recall Definition~\ref{policy_defn}), $J^{(1)}_{\pi_1^*,\pi_2^*}(x')$ is not a function of
the stage index $K$, which enables us to proceed to $o$. 
Finally,
using the above in
(\ref{limit_ineq_equn}) we obtain, $J^{(1)}_{\pi_1^*,\pi_2^*}(x)\le J^{(1)}_{\pi_1,\pi_2^*}(x)$.

Similarly, for $\mathscr{F}_2$ it can be shown that $J^{(2)}_{\pi_1^*,\pi_2^*}(x)\le J^{(2)}_{\pi_1^*,\pi_2}(x)$
for any $\pi_2$ and $x=(r_i,r_j)$.
\end{IEEEproof}

\section{Proof of Lemma~\ref{costs_ordered_corollary}}
Lemma~\ref{costs_ordered_corollary} will be an immediate consequence of the following result.
\label{costs_ordered_lemma_appendix}
\begin{lemma}
\label{costs_ordered_lemma}
Given an NEPP $(\pi_1^*,\pi_2^*)$, for any $(r_i,r_j)\in\mathcal{X}$ we have,
\begin{eqnarray}
\label{costs1_ordered_equn}
J^{(1)}_{\pi_1^*,\pi_2^*}(r_i,\ts) &\le& J^{(1)}_{\pi_1^*,\pi_2^*}(r_i,r_j),  \\
\label{costs2_ordered_equn}
J^{(2)}_{\pi_1^*,\pi_2^*}(\ts,r_j) &\le& J^{(2)}_{\pi_1^*,\pi_2^*}(r_i,r_j). 
\end{eqnarray}
\end{lemma}
\begin{IEEEproof}
We will prove only (\ref{costs1_ordered_equn}); the proof of (\ref{costs2_ordered_equn}) is 
along similar lines.
Since $(\pi_1^*,\pi_2^*)$ is an NEPP, it follows that the policy $\pi_1^*$ is the best response for $\mathscr{F}_1$ 
against the policy $\pi_2^*$ of $\mathscr{F}_2$,
i.e., for any $x\in\mathcal{X}$, $J^{(1)}_{{\pi}_1^*,\pi_2^*}(x)=\inf_{\pi_1}J^{(1)}_{\pi_1,\pi_2^*}(x)$.
Thus $J^{(1)}_{{\pi}_1^*,\pi_2^*}(x)$ can be 
regarded as the optimal cost of
the MDP problem, $MDP_1(\pi_2^*)$, obtained by fixing the policy $\pi_2^*$ of
$\mathscr{F}_2$. For simplicity of notation we will denote $J^{(1)}_{{\pi}_1^*,\pi_2^*}(x)$
as $H^*(x)$. Thus for the states of the form $(r_i,\ts)$,
$H^*(r_i,\ts)$ satisfies the following Bellman equation (this expression is same as the one in (\ref{bellman_J1_equn}))
\begin{eqnarray}
\label{H_single_equn}
H^*(r_i,\ts)=\min\Big\{-\eta_1 r_i,C_\textsf{c}(r_i,\ts)\Big\},
\end{eqnarray}
where 
\begin{eqnarray}
\label{alone_continuing_cost}
C_\textsf{c}(r_i,\ts)=\tau+\sum_{i'}p^{(1)}_{i'} H^*(r_{i'},\ts)
\end{eqnarray}
is the expected cost of continuing, and $-\eta_1 r_i$ is the cost of stopping.

However, for states of the form $(r_i,r_j)$ (where $\mathscr{F}_2$ is also competing for a relay), the 
optimality equation is more involved since the actions of $\mathscr{F}_2$ will now affect both costs
(stopping and continuing) 
of $\mathscr{F}_1$. Defining $\epsilon=\pi_2^*(r_i,r_j,\textsf{s})$ ($\epsilon$ is the probability 
with which $\mathscr{F}_2$ will stop when the state is $(r_i,r_j)$), Bellman equation for 
states of the form $(r_i,r_j)$ can be written as
\begin{eqnarray}
\label{H_competing_equn}
H^*(r_i,r_j)=\min\Big\{C_\textsf{s}(r_i,r_j),C_\textsf{c}(r_i,r_j)\Big\},
\end{eqnarray}
where $C_\textsf{s}(r_i,r_j)$ is the expected cost incurred by $\mathscr{F}_1$ for 
stopping when the state is $(r_i,r_j)$, and 
$C_\textsf{c}(r_i,r_j)$ is the expected cost of continuing. 

The expression for $C_\textsf{s}(r_i,r_j)$ is (recall that $\nu_\rho$, $\rho=1,2$, is the probability that 
$\mathscr{F}_\rho$ gets the relay if both forwarders simultaneously choose to stop),
\begin{eqnarray}
\label{stopping_cost_equn}
C_\textsf{s}(r_i,r_j)
&=&\epsilon\Big(\nu_1 (-\eta_1 r_i)+\nu_2 C_\textsf{c}(r_i,\ts)\Big)
+(1-\epsilon)\Big(-\eta_1 r_i\Big).
\end{eqnarray}
The first term in the RHS of the above expression
is the expected stopping cost incurred by $\mathscr{F}_1$  conditioned on the event that $\mathscr{F}_2$ also decides to stop.
This can be understood as follows: suppose $\mathscr{F}_2$ also decides to stop
(probability of which is $\epsilon$), then w.p.\ $\nu_1$,
$\mathscr{F}_1$ gets the relay incurring a termination cost of $-\eta_1 r_i$, otherwise $\mathscr{F}_2$ gets the relay in which case
$\mathscr{F}_1$ has to continue alone, the expected cost of which is $C_\textsf{c}(r_i,\ts)$. 
The remaining term, 
$(1-\epsilon)(-\eta_1 r_i)$,
in (\ref{stopping_cost_equn}) is the stopping cost incurred to $\mathscr{F}_1$ 
when the action of $\mathscr{F}_2$ is to continue 
(which happens with probability $(1-\epsilon)$).

Similarly, the cost incurred by 
$\mathscr{F}_1$ for continuing, $C_\textsf{c}(r_i,r_j)$, can be written as,
\begin{eqnarray}
\label{continue_cost_equn}
C_\textsf{c}(r_i,r_j)&=&\epsilon\Big(\tau+\sum_{i'}p^{(1)}_{i'}H^*(r_{i'},\ts)\Big)
+ (1-\epsilon)\Big(\tau+\sum_{i',j'}p_{i',j'}H^*(r_{i'},r_{j'})\Big).
\end{eqnarray}

Now, returning to (\ref{H_single_equn}) and (\ref{H_competing_equn}), $H^*$
can be expressed as the fixed point of a mapping $T$ which is, for a function $H(\cdot,\cdot)$,
given by,
\begin{eqnarray*}
TH(r_i,\ts)&=&\min\Big\{-\eta_1 r_i,C_\textsf{c}^H(r_i,\ts)\Big\} \\
TH(r_1,r_2)&=&\min\Big\{C_\textsf{s}^H(r_i,r_j),C_\textsf{c}^H(r_i,r_j)\Big\},
\end{eqnarray*}
where the expressions for $C_\textsf{c}^H(r_i,\ts)$, $C_\textsf{s}^H(r_i,r_j)$ and $C_\textsf{c}^H(r_i,r_j)$ is similar to that of 
$C_\textsf{c}(r_i,\ts)$, $C_\textsf{s}(r_i,r_j)$ and $C_\textsf{c}(r_i,r_j)$ 
(in (\ref{alone_continuing_cost}) (\ref{stopping_cost_equn}) and (\ref{continue_cost_equn}), respectively) with 
$H^*$ replaced by the given function $H$. 
Inductively define $H_k=TH_{k-1}$ with $H_0\equiv0$ (i.e., $H_0(x)=0$ for all $x\in\mathcal{X}$). Since $MDP_1(\pi_2^*)$ is an optimal stopping problem 
\cite{bertsekas-tsitsiklis91stochastic-shortest-path} it follows that
$H_k\rightarrow H^*$ (this is the value iteration algorithm). Hence, to complete the proof we will 
show that $H_k(r_i,\ts)\le H_k(r_i,r_j)$ whenever
$H_{k-1}(r_i,\ts)\le H_{k-1}(r_i,r_j)$. 

Suppose, for some $k\ge1$, $H_{k-1}(r_i,\ts)\le H_{k-1}(r_i,r_j)$  
for all $(r_i,r_j)\in\mathcal{X}$ (this holds trivially for $k=1$). 
First consider the case
where, $-\eta_1 r_i\le C_\textsf{c}^{H_{k-1}}(r_i,\ts)$ 
(i.e., it is optimal to stop when the state is $(r_i,\ts)$). 
\begin{itemize}
\item Then from (\ref{stopping_cost_equn}) we obtain $-\eta_1 r_i\le C_\textsf{s}^{H_{k-1}}(r_i,r_j)$.
\item Also, from the induction hypothesis we have 
\begin{eqnarray*}
\sum_{i'}p^{(1)}_{i'}H_{k-1}(r_{i'},\ts) &=& \sum_{i',j'}p_{i',j'}H_{k-1}(r_{i'},\ts)\\
&\le&\sum_{i',j'}p_{i',j'}H_{k-1}(r_{i'},r_{j'}).
\end{eqnarray*}
Using the above in (\ref{continue_cost_equn}) and recalling (\ref{alone_continuing_cost}) 
we can write 
\begin{eqnarray*}
C_\textsf{c}^{H_{k-1}}(r_i,r_j)
&\ge&\tau+\sum_{i'}p^{(1)}_{i'}H^*(r_{i'},\ts) \\
&=& C_\textsf{c}(r_i,\ts) \\
&\ge& -\eta_1 r_i.
\end{eqnarray*}
\end{itemize}

Thus we have,
\begin{eqnarray*}
H_k(r_i,\ts)&=& \min\Big\{-\eta_1 r_i,C_\textsf{c}^{H_{k-1}}(r_i,\ts)\Big\} \\
&=& -\eta_1 r_i\\
&\le& \min\Big\{C_\textsf{s}^{H_{k-1}}(r_i,r_j),C_\textsf{c}^{H_{k-1}}(r_i,r_j)\Big\}\\
&=& H_k(r_i,r_j).
\end{eqnarray*}

Similarly for the other case, i.e., when $-\eta_1 r_i> C_\textsf{c}^H(r_i,\ts)$,
we can show that both the costs, $C_\textsf{s}^{H_{k-1}}(r_i,r_j)$ 
and $C_\textsf{s}^{H_{k-1}}(r_i,r_j)$, are less than 
$C_\textsf{c}^H(r_i,\ts)$ again yielding $H_k(r_i,\ts)\le H_k(r_i,r_j)$.
\end{IEEEproof}

\verb11\\
\noindent
We are now ready to prove Lemma~\ref{costs_ordered_corollary}.

\emph{Lemma}~\ref{costs_ordered_corollary}:
For an NEPP, $(\pi_{1}^*,\pi_2^*)$, the various costs are ordered as follows:
\begin{eqnarray*}
D^{(1)}\le C^{(1)}_{\pi_1^*,\pi_2^*} \mbox{ and } D^{(2)}\le C^{(2)}_{\pi_1^*,\pi_2^*}.
\end{eqnarray*}

\begin{IEEEproof}
Recalling the cost expressions of $D^{(1)}$ and $C^{(1)}_{\pi_1^*,\pi_2^*}$
(from (\ref{d1_cost_equn}) and (\ref{c1_cost_equn}), respectively)  we can write,
\begin{eqnarray*}
D^{(1)} &=& \tau + \sum_{i'}p^{(1)}_{i'}J^{(1)}_{\pi_1^*,\pi_2^*}(r_{i'},\ts)\\
&=& \tau + \sum_{i',j'} p_{i',j'}J^{(1)}_{\pi_1^*,\pi_2^*}(r_{i'},\ts)\\
&\overset{*}{\le}& \tau + \sum_{i',j'} p_{i',j'}J^{(1)}_{\pi_1^*,\pi_2^*}(r_{i'},r_{j'})\\
&=& C^{(1)}_{\pi_1^*,\pi_2^*},
\end{eqnarray*}
where $*$ is due to Lemma~\ref{costs_ordered_lemma}. Similarly, one can show
that $D^{(2)}\le C^{(2)}_{\pi_1^*,\pi_2^*}$.
\end{IEEEproof}

\section{Obtaining NE Strategies for the Static Game in Table~\ref{bimatrix_game_table}}
\label{obtaining_NE_appendix}
For convenience, let us first recall the game in Table~\ref{bimatrix_game_table}.
\begin{table}[h]
\centering
\begin{tabular}{|c||c|c|}
\hline
	   & \textsf{c} & \textsf{s} \\
\hline
\hline
\textsf{c} & $C^{(1)}_{\pi_1^*,\pi_2^*},C^{(2)}_{\pi_1^*,\pi_2^*}$ & $D^{(1)},-\eta_2 r_j$ \\
\hline
\textsf{s} & $-\eta_1 r_i,D^{(2)}$ & 
$E^{(1)}(r_i),E^{(2)}(r_j)$\\
\hline 
\end{tabular} 
\vspace{1mm}
\caption{Static stage game.}
\vspace{-4mm}
\end{table}

Since only two actions (namely $\textsf{s}$ and $\textsf{c}$) are available to each forwarder,
a strategy used by $\mathscr{F}_1$ can be conveniently represented by $\sigma_1\in[0,1]$, where 
$\sigma_1$ is the probability that $\mathscr{F}_1$ will choose action $\textsf{s}$. Similarly, 
$\sigma_2\in[0,1]$ is the probability that $\mathscr{F}_2$ will choose action $\textsf{s}$.
Given a strategy pair $(\sigma_1,\sigma_2)$ the expected cost 
(obtained from Table~\ref{bimatrix_game_table}) incurred by $\mathscr{F}_1$
can be expressed as 
\begin{eqnarray}
U_1(\sigma_1,\sigma_2) = \sigma_1 A_{\sigma_2} + B_{\sigma_2},
\end{eqnarray}
where 
\begin{eqnarray}
A_{\sigma_2} = (1-\sigma_2)\Big(-\eta_1 r_i-C^{(1)}\Big) + 
\sigma_2\Big(E^{(1)}(r_i) - D^{(1)}\Big)
\end{eqnarray}
and
\begin{eqnarray*}
B_{\sigma_2} = (1-\sigma_2) C^{(1)} + \sigma_2 D^{(1)}. 
\end{eqnarray*}
Let $\sigma_1^*(\sigma_2)$ denote the set of all best responses of $\mathscr{F}_1$ to the strategy $\sigma_2$
of $\mathscr{F}_2$, i.e.,
\begin{eqnarray}
\sigma_1^*(\sigma_2) = \argmin_{\sigma_1\in[0,1]}U_1(\sigma_1,\sigma_2). 
\end{eqnarray}
Since $U_1(\sigma_1,\sigma_2)$ is linear in $\sigma_1$ it follows that,
 $\sigma_1^*(\sigma_2)=\{0\}$ whenever $A_{\sigma_2}>0$,
$\sigma_1^*(\sigma_2)=\{1\}$ whenever $A_{\sigma_2}<0$, 
and  $\sigma_1^*(\sigma_2)=[0,1]$ whenever $A_{\sigma_2}=0$. 
We make use of these observations in the proof of our next lemma. First for convenience
let us denote the thresholds $\frac{C^{(1)}}{-\eta_1}$ and $\frac{C^{(2)}}{-\eta_2}$
by $\zeta^{(1)}$ and $\zeta^{(2)}$, respectively. Recall that we already have,
$\alpha^{(1)}=\frac{D^{(1)}}{-\eta_1}$ and $\alpha^{(2)}=\frac{D^{(2)}}{-\eta_2}$.
The inequalities in (\ref{costs_ordered_equn}) would (since, $-\eta<0$) imply that 
$\zeta^{(1)}\le\alpha^{(1)}$ and $\zeta^{(2)}\le\alpha^{(2)}$.

\begin{lemma}
\label{best_response_lemma}
Suppose $\nu_1\in(0,1)$ and $D^{(1)}<C^{(1)}$, then
\begin{enumerate}
\item If $r_i<\zeta^{(1)}$ then $\sigma_1^*(\sigma_2)=\{0\}$ for all $\sigma_2\in[0,1]$.
\item If $r_i>\alpha^{(1)}$ then $\sigma_1^*(\sigma_2)=\{1\}$ for all $\sigma_2\in[0,1]$. 
\item If $\zeta^{(1)}\le r_i\le \alpha^{(1)}$ then 
defining 
\begin{eqnarray}
\Gamma_{2}&=&\frac{-\eta_1 r_i -C^{(1)}}{\Big(-\eta_1 r_i -C^{(1)}\Big)-\Big(E^{(1)}(r_i)-D^{(1)}\Big)}
\end{eqnarray}
we have:
(i) $\sigma_1^*(\sigma_2)=\{1\}$ for $\sigma_2<\Gamma_{2}$,
(ii) $\sigma_1^*(\sigma_2)=\{0\}$ for $\sigma_2>\Gamma_{2}$, and
(iii) $\sigma_1^*(\Gamma_{2})=[0,1]$.
\end{enumerate}
\end{lemma}
\begin{IEEEproof}[Proof of Part~1]
We will show that $A_{\sigma_2}>0$ for any $\sigma_2\in[0,1]$. Then the proof follows immediately
since, $U_1(\sigma_1,\sigma_2)=\sigma_1 A_{\sigma_2} +B_{\sigma_2}$, is linear in $\sigma_1$.

Let us recall the expression for $A_{\sigma_2}$,
\begin{eqnarray}
\label{Adel_recalled_equn}
A_{\sigma_2}=(1-\sigma_2)\Big(-\eta_1 r_i - C^{(1)}\Big) + \sigma_2 \Big(C^{(1)}(r_i)-D^{(1)}\Big),
\end{eqnarray}
where $E^{(1)}(r_i)=\nu_1(-\eta_1 r_i)+\nu_2 D^{(1)}$ (see (\ref{e1_cost_equn})). 
It is already given that $r_i<\frac{C^{(1)}}{-\eta_1}$, or
\begin{eqnarray}
\label{positive_equn}
\Big(-\eta_1 r_i - C^{(1)}\Big)>0.
\end{eqnarray}
Since $D^{(1)}<C^{(1)}$ we also have $-\eta_1 r_i>D^{(1)}$
which gives $E^{(1)}(r_i)>D^{(1)}$ (this is where $\nu_1\in(0,1)$ is required), i.e, $\Big(E_{(1)}(r_i)-D^1\Big)>0$.
Using this along with inequality (\ref{positive_equn}) 
we obtain the desired result.\\

\emph{(Proof of Part~2)}
Similar to the previous part, the proof follows once we show that $A_{\sigma_2}<0$ 
for all $\sigma_2\in[0,1]$. Since $r_i>\frac{D^{(1)}}{-\eta_1}$ and $D^{(1)}< C^{(1)}$,
we obtain $\Big(-\eta_1 r_i - C^{(1)}\Big)<0$ and $\Big(E^{(1)}(r_i)-D^{(1)}\Big)<0$. Using these in (\ref{Adel_recalled_equn}) we obtain
$A_{\sigma_2}<0$.\\

\emph{(Proof of Part~3)}
Again, since $U_1(\sigma_1,\sigma_2)$ is linear in $\sigma_1$, 
we have to show that $A_{\sigma_2}<0$ whenever $\sigma_2<\Gamma_2$,
$A_{\sigma_2}>0$ whenever $\sigma_2>\Gamma_2$, and $A_{\Gamma_2}=0$.

Suppose $\sigma_2\in[0,1]$ is such that $\sigma_2<\Gamma_2$ 
(thus $\Gamma_2\in(0,1]$),
then recalling the expression for $\Gamma_2$ we can write,
\begin{eqnarray}
\label{del_rearrange_equn}
\sigma_2< \frac{-\eta_1 r_i -C^{(1)}}{\Big(-\eta_1 r_i -C^{(1)}\Big)-\Big(E^{(1)}(r_i)-D^{(1)}\Big)}.
\end{eqnarray}
It is important to note that, since $\frac{C^{(1)}}{-\eta_1}\le r_i\le \frac{D^{(1)}}{-\eta_1}$ 
with $D^{(1)}<C^{(1)}$ 
and $\nu_1\in(0,1)$, the denominator in the RHS of the above expression is strictly negative.   
Thus, rearranging (\ref{del_rearrange_equn}) we 
obtain $A_{\sigma_2}<0$ so that $\sigma_1^*(\sigma_2)=\{1\}$.

Similarly, when $\sigma_2\in[0,1]$ is such that $\sigma_2>\Gamma_2$ (in which case $\Gamma_2\in[0,1)$), then
reversing the inequality in (\ref{del_rearrange_equn}) and rearranging we obtain $A_{\sigma_2}>0$ so that 
$\sigma_1^*(\sigma_2)=\{0\}$. 

Finally, substituting for $\Gamma_2$ in the expression for $A_{\Gamma_2}$ will yield $A_{\Gamma_2}=0$ implying
that any $\sigma_1\in[0,1]$ is a best response against the strategy $\Gamma_2$ played by $\mathscr{F}_2$. Hence
$\sigma_1^*(\Gamma_2)=[0,1]$.
\end{IEEEproof}
\verb11

\begin{figure}[t]
\centering
\subfigure[]{
\includegraphics[scale=0.45]{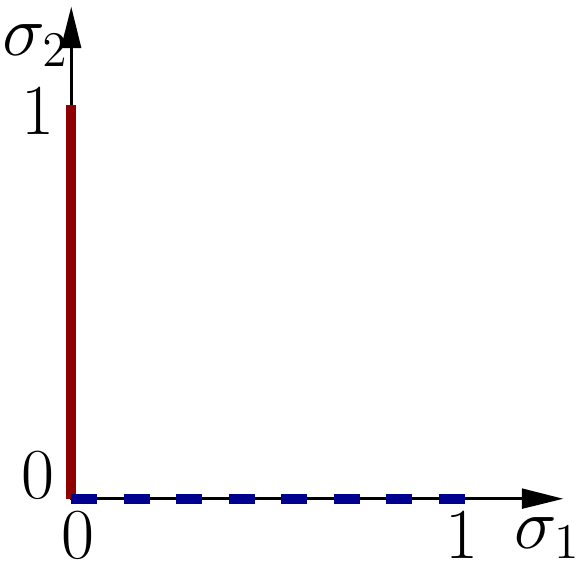}
\label{region1_figure}
}
\subfigure[]{
\includegraphics[scale=0.45]{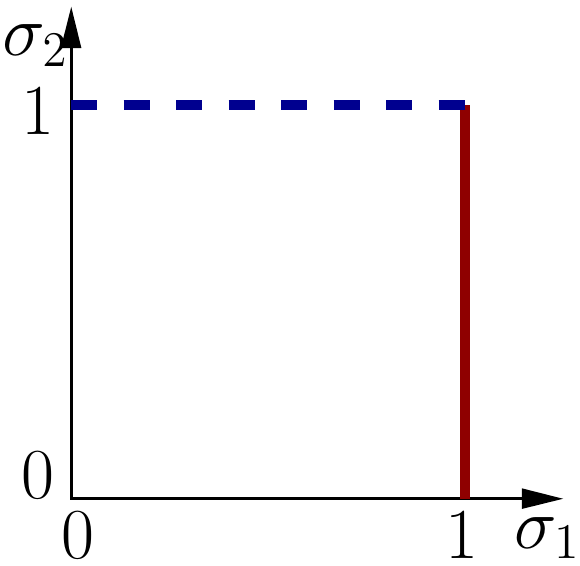}
        \label{region4_figure}
}
\subfigure[]{
\includegraphics[scale=0.45]{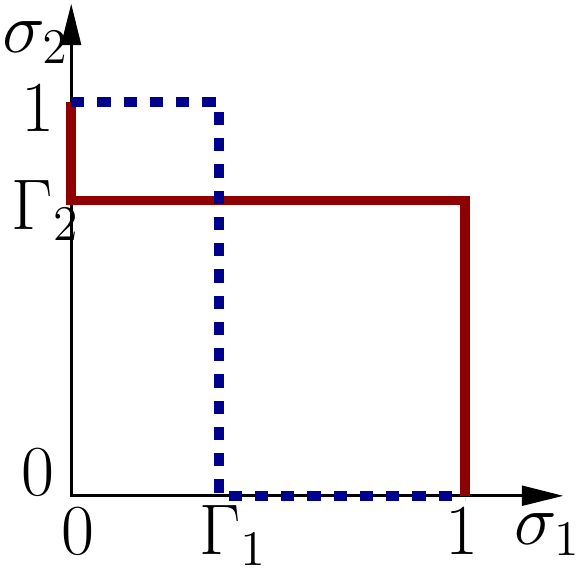}
        \label{region5_figure} 
}
\subfigure[]{
\includegraphics[scale=0.45]{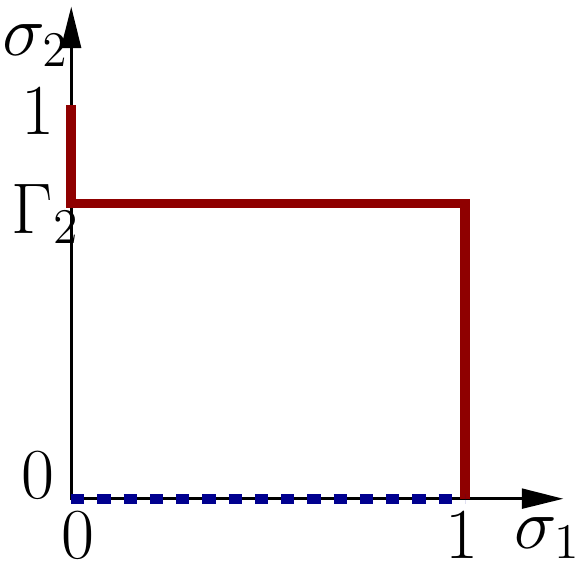}
        \label{region21_figure}
}
\subfigure[]{
\includegraphics[scale=0.45]{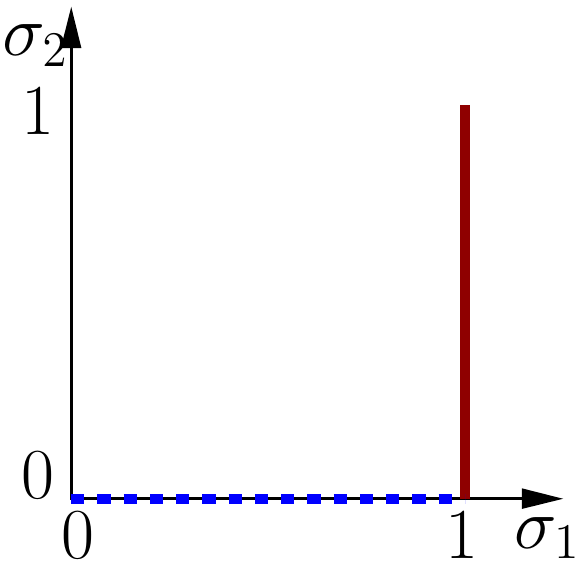}
        \label{region22_figure}
}
\subfigure[]{
\includegraphics[scale=0.45]{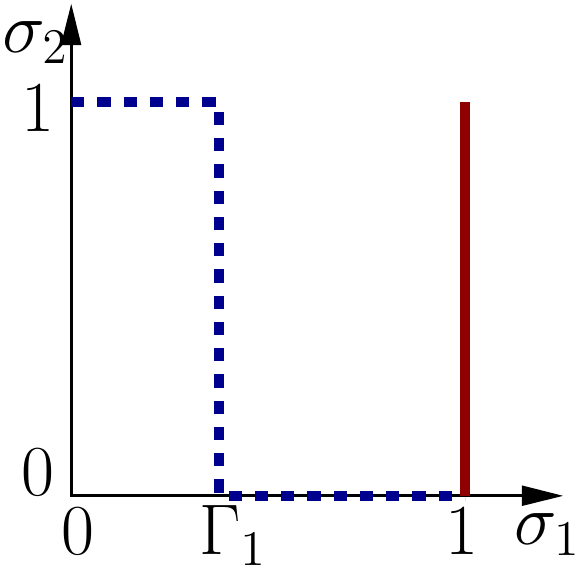}
        \label{region23_figure} 
}
\caption{\label{region2_figure} Plot of best response curves,
$\sigma_1^*(\sigma_2)$ and $\sigma_2^*(\sigma_1)$,
 for $(r_i,r_j)$ in different regions.
In each of these figures, the solid red curve is 
$\sigma_1^*(\sigma_2)$ and the dashed blue curve is $\sigma_2^*(\sigma_1)$.
\subref{region1_figure} $(r_i,r_j)\in\mathcal{R}_1$,  
\subref{region4_figure} $(r_i,r_j)\in\mathcal{R}_5$,
\subref{region5_figure} $(r_i,r_j)\in\mathcal{R}_4$,
\subref{region21_figure} $(r_i,r_j)\in\mathcal{R}_{2a}$,
\subref{region22_figure} $(r_i,r_j)\in\mathcal{R}_{2b}$, and
\subref{region23_figure} $(r_i,r_j)\in\mathcal{R}_{2c}$.}
\vspace{-4mm}
\end{figure}

\emph{{Remark:}} The condition imposed on $\nu_1$ and $C^{(1)}$ in the above lemma 
is only to avoid the less interesting boundary cases.
Also, note that $\Gamma_2$ is a function of the reward $r_i$ to $\mathscr{F}_1$. For notational simplicity 
we do not show $r_i$ as an argument of $\Gamma_2$.

Similarly, for $\mathscr{F}_2$ we can define $\sigma_2^*(\sigma_1)$ as the set of all best responses 
against the strategy $\sigma_1$ played by $\mathscr{F}_1$, and obtain a result analogous to that in 
Lemma~\ref{best_response_lemma}, but with quantities corresponding to $\mathscr{F}_1$ replaced by that
corresponding to $\mathscr{F}_2$, e.g., for instance, $\zeta^{(1)}$ replaced by $\zeta^{(2)}$, $\alpha^{(1)}$ by $\alpha^{(2)}$,
$\Gamma_{2}$ by $\Gamma_{1}$ where
\begin{eqnarray}
\Gamma_{1} = \frac{-\eta_2 r_j -C^{(2)}}{\Big(-\eta_2 r_j -C^{(2)}\Big)- \Big(E^{(2)}(r_j)-D^{(2)}\Big)}.
\end{eqnarray}

\begin{figure}[t]
\centering
\includegraphics[scale=0.5]{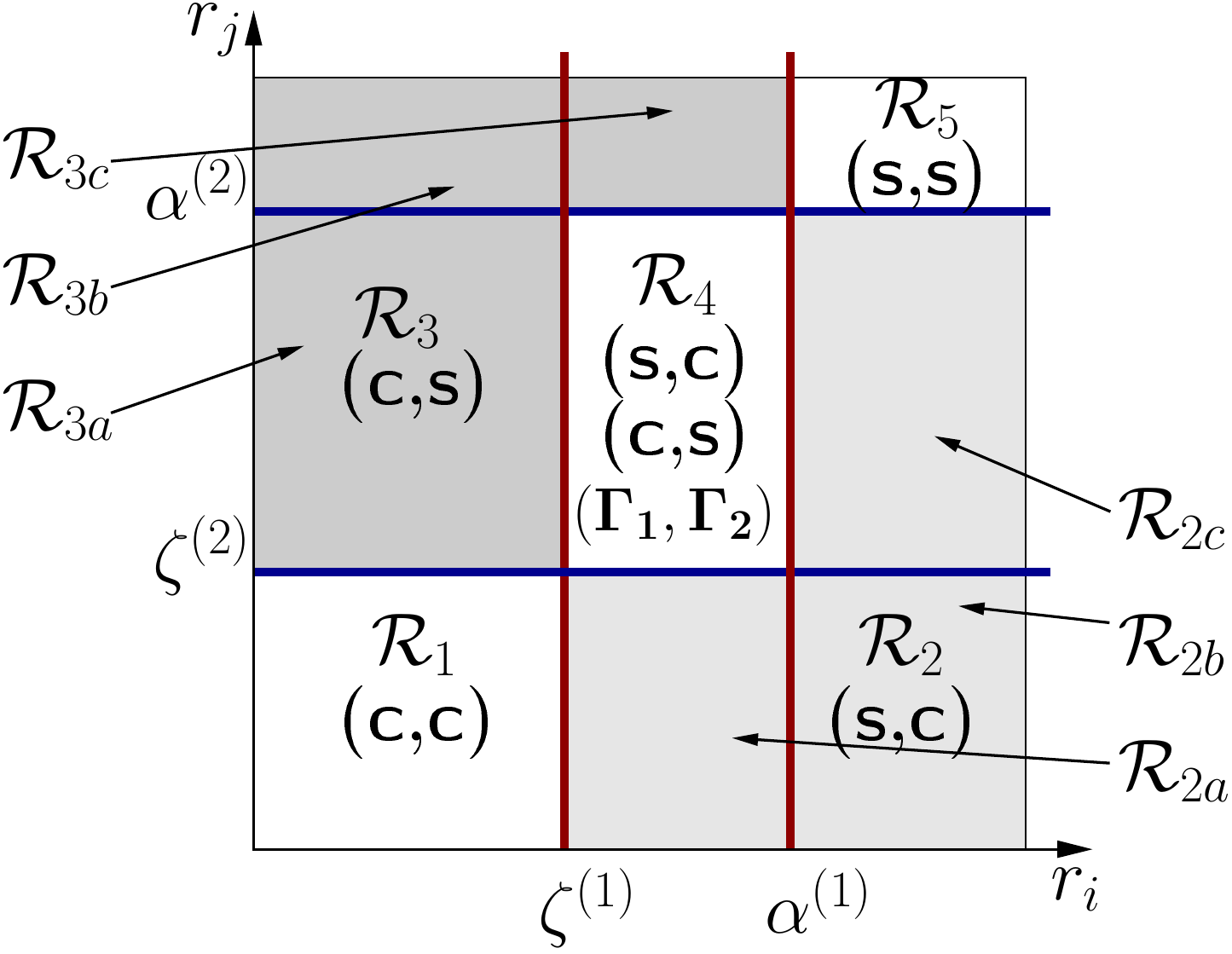}
\caption{\label{partition_figure} Illustration of the various regions
along with the NE strategy corresponding to these regions.}
\end{figure}

Now, for any $(r_i,r_j)$ the points of intersection between
the best response curves $\sigma_1^*(\sigma_2)$ and $\sigma_2^*(\sigma_1)$ constitutes
the NE strategies of the game in Table~\ref{bimatrix_game_table}. For instance, as
shown in Fig.~\ref{region1_figure}, when $(r_i,r_j)$ is such that $r_i<\zeta^{(1)}$ and 
$r_j<\zeta^{(2)}$ (i.e., $(r_i,r_j)\in\mathcal{R}_1$; 
see Fig.~\ref{partition_figure})
then the only point of intersection is $(0,0)$ so that
$(\textsf{c},\textsf{c})$ is the only NE strategy in this region. 
Similarly when 
$(r_i,r_j)\in\mathcal{R}_5$
then $(\textsf{s},\textsf{s})$
is the only NE strategy (see Fig.~\ref{region4_figure}).
An interesting case 
is when 
$(r_i,r_j)\in\mathcal{R}_4$
(see Fig.~\ref{region5_figure})
where there are multiple NE strategies, namely, $(\textsf{s},\textsf{c})$, $(\textsf{c},\textsf{s})$
and the mixed strategy $(\Gamma_{1},\Gamma_{2})$ 
(which depends on the reward pair $(r_i,r_j)$; see remark following Lemma~\ref{best_response_lemma}).

The region $\mathcal{R}_2$ is written as a union of three disjoint regions, 
$\mathcal{R}_{2a}$, $\mathcal{R}_{2b}$
and $\mathcal{R}_{2c}$. However, as shown in 
Fig.~\ref{region21_figure} to \ref{region23_figure}, the best response curves 
for $(r_i,r_j)$ in each of these sub-regions intersect at $(1,0)$. Hence
$(\textsf{s},\textsf{c})$ is the NE strategy in the union region $\mathcal{R}_{2}$.
Similarly, $(\textsf{c},\textsf{s})$ is the NE strategy in region $\mathcal{R}_3$
which is also composed of three sub-regions.
%%%%%%%%%%%%%%%%%%%%%%%%%%%%%%%%%%%%%%%%%%%%%%%%%%%%%%%%%%%%%%%%%%%%%%%%%%%%%%%%%%%%%%%%%%%%%%%%%%%%%
%%%%%%%%%%%%%%%%%%%%%%%%%%%%%%%%%%%%%%%%%%%%%%%%%%%%%%%%%%%%%%%%%%%%%%%%%%%%%%%%%%%%%%%%%%%%%%%%%%%%%%

\begin{table}[h]
\centering
\begin{tabular}{|l|}
\hline
$\mathcal{R}_1=\Big\{(r_i,r_j): r_i<\zeta^{(1)}, r_j<\zeta^{(2)}\Big\}$ \\
\hline
\hline
$\mathcal{R}_2=\mathcal{R}_{2a}\cup\mathcal{R}_{2b}\cup\mathcal{R}_{2c}$ where \\ 
$\mathcal{R}_{2a}=\Big\{(r_i,r_j):\zeta^{(1)}\le r_i\le\alpha^{(1)}, r_j<\zeta^{(2)}\Big\}$ \\
$\mathcal{R}_{2b}=\Big\{(r_i,r_j):r_i>\alpha^{(1)}, r_j<\zeta^{(2)}\Big\}$ \\%& \\
$\mathcal{R}_{2c}=\Big\{(r_i,r_j):r_i>\alpha^{(1)}, \zeta^{(2)}\le r_j\le\alpha^{(2)}\Big\}$ \\
\hline
\hline
$\mathcal{R}_3=\mathcal{R}_{3a}\cup\mathcal{R}_{3b}\cup\mathcal{R}_{3c}$ where \\
$\mathcal{R}_{3a}=\Big\{(r_i,r_j):r_i<\zeta^{(1)},\zeta^{(2)}\le r_j\le\alpha^{(2)}\Big\}$ \\
$\mathcal{R}_{3b}=\Big\{(r_i,r_j):r_i<\zeta^{(1)}, r_j>\alpha^{(2)}\Big\}$ \\%& \\
$\mathcal{R}_{3c}=\Big\{(r_i,r_j):\zeta^{(1)}\le r_i\le\alpha^{(1)},r_j>\alpha^{(2)}\Big\}$ \\
\hline
\hline
$\mathcal{R}_4=\Big\{(r_i,r_j):\zeta^{(1)}\le r_i\le\alpha^{(1)},\zeta^{(2)}\le r_j\le\alpha^{(2)}\Big\}$\\
\hline
\hline
$\mathcal{R}_5=\Big\{(r_i,r_j):r_i>\alpha^{(1)},r_j>\alpha^{(2)}\Big\}$\\
\hline	
\end{tabular} 
\caption{\label{region_table} Formal definition of various regions depicted in 
Fig.~\ref{partition_figure}.}
\vspace{-5mm}
\end{table}

We have thus identified a partition of the set 
 $\{(r_i,r_j): i,j\in[n]\}$ into five regions such that the set of NE strategies
corresponding to each region are different. 
These regions along with the corresponding NE strategies are depicted in Fig.~\ref{partition_figure}.
A formal definition of the various regions is available in Table~\ref{region_table}.
Note that, these regions depend on the cost pair $\mathbf{C}=(C^{(1)},C^{(2)})$; for simplicity we have
not shown this explicitly in Fig.~\ref{partition_figure} and in Table~\ref{region_table}.

\section{Proof of Theorem~\ref{filar_theorem_PO}}
\label{filar_theorem_PO_appendix}
\vspace{2mm}
\emph{Theorem}~\ref{filar_theorem_PO}:
Given a PO policy pair $(\overline{\pi}_1^*,\overline{\pi}_2^*)$,
construct a strategy vector pair $\{({f}_\ell^*,{g}_\ell^*):\ell\in\mathcal{L}\}$ as follows: 
$f_\ell^*(r_i)=\overline{\pi}_1^*(r_i,\ell)$ and 
${g}_\ell^*(r_j)=\overline{\pi}_2^*(\ell,r_j)$ for all $i,j\in[n]$.
Now, suppose for each $\ell$, $({f}_\ell^*,{g}_\ell^*)$ is a NE vector 
for 
$\mathcal{G}(\overline{\pi}_1^*,\overline{\pi}_2^*)$
such that, 
\begin{eqnarray*}
\min\Big\{C^{(1)}_{\st,{g}_\ell^*}(r_i),C^{(1)}_{\ct,{g}_\ell^*}\Big\}&=&
G^{(1)}_{\overline{\pi}_1^*,\overline{\pi}_2^*}(r_i,\ell), \mbox{ and }\\
\min\Big\{C^{(2)}_{\st,f_\ell^*}(r_j),C^{(2)}_{\ct,f_\ell^*}\Big\}&=&
G^{(2)}_{\overline{\pi}_1^*,\overline{\pi}_2^*}(\ell,r_j).
\end{eqnarray*}
Then $(\overline{\pi}_1^*,\overline{\pi}_2^*)$ is a PO-NEPP.

\begin{IEEEproof}
Given the policy pair $(\overline{\pi}_1^*,\overline{\pi}_2^*)$ as in the 
hypothesis, let $\overline{\pi}_1$ be any PO policy. We will show that 
$G^{(1)}_{\overline{\pi}_1^*,\overline{\pi}_2^*}(r_i,\ell)\le
G^{(1)}_{\overline{\pi}_1,\overline{\pi}_2^*}(r_i,\ell)$; 
the proof of, $G^{(1)}_{\overline{\pi}_1^*,\overline{\pi}_2^*}(r_i,\ell)\le
G^{(1)}_{\overline{\pi}_1^*,\overline{\pi}_2}(r_i,\ell)$ for any $\overline{\pi}_2$,
is along similar lines.

Since $f_\ell^*=BR_1(g_\ell^*)$, for the Bayesian game $\mathcal{G}(\overline{\pi}_1^*,\overline{\pi}_2^*)$, 
the expected cost incurred to $\mathscr{F}_1$ 
when its observation is $(r_i,\ell)$ is $\min\Big\{C^{(1)}_{\st,g_\ell^*}(r_i),C^{(1)}_{\ct,g_\ell^*}\Big\}$.
Hence, using (\ref{min_costs1_equn}) we can write
\begin{eqnarray*}
G^{(1)}_{\overline{\pi}_1^*,\overline{\pi}_2^*}(r_i,\ell)
&\le& C^{(1)}_{\overline{\pi}_1(r_i,\ell),g_\ell^*}(r_i)\\
&=&\mathbb{E}_{\overline{\pi}_1,\overline{\pi}_2^*}^{(r_i,\ell)}
\Big[g_1(X_1,(A_{1,1},A_{2,1}))\Big]
+\mathbb{E}_{\overline{\pi}_1,\overline{\pi}_2^*}^{(r_i,\ell)}
\Big[G^{(1)}_{\overline{\pi}_1^*,\overline{\pi}_2^*}(O_{1,2})\Big].
\end{eqnarray*}
Applying the above inequality $K$ times we obtain
\begin{eqnarray}
\label{PO_limit_equn}
G^{(1)}_{\overline{\pi}_1^*,\overline{\pi}_2^*}(r_i,\ell)&\le&
\sum_{k=1}^K \mathbb{E}_{\overline{\pi}_1,\overline{\pi}_2^*}^{(r_i,\ell)}
\Big[g_1(X_k,(A_{1,k},A_{2,k}))\Big] +
\mathbb{E}_{\overline{\pi}_1,\overline{\pi}_2^*}^{(r_i,\ell)}
\Big[G^{(1)}_{\overline{\pi}_1^*,\overline{\pi}_2^*}(O_{1,K+1})\Big].
\end{eqnarray}
Again, as in the proof of Theorem~\ref{filar_theorem} Part-(b), we will assume
that the PO policy pair $(\overline{\pi}_1,\overline{\pi}_2^*)$ is such that
using this policy pair $\mathscr{F}_1$ will eventually terminate starting from any observation 
$o_1$, i.e., 
\begin{eqnarray}
\label{PO_termination_equn}
\lim_{K\rightarrow\infty}\mathbb{P}_{\overline{\pi}_1,\overline{\pi}_2^*}^{o_1}(O_{1,K}=\ts)&=&1.
\end{eqnarray}
Hence we have
\begin{eqnarray*}
\lim_{K\rightarrow\infty}\mathbb{E}_{\overline{\pi}_1,\overline{\pi}_2^*}^{(r_i,\ell)}
\Big[G^{(1)}_{\overline{\pi}_1^*,\overline{\pi}_2^*}(O_{1,K+1})\Big]&=&
G^{(1)}_{\overline{\pi}_1^*,\overline{\pi}_2^*}(\ts)\\
&=&0.
\end{eqnarray*}
Using the above and recalling (\ref{PO_total_cost1_equn}) while
taking $\lim_{K\rightarrow\infty}$ in (\ref{PO_limit_equn}) we obtain
$G^{(1)}_{\overline{\pi}_1^*,\overline{\pi}_2^*}(r_i,\ell)\le 
G^{(1)}_{\overline{\pi}_1,\overline{\pi}_2^*}(r_i,\ell)$.

Finally, suppose the PO policy pair $(\overline{\pi}_1,\overline{\pi}_2^*)$ does not satisfy
(\ref{PO_termination_equn}), then there is a positive probability that $\mathscr{F}_1$ will continue 
forever yielding $G^{(1)}_{\overline{\pi}_1,\overline{\pi}_2^*}(r_i,\ell)=\infty$. Thus for this case,
$G^{(1)}_{\overline{\pi}_1^*,\overline{\pi}_2^*}(r_i,\ell)\le 
G^{(1)}_{\overline{\pi}_1,\overline{\pi}_2^*}(r_i,\ell)$, trivially holds.
\end{IEEEproof}

\section{Proof of Lemma~\ref{best_response_ordering_lemma}}
\label{best_response_ordering_lemma_appendix}

\emph{Lemma}~\ref{best_response_ordering_lemma}:
(1) Let $\Psi_\ell,\Psi_\ell^{o}\in\mathcal{A}_0$ be two thresholds of $\mathscr{F}_2$ such that $\Psi_\ell<\Psi_\ell^{o}$,
then the best response of $\mathscr{F}_1$ to these are ordered as, $BR_1(\Psi_\ell)\ge BR_1(\Psi_\ell^{o})$.
(2) Similarly, if $\Phi_\ell,\Phi_\ell^{o}\in\mathcal{A}_0$ are two thresholds of $\mathscr{F}_1$ such that $\Phi_\ell<\Phi_\ell^{o}$
then $BR_2(\Phi_\ell)\ge BR_2(\Phi_\ell^{o})$.

\begin{IEEEproof}
For convenience, first let us recall the expressions of the costs $C_{\st,g_\ell}^{(1)}(r_i)$
and $C_{\ct,g_\ell}^{(1)}$ from (\ref{PO_stopping_cost1_equn}) and (\ref{PO_continuing1_equn}) 
(since the given PO policy pair is 
$(\overline{\pi}_1^*,\overline{\pi}_2^*)$, these costs correspond to the Bayesian game 
$\mathcal{G}(\overline{\pi}_1^*,\overline{\pi}_2^*)$):
\begin{eqnarray}
\label{stopping_help_equn}
C_{\st,g_\ell}^{(1)}(r_i) &=& \widetilde{g}_\ell (-\eta_1 r_i) + 
(1-\widetilde{g}_\ell) E^{(1)}(r_i)\\
\label{continuing_help_equn}
C_{\ct,{g}_\ell}^{(1)} &=& \widetilde{g}_\ell \overline{C}^{(1)}_{\overline{\pi}_1^*,\overline{\pi}_2^*} + 
(1-\widetilde{g}_\ell) D^{(1)}.
\end{eqnarray}
Also, recall from (\ref{PO_costs_ordered_equn}) that the cost of continuing alone is 
less than the cost of continuing along with the competing forwarder, i.e.,
\begin{eqnarray}
\label{help_ineq_equn}
{D}^{(1)}\le\overline{C}^{(1)}_{\overline{\pi}_1^*,\overline{\pi}_2^*}
\end{eqnarray}

We will only prove Part-(1); the proof of Part-(2) is similar.
Let $g_\ell$ and $g_\ell^{o}$ be the threshold vectors of $\mathscr{F}_2$ whose corresponding 
thresholds are $\Psi_\ell$ and $\Psi_\ell^{o}$, respectively. 
Given that $\Psi_\ell<\Psi_\ell^{o}$, to prove $BR_1(\Psi_\ell)\ge BR_1(\Psi_\ell^{o})$
it is sufficient to show that, for any $r_i$, $C_{\st,g_\ell}^{(1)}(r_i)\le C_{\ct,g_\ell}^{(1)}$ implies
$C_{\st,g_\ell^{o}}^{(1)}(r_i)\le C_{\ct,g_\ell^{o}}^{(1)}$. 

Let us begin with an $r_i$ such that $C_{\st,g_\ell}^{(1)}(r_i)\le C_{\ct,g_\ell}^{(1)}$, or alternatively, 
(recall (\ref{stopping_help_equn}) and (\ref{continuing_help_equn})) $r_i$ is such that,
\begin{eqnarray*}
\widetilde{g}_\ell (-\eta_1 r_i) + (1-\widetilde{g}_\ell) E^{(1)}(r_i)
&\le& \widetilde{g}_\ell \overline{C}^{(1)}_{\overline{\pi}_1^*,\overline{\pi}_2^*} + 
(1-\widetilde{g}_\ell) D^{(1)}. 
\end{eqnarray*}
Substituting $E^{(1)}(r_i)=\nu_1 (-\eta_1 r_i) + \nu_2 D^{(1)}$ in the above expression, and 
then simplifying we obtain, 
\begin{eqnarray*}
-\eta_1 r_i &\le& \frac{\widetilde{g}_\ell \overline{C}^{(1)}_{\overline{\pi}_1^*,\overline{\pi}_2^*} +
(1-\widetilde{g}_\ell) \nu_1 D^{(1)}}
{\widetilde{g}_\ell + (1-\widetilde{g}_\ell) \nu_1}
\end{eqnarray*}
Thus $-\eta_1 r_i$, being less than the convex combination of the costs 
$\overline{C}^{(1)}_{\overline{\pi}_1^*,\overline{\pi}_2^*}$ and $D^{(1)}$, is less than
both of these. Further since 
${D}^{(1)}\le\overline{C}^{(1)}_{\overline{\pi}_1^*,\overline{\pi}_2^*}$, there are only
two cases which are possible: 
$-\eta_1 r_i<D^{(1)}\le \overline{C}^{(1)}_{\overline{\pi}_1^*,\overline{\pi}_2^*}$, and
$D^{(1)}\le -\eta_1 r_i \le \overline{C}^{(1)}_{\overline{\pi}_1^*,\overline{\pi}_2^*}$.
We will consider these two cases separately below.

\textbf{\emph{Case-1:}} Suppose 
$-\eta_1 r_i<D^{(1)}\le 
\overline{C}^{(1)}_{\overline{\pi}_1^*,\overline{\pi}_2^*}$,
then
\begin{eqnarray*}
E^{(1)}(r_i) 
&=& \nu_1 (-\eta_1 r_i) + \nu_2 D^{(1)}\nonumber\\
&\le& D^{(1)}.
\end{eqnarray*}
Using the above two inequalities in the expression of $C_{\st,g_\ell^{o}}^{(1)}(r_i)$, and then comparing
with $C_{\ct,g_\ell^{o}}^{(1)}$ we obtain $C_{\st,g_\ell^{o}}^{(1)}(r_i)\le C_{\ct,g_\ell^{o}}^{(1)}$.\\

\textbf{\emph{Case-2:}} 
Suppose $D^{(1)}\le -\eta_1 r_i \le \overline{C}^{(1)}_{\overline{\pi}_1^*,\overline{\pi}_2^*}$. Then we have
$E^{(1)}(r_i)\ge D^{(1)}$.
Define $\kappa(p)$ for $p\in[0,1]$ as,
\begin{eqnarray}
\label{kappa_equn}
\kappa(p)&=& p\Big(-\eta_1 r_i-E^{(1)}(r_i) -\overline{C}^{(1)}_{\overline{\pi}_1^*,\overline{\pi}_2^*}+D^{(1)}\Big)
+ \Big(E^{(1)}(r_i)-D^{(1)}\Big).
\end{eqnarray}
Since $-\eta_1 r_i \le \overline{C}^{(1)}_{\overline{\pi}_1^*,\overline{\pi}_2^*}$ and $E^{(1)}(r_i)\ge D^{(1)}$ we have, 
$\kappa(p)$ is decreasing in $p$.
Hence we can write $\kappa(\widetilde{g}_\ell^{o})\le \kappa(\widetilde{g}_\ell)$ because,
with $\Psi_\ell<\Psi_\ell^{o}$ we have,
\begin{eqnarray}
\widetilde{g}_\ell = \sum_{j=1}^{\Psi_\ell} p^{(2)}_{j|\ell} \le 
\sum_{j=1}^{\Psi_\ell^{o}} p^{(2)}_{j|\ell} = \widetilde{g}_\ell^{o}.
\end{eqnarray}
Finally, rearranging the terms in (\ref{kappa_equn}) one can obtain,
\begin{eqnarray*}
C_{\st,g_\ell^{o}}^{(1)}(r_i)- C_{\ct,g_\ell^{o}}^{(1)} 
&=& \kappa(\widetilde{g}_\ell^{o}) \\
&\le& \kappa(\widetilde{g}_\ell) \\
&=&  C_{\st,g_\ell}^{(1)}(r_i)- C_{\ct,g_\ell}^{(1)} \\
&\overset{*}{\le}& 0,
\end{eqnarray*}
where $*$ is because we started with an $r_i$ such that,  $C_{\st,g_\ell}^{(1)}(r_i)\le C_{\ct,g_\ell}^{(1)}$.
\end{IEEEproof}

\section{Obtaining Cooperative Optimal Policy Pair}
\label{cooperative_case_section_appendix}
We will first prove Lemma~\ref{pareto_lemma}.

\emph{Lemma}~\ref{pareto_lemma}:
The policy pair $(\pi_1^\gamma,\pi_2^\gamma)$ is \emph{Pareto optimal}, i.e.,
for any other policy $(\pi_1,\pi_2)$,
\begin{itemize}
\item[(1)] if $C^{(1)}_{\pi_1,\pi_2}<C^{(1)}_{\pi_1^\gamma,\pi_2^\gamma}$ then 
$C^{(2)}_{\pi_1^\gamma,\pi_2^\gamma}<C^{(2)}_{\pi_1,\pi_2}$, and
\item[(2)] if $C^{(2)}_{\pi_1,\pi_2}<C^{(2)}_{\pi_1^\gamma,\pi_2^\gamma}$ then 
$C^{(1)}_{\pi_1^\gamma,\pi_2^\gamma}<C^{(1)}_{\pi_1,\pi_2}$.
\end{itemize}
\begin{IEEEproof}
We will prove Part-(1); the proof of Part-(2) is similar.
Since $(\pi_1^\gamma,\pi_2^\gamma)$ is optimal for the problem in (\ref{cooperative_problem_equn})
we can write, for any policy pair $(\pi_1,\pi_2)$,
\begin{eqnarray*}
\gamma C^{(1)}_{\pi_1^\gamma,\pi_2^\gamma} + (1-\gamma) C^{(2)}_{\pi_1^\gamma,\pi_2^\gamma} &\le&
\gamma C^{(1)}_{\pi_1,\pi_2} + (1-\gamma) C^{(2)}_{\pi_1,\pi_2},
\end{eqnarray*}
rewriting which we obtain
\begin{eqnarray*}
C^{(2)}_{\pi_1^\gamma,\pi_2^\gamma} &\le&  \frac{\gamma}{1-\gamma}
\Big(C^{(1)}_{\pi_1,\pi_2} -C^{(1)}_{\pi_1^\gamma,\pi_2^\gamma}\Big) +  C^{(2)}_{\pi_1,\pi_2}.
\end{eqnarray*}
Now, if $C^{(1)}_{\pi_1,\pi_2}<C^{(1)}_{\pi_1^\gamma,\pi_2^\gamma}$ then from the above expression we have
$C^{(2)}_{\pi_1^\gamma,\pi_2^\gamma}<C^{(2)}_{\pi_1,\pi_2}$.
\end{IEEEproof}

We now proceed to obtain $(C^{(1)}_{\pi_1^\gamma,\pi_2^\gamma},C^{(2)}_{\pi_1^\gamma,\pi_2^\gamma})$
by formulating the problem in (\ref{cooperative_problem_equn}) as an MDP.
The state space, action space and the state transitions remain same 
as in Section~\ref{completely_observable_case_section}. However the one-step costs have 
to be appropriately modified to take into account the multiplier $\gamma$. Without writing
down all the details
we will proceed to the Bellman equation. The one-step costs will be evident
from these.

For states of the form $(r_i,\ts)$
\begin{eqnarray}
\label{coop_bellman_ri_equn}
J^*(r_i,\ts)
&=&\min\Big\{-\gamma \eta_1 r_i, \gamma \tau + \sum_{i'} p^{(1)}_{i'} J^*(r_i',\ts)\Big\}. 
\end{eqnarray}
The first term in the $\min$ expression above corresponds to the cost of the joint-action 
$(\textsf{s},\textsf{s})$ or $(\textsf{s},\textsf{c})$ (since $\mathscr{F}_2$ has terminated, its action
is irrelavent), and the second term is the expected cost of choosing 
$(\textsf{c},\textsf{s})$ (or $(\textsf{c},\textsf{c})$). Similarly, when the 
state is $(\ts,r_j)$ we can write
\begin{eqnarray}
\label{coop_bellman_rj_equn}
J^*(\ts,r_j)
&=& \min\Big\{-(1-\gamma) \eta_2 r_j, (1-\gamma)\tau + \sum_{j'} p^{(2)}_{j'} J^*(\ts,r_j')\Big\}. 
\end{eqnarray}

The more interesting case is when both forwarders are still competing, i.e., when the state is of the form
$(r_i,r_j)$, where the optimality equation is
\begin{eqnarray}
\label{coop_bellman_equn}
J^*(r_i,r_j)=\min\Big\{C_{\textsf{s},\textsf{c}},C_{\textsf{c},\textsf{s}},
C_{\textsf{s},\textsf{s}}, C_{\textsf{c},\textsf{c}} \Big\};
\end{eqnarray}
$C_{a_1,a_2}$ is the expected cost (one-step $+$ future cost-to-go) of choosing 
the joint-action $(a_1,a_2)$. When the joint-action chosen is $(\textsf{s},\textsf{c})$,
since $\mathscr{F}_1$ stops and $\mathscr{F}_2$ continues the one-step cost is 
$\Big(-\gamma \eta_1 r_i + (1-\gamma)\tau\Big)$. The subsequent state is of the form $(\ts,r_{j'})$ 
w.p.\ $p^{(2)}_{j'}$. Hence the expression for $C_{\textsf{s},\textsf{c}}$ can be written as
\begin{eqnarray}
C_{\textsf{s},\textsf{c}} &=& -\gamma \eta_1 r_i + (1-\gamma)\tau + 
\sum_{j'}p^{(2)}_{j'} J^*(\ts,r_{j'}).
\end{eqnarray}
Similarly $C_{\textsf{c},\textsf{s}}$ can be written as
\begin{eqnarray}
C_{\textsf{c},\textsf{s}} &=& \gamma \tau - (1-\gamma) \eta_2 r_j +
\sum_{i'} p^{(1)}_{i'} J^*(r_{i'},\ts).
\end{eqnarray}

When both forwarders decide to stop then w.p.\ $\nu_1$, $\mathscr{F}_1$ gets the relay
in which case the cost incurred is $C_{\textsf{s},\textsf{c}}$; otherwise, w.p.\ 
$\nu_2$, $\mathscr{F}_2$ get the relay and the cost incurred is $C_{\textsf{c},\textsf{s}}$. Hence
\begin{eqnarray}
\label{css_cost_equn}
C_{\textsf{s},\textsf{s}} &=& \nu_1 C_{\textsf{s},\textsf{c}} + 
\nu_2 C_{\textsf{c},\textsf{s}}.
\end{eqnarray}

Finally, when both forwarders continue the one-step cost is simply 
$\Big(\gamma \tau + (1-\gamma)\tau\Big)=\tau$ and the subsequent state is still of the form 
$(r_{i'},r_{j'})$. Thus we can write
\begin{eqnarray}
C_{\textsf{c},\textsf{c}} &=&  \tau  +\sum_{i',j'} p_{i',j'} J^*(r_{i'},r_{j'}).
\end{eqnarray}

From (\ref{css_cost_equn}) note that 
$C_{\textsf{s},\textsf{s}}\ge\min\{C_{\textsf{s},\textsf{c}},C_{\textsf{c},\textsf{s}}\}$
which means that the joint-action $(\textsf{s},\textsf{s})$ can never be optimal.
Thus, under cooperation the forwarders never compete for a relay;
either $\mathscr{F}_1$ will choose the relay, or $\mathscr{F}_2$ will choose, or both continue. 
Expression (\ref{coop_bellman_equn}) can therefore be simplified as
\begin{eqnarray}
\label{coop_bellman_simplified_equn}
J^*(r_i,r_j)=\min\Big\{C_{\textsf{s},\textsf{c}},C_{\textsf{c},\textsf{s}},
C_{\textsf{c},\textsf{c}} \Big\}.
\end{eqnarray}

One can perform value iteration to solve for $J^*$ in
(\ref{coop_bellman_ri_equn}), (\ref{coop_bellman_rj_equn}) and
(\ref{coop_bellman_simplified_equn}). Given $J^*$ it is easy to obtain the 
optimal policy $(\pi_1^\gamma,\pi_2^\gamma)$ (simply choose the joint-action that
minimizes the RHS of these expressions, breaking ties arbitrarily) and then the 
cost pair, $(C^{(1)}_{\pi_1^\gamma,\pi_2^\gamma},
C^{(2)}_{\pi_1^\gamma,\pi_2^\gamma})$.

\end{document}